\documentclass[twocolumn, times]{aastex63}

\hypersetup{linkcolor=teal, citecolor=teal, filecolor=teal, urlcolor=teal}

\newcommand{\mgii}{\ion{Mg}{2}}
\newcommand{\ciii}{\ion{C}{3}\rbrack}
\newcommand{\civ}{\ion{C}{4}}

\newcommand{\oh}{OH}
\newcommand{\oiii}{\lbrack\ion{O}{3}\rbrack}

\newcommand{\co}{CO}

\newcommand{\cimu}{\lbrack C\ensuremath{\,\textsc{i}}\rbrack\,369\,$\mu$m}
\newcommand{\ciimu}{\lbrack C\ensuremath{\,\textsc{ii}}\rbrack\,158\,$\mu$m}
\newcommand{\oimu}{\lbrack O\ensuremath{\,\textsc{i}}\rbrack\,146\,$\mu$m}
\newcommand{\oiiimu}{\lbrack O\ensuremath{\,\textsc{iii}}\rbrack\,88\,$\mu$m}
\newcommand{\lya}{Ly\ensuremath{\alpha}}
\newcommand{\hi}{H\ensuremath{\,\textsc{i}}}

\newcommand{\ha}{H\ensuremath{\alpha}}
\newcommand{\hb}{H\ensuremath{\beta}}

\newcommand{\xs}{X--shooter}

\def\lsim{\mathrel{\rlap{\lower 3pt \hbox{$\sim$}} \raise 2.0pt \hbox{$<$}}}
\def\gsim{\mathrel{\rlap{\lower 3pt \hbox{$\sim$}} \raise 2.0pt \hbox{$>$}}}

\def\lsun{L_{\rm \odot}}
\def\msun{M_{\rm \odot}}
\def\msunyr{M_{\rm \odot}\,{\rm yr}^{-1}}
\def\kms{{\rm km}\,{\rm s}^{-1}}

\def\ergs{{\rm erg}\,{\rm s}^{-1}}
\def\ergscm{{\rm erg}\,{\rm s}^{-1}\,{\rm cm}^{-2}}

\def\magnitude{\textrm{mag}}

\def\Lbol{L_{\mathrm{bol}}}
\def\Ledd{L_{\mathrm{Edd}}}
\def\mbh{M_{\mathrm{BH}}}
\def\mdyn{M_{\mathrm{Dyn}}}
\def\MMCDL{M_{\rm 1450}}
\def\lrest{\lambda_\mathrm{rest}}
\def\Eratio{\lambda_{\mathrm{Edd}}}

\def\NQSO{38}
\def\NQSOLIT{97}

\usepackage{graphicx}
\usepackage{natbib}
\usepackage{relsize}
\usepackage{textcomp}
\usepackage{upgreek}
\usepackage{amsmath}
\usepackage{amssymb}
\usepackage{calrsfs}
\usepackage{afterpage}
\usepackage{float}
\usepackage{makecell}
\usepackage[T1]{fontenc}

\setcitestyle{notesep={}}

\defcitealias{Schindler2020}{Paper~I}

\defcitealias{Banados2014}{B14}
\defcitealias{Banados2016}{B16}
\defcitealias{Banados2018}{B18}
\defcitealias{Banados2019Pisco}{B19}
\defcitealias{Chehade2018}{C18}
\defcitealias{Decarli2018}{D18}
\defcitealias{DeRosa2011}{D11}
\defcitealias{Eilers2020}{E20}
\defcitealias{Fan2000}{F00}
\defcitealias{Fan2001}{F01}
\defcitealias{Farina2019}{F19}
\defcitealias{Jiang2008}{J08}
\defcitealias{Jiang2015}{J15}
\defcitealias{Jiang2016}{J16}
\defcitealias{Mazzucchelli2017}{M17}
\defcitealias{Mortlock2009}{M09}
\defcitealias{Mortlock2011}{M11}
\defcitealias{Venemans2013}{V13}
\defcitealias{Venemans2015}{V15}
\defcitealias{Venemans2019}{V19}
\defcitealias{Venemans2020}{V20}
\defcitealias{WangFeige2016}{W16}
\defcitealias{WangFeige2017}{W17}
\defcitealias{WangRan2013}{W13}
\defcitealias{Willott2007}{W07}
\defcitealias{Willott2010discovery}{W10}
\defcitealias{Willott2015}{W15}
\defcitealias{Wu2015}{W15}

\defcitealias{York2000}{SDSS}

\received{June 8, 2022}
\revised{September 8, 2022}
\accepted{September 14, 2022}

\submitjournal{AJ}

\shorttitle{Supermassive Black Holes at the Epoch of Reionization. II.}

\shortauthors{Farina et al.}

\begin{document}

\title{The X--shooter/ALMA Sample of Quasars in the Epoch of Reionization. II.\\
Black Hole Masses, Eddington Ratios, and the Formation of the First Quasars}

\correspondingauthor{Emanuele Paolo Farina}
\email{emanuele.paolo.farina@gmail.com}


\author[0000-0002-6822-2254]{Emanuele Paolo Farina}
\affiliation{Gemini Observatory, NSF's NOIRLab, 670 N A'ohoku Place, HI-96720, Hilo, USA}
\affiliation{Max Planck Institut f\"ur Astronomie, K\"onigstuhl 17, D-69117, Heidelberg, Germany}
\affiliation{Max Planck Institut f\"ur Astrophysik, Karl--Schwarzschild--Stra{\ss}e 1, D-85748, Garching bei M\"unchen, Germany}

\author[0000-0002-4544-8242]{Jan--Torge Schindler}
\affiliation{Max Planck Institut f\"ur Astronomie, K\"onigstuhl 17, D-69117, Heidelberg, Germany}
\affiliation{Leiden Observatory, Leiden University, Niels Bohrweg 2, NL-2333 CA Leiden, Netherlands}


\author[0000-0003-4793-7880]{Fabian Walter}
\affiliation{Max Planck Institut f\"ur Astronomie, K\"onigstuhl 17, D-69117, Heidelberg, Germany}


\author[0000-0002-2931-7824]{Eduardo Ba{\~n}ados} 
\affiliation{Max Planck Institut f\"ur Astronomie, K\"onigstuhl 17, D-69117, Heidelberg, Germany}

\author[0000-0003-0821-3644]{Frederick B.\ Davies}
\affiliation{Max Planck Institut f\"ur Astronomie, K\"onigstuhl 17, D-69117, Heidelberg, Germany}

\author[0000-0002-2662-8803]{Roberto Decarli}
\affiliation{INAF --- Osservatorio di Astrofisica e Scienza dello Spazio di Bologna, via Gobetti 93/3, I-40129, Bologna, Italy}

\author[0000-0003-2895-6218]{Anna-Christina Eilers}\thanks{NASA Hubble Fellow}
\affiliation{MIT Kavli Institute for Astrophysics and Space Research, 77 Massachusetts Ave., Cambridge, MA 02139, USA}

\author[0000-0003-3310-0131]{Xiaohui Fan}
\affiliation{Steward Observatory, University of Arizona, 933 N Cherry Ave, Tucson, AZ 85721, USA}

\author[0000-0002-7054-4332]{Joseph F.\ Hennawi}
\affiliation{Department of Physics, University of California, Santa Barbara, CA 93106-9530, USA}
\affiliation{Leiden Observatory, Leiden University, Niels Bohrweg 2, NL-2333 CA Leiden, Netherlands}

\author[0000-0002-5941-5214]{Chiara Mazzucchelli}
\affiliation{European Southern Observatory, Alonso de C\'ordova 3107, Vitacura, Regi\'on Metropolitana, Chile}
\affiliation{N\'ucleo de Astronom\'ia de la Facultad de Ingenier\'ia, Universidad Diego Portales, Av. Ej\'ercito Libertador 441, Santiago, Chile}

\author[0000-0001-5492-4522]{Romain A. Meyer}
\affiliation{Max Planck Institut f\"ur Astronomie, K\"onigstuhl 17, D-69117, Heidelberg, Germany}

\author[0000-0002-3683-7297]{Benny Trakhtenbrot}
\affiliation{School of Physics and Astronomy, Tel Aviv University, Tel Aviv, 69978, Israel}

\author[0000-0002-3216-1322]{Marta Volonteri}
\affiliation{Institut d'Astrophysique de Paris, Sorbonne Universit{\'e}, CNRS, UMR 7095, 98 bis bd Arago, F-75014, Paris, France}

\author[0000-0002-7633-431X]{Feige Wang}\thanks{NASA Hubble Fellow}
\affiliation{Steward Observatory, University of Arizona, 933 N Cherry Ave, Tucson, AZ 85721, USA}

\author[0000-0003-0960-3580]{G\'abor Worseck}
\affiliation{Institut f\"ur Physik und Astronomie, Universit\"at Potsdam, Karl-Liebknecht-Str. 24/25, D-14476 Potsdam, Germany}

\author[0000-0001-5287-4242]{Jinyi Yang}\thanks{Strittmatter Fellow}
\affiliation{Steward Observatory, University of Arizona, 933 N Cherry Ave, Tucson, AZ 85721, USA}


\author[0000-0001-6179-7701]{Thales A.\ Gutcke}\thanks{NASA Hubble Fellow}
\affiliation{Department of Astrophysical Sciences, Princeton University, Peyton Hall, Princeton, NJ 08544, USA}

\author[0000-0001-9024-8322]{Bram P.\ Venemans}
\affiliation{Leiden Observatory, Leiden University, Niels Bohrweg 2, NL-2333 CA Leiden, Netherlands}


\author[0000-0001-8582-7012]{Sarah E.\ I.\ Bosman}
\affiliation{Max Planck Institut f\"ur Astronomie, K\"onigstuhl 17, D-69117, Heidelberg, Germany}

\author[0000-0002-6748-2900]{Tiago Costa}
\affiliation{Max Planck Institut f\"ur Astrophysik, Karl--Schwarzschild--Stra{\ss}e 1, D-85748, Garching bei M\"unchen, Germany}

\author[0000-0003-3242-7052]{Gisella De Rosa}
\affiliation{Space Telescope Science Institute, 3700 San Martin Drive, Baltimore, MD 21218, USA}

\author[0000-0002-0174-3362]{Alyssa B.\ Drake}
\affiliation{Centre for Astrophysics Research, Department of Physics, Astronomy and Mathematics, University of Hertfordshire, Hatfield AL10 9AB, UK}
\affiliation{Max Planck Institut f\"ur Astronomie, K\"onigstuhl 17, D-69117, Heidelberg, Germany}

\author[0000-0003-2984-6803]{Masafusa Onoue}
\affiliation{Max Planck Institut f\"ur Astronomie, K\"onigstuhl 17, D-69117, Heidelberg, Germany}
\affiliation{Kavli Institute for Astronomy and Astrophysics, Peking University, Beijing 100871, China}
\affiliation{Kavli Institute for the Physics and Mathematics of the Universe (Kavli IPMU, WPI), The University of Tokyo, Chiba 277-8583, Japan}

\begin{abstract}
We present measurements of black hole masses and Eddington ratios ($\Eratio{}$) for a sample of \NQSO{} bright ($\MMCDL < -24.4$\,mag) quasars at $5.8\lesssim z\lesssim7.5$, derived from \textit{VLT}/\xs{} near--IR spectroscopy of their broad \civ{} and \mgii{} emission lines.
The black hole masses (on average $\mbh{}\sim4.6\times10^9\,\msun$) and accretion rates ($0.1 \lesssim \Eratio{} \lesssim 1.0$) are broadly consistent with that of similarly luminous $0.3 \lesssim z \lesssim 2.3$ quasars, but there is evidence for a mild increase in the Eddington ratio above $z \gtrsim 6$.
Combined with deep ALMA observations of the \ciimu{} line from the host galaxies and \textit{VLT}/MUSE investigations of the extended \lya{} halos, this study provides fundamental clues to models of the formation and growth of the first massive galaxies and black holes.
Compared to local scaling relations, $z \gtrsim 5.7$ black holes appear to be over--massive relative to their hosts, with accretion properties that do not change with host galaxy morphologies.
Assuming that the kinematics of the $T\sim10^{4}$\,K gas, traced by the extended \lya{} halos, are dominated by the gravitational potential of the dark matter halo, we observe a similar relation between black hole mass and circular velocity as reported for $z \sim 0$ galaxies.
These results paint a picture where the first supermassive black holes reside in massive halos at $z \gtrsim 6$ and lead the first stages of galaxy formation by rapidly growing in mass with a duty cycle of order unity.
The duty cycle needs to drastically drop towards lower redshifts, while the host galaxies continue forming stars at a rate of hundreds of solar masses per year, sustained by the large reservoirs of cool gas surrounding them.
\end{abstract}


\section{Introduction}\label{sec:intro}

%
One year after the first identification of a quasar at high--redshift \citep[3C 273 at $z=0.158$,][]{Schmidt1963}, it was already clear that the observed luminosities (and variability timescales) require the presence of a super--massive black hole (SMBH) located at the center of the galaxy \citep[][]{Salpeter1964, Zeldovich1964}.
This argument was refined a few years later by \citet{Lynden1969}.
It is now well established that virtually all massive galaxies in the (local) Universe host a SMBH \citep[e.g.,][]{Kormendy1995, Richstone1998, Kormendy2001, Kormendy2013}.
If, during an active phase, a fraction of their energy and/or momentum output couples with the gas in the galaxy, the SMBHs will affect both their own growth and the star--formation history of their hosts, at scales much larger than the black hole sphere of influence \citep[e.g.,][]{Silk1998, Dimatteo2005}.
This co--evolutionary scenario is supported by the observed correlations between the mass of the black hole and host properties such as mass/luminosity \citep[e.g.,][]{Dressler1989, McLure2002, Marconi2003, Haring2004, McConnell2013}, velocity dispersion \citep[e.g.,][]{Ferrarese2000, Gebhardt2000, Tremaine2002}, and even with the circular velocity or dynamical mass of the dark matter halo \citep[e.g.,][]{Ferrarese2002, Volonteri2011}. [The interested reader can refer to \citet{Kormendy2013} for a review on the different correlations and to \citet{Jahnke2011} for concerns on correlations vs. causation.]
Intriguingly, some of these correlations evolve with cosmic time \citep[but see, e.g.,][]{Shen2015, Suh2020}.
In particular, the mass ratios between black holes and host galaxies and that between black holes and bulges increase with redshift, potentially following different evolutionary paths \citep[e.g.,][]{Kormendy2013, Ding2020}.
Albeit biases in the study of flux--limited samples can contribute to the observed evolution \citep[e.g.,][]{Lauer2007, Schulze2011, Schulze2014, Izumi2021}, black holes at higher redshift are generally more massive with respect to their host than they are in the local Universe \citep[e.g.,][]{Walter2004, Peng2006, Merloni2010, Decarli2010Masses, Decarli2010Evolution, Bennert2011, Targett2012, Trakhtenbrot2015, Venemans2016, Willott2017, Pensabene2020, Neeleman2021}.
This suggests that, in the initial stages, black holes may have dominated the symbiotic evolution with their host galaxies \citep[see][ for a review on the topic]{Volonteri2012}.
A detailed characterization of the general physical properties of the first quasars is thus fundamental to shed light on massive galaxy formation in the early universe.
The last years have seen a surge in the number of quasars known at $z > 5.5$ \citep[e.g.,][]{Fan2001, Fan2003, Fan2006, Willott2007, Willott2010discovery, Mortlock2011, Venemans2007, Venemans2013, Wu2015, Jiang2016, Banados2016, Banados2019Pisco, Matsuoka2016, Matsuoka2018, Matsuoka2019, Matsuoka2021, Mazzucchelli2017, Reed2019, Yang2019z65, Pons2019, WangFeige2020z7, WangFeige2021, Onoue2021, Wenzl2021, Wagenveld2022}.
This large sample opens a new era for statistical studies of the population of SMBHs emerging from the reionization epoch. 
Deep near--IR spectroscopy campaigns revealed that, already at \(z > 7\), luminous quasars are powered by massive black holes \citep[with \(\sim 10^{9}\,\msun{}\),][]{Banados2018, Yang2020z75, WangFeige2021} with metal--enriched (solar metallicity) broad line regions \citep[BLR, e.g.,][, see also \citealt{WangShu2021, Lai2022}]{Onoue2020, Yang2021}.
Furthermore, there is no apparent redshift evolution in the continuum slope or in the shape of low ionization broad emission lines at this redshift when comparing the composite spectra of typical SDSS quasars \citep[at \(0.04 < z < 4.79\),][]{VandenBerk2001} to bright $z \sim 6$ quasars \citep[][]{Shen2019, Yang2021}.
This suggests that, to first order, the same physical mechanisms are powering the quasar emission at different redshifts.
Yet, the observed increase in the average blue--shift of the \civ{}--\mgii{} velocity difference \citep[e.g.,][]{Venemans2016, Mazzucchelli2017, Meyer2019, Schindler2020, Yang2021} and in the fraction of broad absorption (BAL) line \citep[e.g.,][]{Yang2021, Bischetti2022} and weak line quasars \citep[WLQ, e.g.,][]{Shen2019} may imply, barring selection effects, a change in the BLR's physical status and/or that strong outflows may have a more relevant role towards the highest redshifts.
Despite our expansion in knowledge about the SMBHs powering the first quasars, still little is known about the rest--frame UV emission of their host galaxies.
Indeed, given the extreme brightness of the quasar emission at these wavelengths, the detection of the stellar component has proven to be prohibitive with large aperture ground--based telescopes \citep[e.g.,][]{Targett2012} and even with HST \citep[e.g.,][]{Mechtley2012, Marshall2020}.
On the other hand, the inter--stellar medium (ISM) of tens of quasar host galaxies has been detected at mm--wavelengths in the dust continuum and in the \ciimu{} emission line \citep[e.g.,][]{WangRan2016, WangRan2019, Decarli2017, Decarli2018, Venemans2018, WangFeige2019, Rojasruiz2021, Khusanova2022} and, in some cases, also in the \oiiimu{}, \oimu{}, \cimu{}, \co{}, and water lines \citep[e.g.,][]{WangRan2011a, WangRan2011b, WangRan2016, Venemans2017CO, Walter2018, WangFeige2019, Shao2019, Novak2019, Yang2019MultiLine, Pensabene2021, Meyer2022, Decarli2022, Li2022, Pensabene2022}.
In particular, ALMA \ciimu{} observations at a resolution of $\lesssim 1$\,kpc are providing detailed information on star--formation rates, the presence of molecular outflows, dynamical masses, and morphology of early massive galaxies hosting SMBHs \citep[][]{Venemans2019, Venemans2020, Novak2020, Neeleman2021, Walter2022}. 
To investigate the link between SMBHs and their hosts at $z \gtrsim 6$, we thus started a \textit{VLT}/\xs{} program targeting quasars already observed with ALMA.
The first results from this program were presented in \citet[][, hereafter \citetalias{Schindler2020}]{Schindler2020}, in which we introduced the sample of \NQSO{} \(5.78<z<7.54\) quasars and analyzed the properties of their broad emission lines.
To summarize, the main results of \citetalias{Schindler2020} are:
\textit{(i)} the broad line region of early quasars is already enriched, with iron at levels comparable to lower redshift systems; 
\textit{(ii)} both the \mgii{} and the \civ{} emission lines are blue--shifted with respect to the host galaxy systemic redshift (traced by the \ciimu{} emission line); 
\textit{(iii)} these \mgii{} and the \civ{} blue--shifts strongly correlate with each other, implying a common physical origin of the velocity displacement; and
\textit{(iv)} the median velocity blue--shift between \civ{} and \mgii{} broad lines is increased by a factor $\sim2.3\times$ with respect to similarly luminous quasars at lower--redshifts.
In this paper, we will focus on black hole masses and accretion rates and connect these with the host galaxy properties \citep[see also,][]{Neeleman2021}.
In addition, we will link the characteristics of the first SMBHs with their large scale ($\gtrsim 10$\,kpc) cool gas reservoirs probed by \textit{VLT}/MUSE observations of the extended \lya{} nebulae surrounding them \citep[namely, the \textit{REQUIEM} survey, ][]{Farina2019}.
These halos show quiescent kinematics and contain a sufficient supply of cool gas ($>10^{9}\,\msun{}$) to fuel the future star formation and black hole mass growth \citep[e.g.,][]{Farina2019, Drake2022}.
By linking the SMBHs with the ISM properties of their hosts and with the cool gas in the circum--galactic medium (CGM), this study aims to provide a comprehensive view of massive galaxy formation at the epoch of reionization.
Throughout this paper, we assume a concordance cos\-mo\-lo\-gy with $H_0$=70\,km\,s$^{-1}$\,Mpc$^{-1}$, $\Omega_{\rm M}$=0.3, and \mbox{$\Omega_\Lambda$=1-$\Omega_{\rm M}$=0.7}, in broad agreement with the Planck cosmological parameters \citep{Planck2020}.
In this cosmology, at $z=6.3$ (the average redshift of our sample) the Universe is 0.859\,Gyr old, and an angular scale $\theta=1$\arcsec\ corresponds to a proper transverse separation of~5.6\,kpc.
All quoted magnitudes are in the standard AB photometric system \citep{Oke1974, Oke1983}.
%

\section{Sample}\label{sec:sample}

\begin{figure}[tb]
    \centering
    \includegraphics[width=0.98\columnwidth]{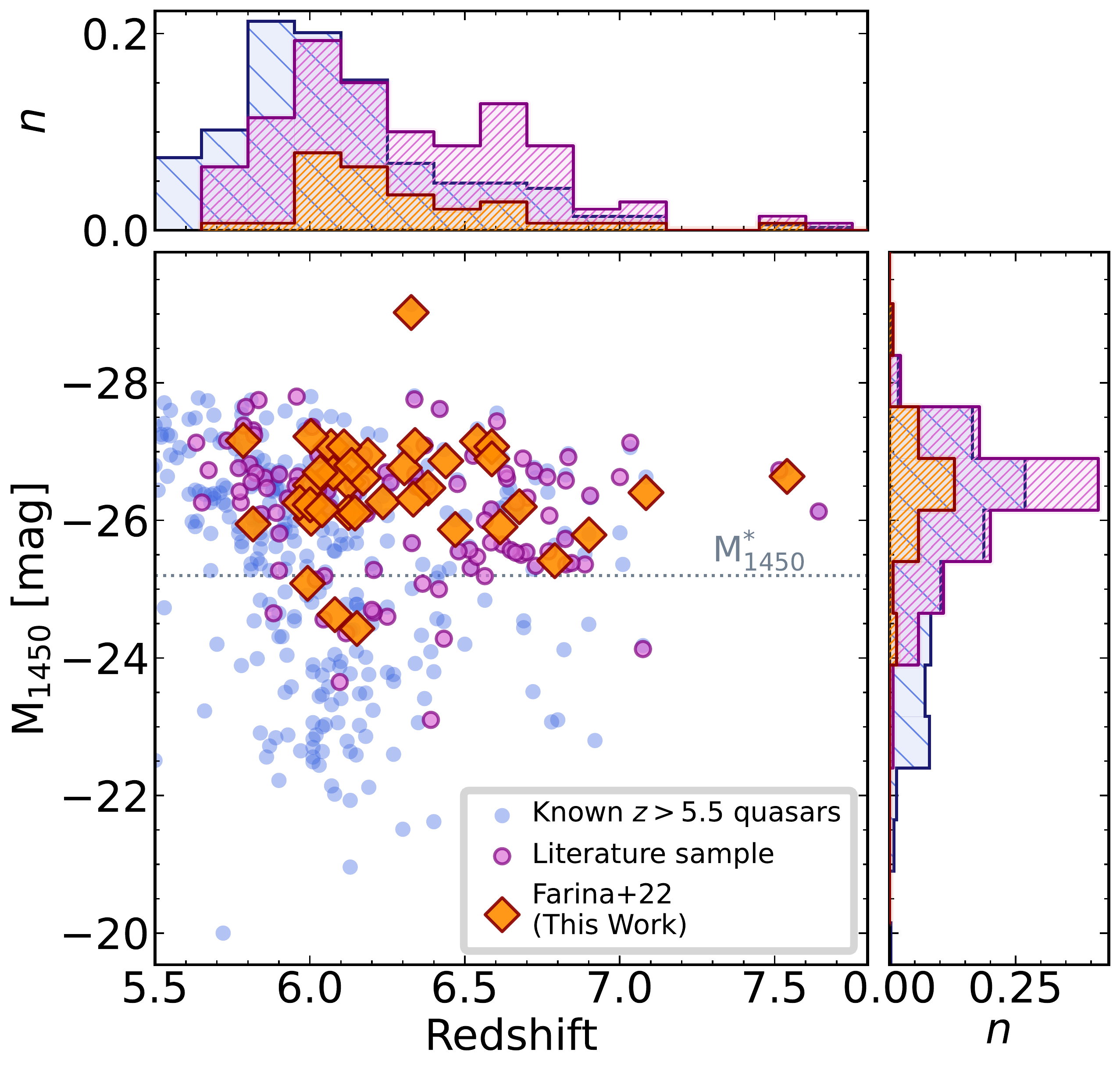}
    \caption{
    Distribution, in the redshift versus absolute magnitude plane, of the \NQSO{} $z\gtrsim5.7$ quasars in our sample (orange diamonds) and of the \NQSOLIT{} quasars for which black hole masses and Eddington ratios have been presented in the literature (violet circles, see \autoref{sec:results} for details).
    For comparison, pale blue circles indicate all high--redshift quasars published at the time of writing (May 2022).
    Data for the quasar J0439$+$1634 (\(z=6.51\)), that is lensed by a foreground galaxy, has been corrected by the magnification factor of \(\sim51\) calculated by \citet{Fan2019}. 
    The horizontal dotted line marks the characteristic magnitude of $z\gtrsim6$ quasars ($\MMCDL^\star\sim-25.2\,\magnitude$).
    Histograms on the top and right panel follow the same color scheme as the central panel and are normalized by the total number of objects (histograms of the quasars part of our sample and from the literature are stacked one above the other with normalization given by the total \NQSO{}+\NQSOLIT{} number of objects).
    Most of the targets from our sample and from the literature have $\MMCDL{}\lesssim-26$\,mag.
    Thus we are probing the bright end of the underlying $z\gtrsim6$ quasar population.
    }
    \label{fig:sample}
\end{figure}

%
Our sample consists of \NQSO{}~bright quasars in the redshift range $5.78<z<7.54$ with an average redshift of $\langle z \rangle=6.3$ (see \autoref{tab:sample} and \autoref{fig:sample}).
For all quasars in our sample we will report \textit{systemic} redshifts ($z_\textrm{sys}$), mainly derived from the \ciimu{} emission from the host galaxy.
Wherever this line has not been detected, we consider as \textit{systemic} the redshift of the extended \lya{} emission or of the \mgii{} line.
These are considered good tracers of systemic redshifts, with an average displacement between the \lya{} halo and the \ciimu{} redshifts of $\sim70\,\kms$ \citep[][]{Farina2019} and between the \mgii{} and the \oiii{} redshifts of roughly $-100\,\kms$ \citep[][]{Richards2002, Hewett2010}.
The absolute magnitudes of the considered quasars vary from $\MMCDL=-24.4\,\magnitude$ to $-29.0\,\magnitude$ (see \autoref{tab:sample} and \autoref{fig:sample}).
For comparison, current constraints on the $z\gtrsim6$ quasar luminosity function suggest a characteristic magnitude of $\MMCDL^\star\sim-25.2\,\magnitude$  \citep[e.g.,][, see \citealt{Kulkarni2019} and \citealt{Shen2020LF} for an extensive discussion on the evolution of the quasar luminosity function at high--redshift]{Willott2010discovery, Kashikawa2015, Jiang2016, Matsuoka2018LF, WangFeige2019LF}.
This is a factor of $\sim4\times$ fainter than the average absolute magnitude of our sample: $\langle\MMCDL\rangle=-26.8\,\magnitude$.
The considered targets are, thus, sampling the bright--end of the highest--redshift quasar population.
Eight out of the \NQSO{} objects in our sample are classified as BAL quasars \citep[i.e., P009$-$10, P065$-$26, J1044$-$0125, P239$-$07, J2211$-$3206, J2310$+$1855, J2318$-$3029, and J2348$-$3054, see:][; \citetalias{Schindler2020}]{Derosa2014, Shen2019, Eilers2020, Bischetti2022}.
The fraction of BAL quasar in our sample ($21^{+10}_{-7}$\,\%) is higher, but nevertheless consistent, with the incidence of such systems reported in lower redshift studies.
Indeed, the fraction of bright ($M_{i}\left[z=2\right]<-27$\,\magnitude{}, corresponding to $\MMCDL{}\lesssim25.3$\,\magnitude{}) BAL quasars in the 14$^\textrm{th}$ data release of the \citetalias{York2000} quasar catalog is $\sim12$\,\% \citep[][, see also, e.g.,  \citealt{Maddox2008, Knigge2008, Allen2011}]{Paris2018}. 

{\begin{deluxetable*}{llllCllCCll}
\rotate
\tabletypesize{\footnotesize} 
\tablecaption{Properties of the quasars observed with \xs\label{tab:sample}}
\tablehead{
\colhead{Quasar Name}                   &
\colhead{ID}                            &
\colhead{RA}                            &
\colhead{Dec}                           &
\colhead{$z_{\textrm{sys}}$}            &
\colhead{$z_{\textrm{sys}}$ Method}     &
\colhead{$z_{\textrm{sys}}$ Ref.}       &
\colhead{$J$}                           &
\colhead{$\MMCDL$}                      &
\colhead{Disc. Ref.}                    &
\colhead{Notes}                         \\
\colhead{}         &
\colhead{}         &
\colhead{(J2000)}  &
\colhead{(J2000)}  &
\colhead{}         &
\colhead{}         &
\colhead{}         &
\colhead{(mag)}    &
\colhead{(mag)}    &
\colhead{}         &
\colhead{}
}
\startdata
PSO~J004.3936$+$17.0862 & P004$+$17    & 00:17:34.467 & $+$17:05:10.70 & 5.8165\pm0.0023 & \ciimu & \citetalias{Eilers2020}       & 20.67\pm0.16 & -25.95_{+0.04}^{-0.05} & \citetalias{Banados2016}                &           \\
PSO~J007.0273$+$04.9571 & P007$+$04    & 00:28:06.560 & $+$04:57:25.68 & 6.0015\pm0.0002 & \ciimu & \citetalias{Venemans2020}     & 19.77\pm0.11 & -26.51_{+0.06}^{-0.05} & \citetalias{Banados2014, Jiang2015}     &  pDLA     \\
PSO~J009.7355$-$10.4316 & P009$-$10    & 00:38:56.522 & $-$10:25:53.90 & 6.0040\pm0.0003 & \ciimu & \citetalias{Venemans2020}     & 19.93\pm0.07 & -26.03_{+0.04}^{-0.04} & \citetalias{Banados2016}                &  BAL      \\
PSO~J011.3898$+$09.0324 & P011$+$09    & 00:45:33.568 & $+$09:01:56.96 & 6.4694\pm0.0025 & \ciimu & \citetalias{Eilers2020}       & 20.80\pm0.13 & -25.87_{+0.02}^{-0.02} & \citetalias{Mazzucchelli2017}           &           \\
VIK~J0046$-$2837        & J0046$-$2837 & 00:46:23.645 & $-$28:37:47.34 & 5.9926\pm0.0028 & \mgii  & \citetalias{Schindler2020}    & 20.96\pm0.09 & -25.09_{+0.24}^{-0.19} & \citetalias{Decarli2018}                &           \\
SDSS~J0100$+$2802       & J0100$+$2802 & 01:00:13.027 & $+$28:02:25.84 & 6.3269\pm0.0002 & \ciimu & \citetalias{Venemans2020}     & 17.64\pm0.02 & -29.02_{+0.00}^{-0.00} & \citetalias{Wu2015}                     &           \\
VIK~J0109$-$3047        & J0109$-$3047 & 01:09:53.131 & $-$30:47:26.31 & 6.7904\pm0.0003 & \ciimu & \citetalias{Venemans2020}     & 21.27\pm0.16 & -25.41_{+0.03}^{-0.03} & \citetalias{Venemans2013}               &           \\
PSO~J036.5078$+$03.0498 & P036$+$03    & 02:26:01.875 & $+$03:02:59.40 & 6.5405\pm0.0001 & \ciimu & \citetalias{Venemans2020}     & 19.51\pm0.03 & -27.15_{+0.01}^{-0.01} & \citetalias{Venemans2015}               &           \\
VIK~J0305$-$3150        & J0305$-$3150 & 03:05:16.916 & $-$31:50:55.90 & 6.6139\pm0.0001 & \ciimu & \citetalias{Venemans2019}     & 20.68\pm0.07 & -25.91_{+0.01}^{-0.01} & \citetalias{Venemans2013}               &           \\
PSO~J056.7168$-$16.4769 & P056$-$16    & 03:46:52.044 & $-$16:28:36.88 & 5.9670\pm0.0023 & \ciimu & \citetalias{Eilers2020}       & 20.25\pm0.10 & -26.26_{+0.02}^{-0.02} & \citetalias{Banados2016}                &  pDLA     \\
PSO~J065.4085$-$26.9543 & P065$-$26    & 04:21:38.049 & $-$26:57:15.61 & 6.1871\pm0.0003 & \ciimu & \citetalias{Venemans2020}     & 19.36\pm0.02 & -26.94_{+0.01}^{-0.01} & \citetalias{Banados2016}                &  pDLA     \\
PSO~J065.5041$-$19.4579 & P065$-$19    & 04:22:00.995 & $-$19:27:28.69 & 6.1247\pm0.0006 & \ciimu & \citetalias{Decarli2018}      & 19.90\pm0.15 & -26.11_{+0.03}^{-0.03} & \citetalias{Banados2016}                &  BAL      \\
SDSS~J0842$+$1218       & J0842$+$1218 & 08:42:29.430 & $+$12:18:50.50 & 6.0754\pm0.0005 & \ciimu & \citetalias{Venemans2020}     & 19.78\pm0.03 & -26.69_{+0.01}^{-0.01} & \citetalias{DeRosa2011, Jiang2015}      &  BAL?     \\
SDSS~J1030$+$0524       & J1030$+$0524 & 10:30:27.098 & $+$05:24:55.00 & 6.3048\pm0.0012 & \lya   & \citetalias{Farina2019}       & 19.79\pm0.08 & -26.76_{+0.02}^{-0.02} & \citetalias{Fan2001}                    &           \\
PSO~J158.69378$-$14.421 & P158$-$14    & 10:34:46.509 & $-$14:25:15.89 & 6.0681\pm0.0024 & \ciimu & \citetalias{Eilers2020}       & 19.19\pm0.06 & -27.07_{+0.03}^{-0.03} & \citetalias{Chehade2018}                &           \\
PSO~J159.2257$-$02.5438 & P159$-$02    & 10:36:54.190 & $-$02:32:37.94 & 6.3809\pm0.0005 & \ciimu & \citetalias{Decarli2018}      & 20.00\pm0.10 & -26.47_{+0.02}^{-0.02} & \citetalias{Banados2016}                &           \\
SDSS~J1044$-$0125       & J1044$-$0125 & 10:44:33.041 & $-$01:25:02.20 & 5.7846\pm0.0005 & \ciimu & \citetalias{Venemans2020}     & 19.25\pm0.05 & -27.16_{+0.03}^{-0.03} & \citetalias{Fan2000}                    &  BAL      \\
VIK~J1048$-$0109        & J1048$-$0109 & 10:48:19.082 & $-$01:09:40.29 & 6.6759\pm0.0002 & \ciimu & \citetalias{Venemans2020}     & 20.65\pm0.17 & -26.20_{+0.03}^{-0.03} & \citetalias{WangFeige2017}              &           \\
ULAS~J1120$+$0641       & J1120$+$0641 & 11:20:01.478 & $+$06:41:24.30 & 7.0848\pm0.0004 & \ciimu & \citetalias{Venemans2020}     & 20.36\pm0.05 & -26.40_{+0.00}^{-0.00} & \citetalias{Mortlock2011}               &           \\
ULAS~J1148$+$0702       & J1148$+$0702 & 11:48:03.286 & $+$07:02:08.33 & 6.3337\pm0.0028 & \mgii  & \citetalias{Schindler2020}    & 20.30\pm0.11 & -26.31_{+0.02}^{-0.01} & \citetalias{Jiang2016}                  &           \\
PSO~J183.1124$+$05.0926 & P183$+$05    & 12:12:26.984 & $+$05:05:33.49 & 6.4386\pm0.0002 & \ciimu & \citetalias{Venemans2020}     & 19.77\pm0.08 & -26.87_{+0.02}^{-0.01} & \citetalias{Mazzucchelli2017}           &  pDLA     \\
SDSS~J1306$+$0356       & J1306$+$0356 & 13:06:08.258 & $+$03:56:26.30 & 6.0330\pm0.0002 & \ciimu & \citetalias{Venemans2020}     & 19.71\pm0.10 & -26.70_{+0.01}^{-0.01} & \citetalias{Fan2001}                    &           \\
ULAS~J1319$+$0950       & J1319$+$0950 & 13:19:11.302 & $+$09:50:51.49 & 6.1347\pm0.0005 & \ciimu & \citetalias{Venemans2020}     & 19.70\pm0.03 & -26.80_{+0.00}^{-0.00} & \citetalias{Mortlock2009}               &           \\
ULAS~J1342$+$0928       & J1342$+$0928 & 13:42:08.105 & $+$09:28:38.61 & 7.5400\pm0.0003 & \ciimu & \citetalias{Banados2019Pisco} & 20.30\pm0.02 & -26.64_{+0.00}^{-0.00} & \citetalias{Banados2018}                &           \\
CFHQS~J1509$-$1749      & J1509$-$1749 & 15:09:41.779 & $-$17:49:26.80 & 6.1225\pm0.0007 & \ciimu & \citetalias{Decarli2018}      & 19.80\pm0.08 & -26.56_{+0.01}^{-0.01} & \citetalias{Willott2007}                &           \\
PSO~J231.6576$-$20.8335 & P231$-$20    & 15:26:37.838 & $-$20:50:00.66 & 6.5869\pm0.0004 & \ciimu & \citetalias{Venemans2020}     & 19.66\pm0.05 & -27.07_{+0.03}^{-0.03} & \citetalias{Mazzucchelli2017}           &  BAL?     \\
PSO~J239.7124$-$07.4026 & P239$-$07    & 15:58:50.991 & $-$07:24:09.59 & 6.1097\pm0.0024 & \ciimu & \citetalias{Eilers2020}       & 19.35\pm0.08 & -27.07_{+0.02}^{-0.01} & \citetalias{Banados2016}                &  BAL      \\
PSO~J308.0416$-$21.2339 & P308$-$21    & 20:32:09.994 & $-$21:14:02.31 & 6.2355\pm0.0003 & \ciimu & \citetalias{Venemans2020}     & 20.17\pm0.11 & -26.27_{+0.01}^{-0.01} & \citetalias{Banados2016}                &           \\
SDSS~J2054$-$0005       & J2054$-$0005 & 20:54:06.490 & $-$00:05:14.80 & 6.0389\pm0.0001 & \ciimu & \citetalias{Venemans2020}     & 20.12\pm0.06 & -26.15_{+0.05}^{-0.04} & \citetalias{Jiang2008}                  &           \\
CFHQS~J2100$-$1715      & J2100$-$1715 & 21:00:54.619 & $-$17:15:22.50 & 6.0807\pm0.0004 & \ciimu & \citetalias{Venemans2020}     & 21.42\pm0.10 & -24.63_{+0.05}^{-0.05} & \citetalias{Willott2010discovery}       &           \\
PSO~J323.1382$+$12.2986 & P323$+$12    & 21:32:33.189 & $+$12:17:55.26 & 6.5872\pm0.0004 & \ciimu & \citetalias{Venemans2020}     & 19.74\pm0.03 & -26.89_{+0.01}^{-0.01} & \citetalias{Mazzucchelli2017}           &           \\
VIK~J2211$-$3206        & J2211$-$3206 & 22:11:12.391 & $-$32:06:12.95 & 6.3394\pm0.0010 & \ciimu & \citetalias{Decarli2018}      & 19.62\pm0.03 & -27.09_{+0.03}^{-0.03} & \citetalias{Decarli2018}                &  BAL      \\
CFHQS~J2229$+$1457      & J2229$+$1457 & 22:29:01.649 & $+$14:57:09.00 & 6.1517\pm0.0005 & \ciimu & \citetalias{Willott2015}      & 21.95\pm0.07 & -24.43_{+0.08}^{-0.07} & \citetalias{Willott2010discovery}       &           \\
PSO~J340.2041$-$18.6621 & P340$-$18    & 22:40:49.001 & $-$18:39:43.81 & 6.0007\pm0.0020 & \lya   & \citetalias{Farina2019}       & 20.28\pm0.08 & -26.23_{+0.02}^{-0.02} & \citetalias{Banados2014}                &  BAL      \\
SDSS~J2310$+$1855       & J2310$+$1855 & 23:10:38.880 & $+$18:55:19.70 & 6.0031\pm0.0002 & \ciimu & \citetalias{WangRan2013}      & 18.88\pm0.05 & -27.22_{+0.02}^{-0.02} & \citetalias{WangRan2013, Jiang2016}     &  BAL/pDLA \\
VIK~J2318$-$3029        & J2318$-$3029 & 23:18:33.103 & $-$30:29:33.36 & 6.1456\pm0.0002 & \ciimu & \citetalias{Venemans2020}     & 20.20\pm0.06 & -26.11_{+0.02}^{-0.02} & \citetalias{Decarli2018}                &  BAL?     \\
VIK~J2348$-$3054        & J2348$-$3054 & 23:48:33.336 & $-$30:54:10.24 & 6.9007\pm0.0005 & \ciimu & \citetalias{Venemans2020}     & 21.14\pm0.08 & -25.79_{+0.03}^{-0.03} & \citetalias{Venemans2013}               &  BAL      \\
PSO~J359.1352$-$06.3831 & P359$-$06    & 23:56:32.452 & $-$06:22:59.26 & 6.1719\pm0.0002 & \ciimu & \citetalias{Venemans2020}     & 19.85\pm0.10 & -26.62_{+0.02}^{-0.02} & \citetalias{Banados2016, WangFeige2016} &           \\
\enddata 
\tablecomments{\scriptsize
References for $z_\textrm{sys}$ and discovery are:
\citetalias{Banados2014} -- \cite{Banados2014};
\citetalias{Banados2016} -- \cite{Banados2016};
\citetalias{Banados2018} -- \cite{Banados2018};
\citetalias{Banados2019Pisco} -- \cite{Banados2019Pisco};
\citetalias{Chehade2018} -- \cite{Chehade2018};
\citetalias{Decarli2018} -- \cite{Decarli2018};
\citetalias{DeRosa2011} -- \cite{DeRosa2011};
\citetalias{Eilers2020} -- \cite{Eilers2020};
\citetalias{Fan2000} -- \cite{Fan2000};
\citetalias{Fan2001} -- \cite{Fan2001};
\citetalias{Farina2019} -- \cite{Farina2019};
\citetalias{Jiang2008} -- \cite{Jiang2008};
\citetalias{Jiang2015} -- \cite{Jiang2015};
\citetalias{Jiang2016} -- \cite{Jiang2016};
\citetalias{Mazzucchelli2017} -- \cite{Mazzucchelli2017};
\citetalias{Mortlock2009} -- \cite{Mortlock2009};
\citetalias{Mortlock2011} -- \cite{Mortlock2011};
\citetalias{Venemans2013} -- \cite{Venemans2013};
\citetalias{Venemans2015} -- \cite{Venemans2015};
\citetalias{Venemans2019} -- \cite{Venemans2019};
\citetalias{Venemans2020} -- \cite{Venemans2020};
\citetalias{WangFeige2016} -- \cite{WangFeige2016};
\citetalias{WangFeige2017} -- \cite{WangFeige2017};
\citetalias{WangRan2013} -- \cite{WangRan2013};
\citetalias{Willott2007} -- \cite{Willott2007};
\citetalias{Willott2010discovery} -- \cite{Willott2010discovery};
\citetalias{Willott2015} -- \cite{Willott2015}; and
\mbox{\citetalias{Wu2015} -- \cite{Wu2015}}
}
\end{deluxetable*}
}

\section{Observations, Data Reduction, and Fitting Procedure}\label{sec:observations}

%
We refer the interested reader to \citetalias{Schindler2020} for an exhaustive description of the data collection and reduction, and of the quasar fitting procedure.
For the sake of completeness, we provide a brief summary of the followed steps in the next sections.
%

\subsection{Observations}

%
Near--IR spectra of the quasars in our sample were collected with the medium resolution spectrograph \xs{} \citep[][]{Vernet2011}, currently mounted on the Cassegrain focus of the ESO/\textit{VLT} Telescope \textit{Melipal}.
Observations were taken with the 0\farcs6, 0\farcs9, or 1\farcs2 slit (delivering a resolution $R=\lambda/\Delta\lambda\sim8100-4300$ in the near--IR) and following the typical ABBA or ABAB dither pattern.
Exposure times range from 40\,min to 22.3\,h per target, with a median of 2\,h \citepalias[see Table~1 in][ for further details]{Schindler2020}. 
The data was collected as part of the following observing programs: 60.A-9418(A, PI:~Ryan--Weber), 
084.A-0360(A, PI:~Hjorth), 
084.A-0390(A, PI:~Ryan--Weber), 
085.A-0299(A, PI:~D'Odorico), 
086.A-0162(A, PI:~D'Odorico), 
087.A-0890(A, PI:~De Rosa), 
088.A-0897(A, PI:~De Rosa), 
089.A-0814(A, PI:~Becker), 
091.C-0934(B, PI:~Kaper), 
093.A-0707(A, PI:~Becker), 
096.A-0095(A, PI:~Pettini), 
096.A-0418(A, PI:~Shanks),
096.A-0418(B, PI:~Shanks),
097.B-1070(A, PI:~Farina), 
098.B-0537(A, PI:~Farina),
0100.A-0625(A, PI:~D'Odorico),
0100.A-0898(A, PI:~Venemans), 
0101.B-0272(A, PI:~Eilers), 
0102.A-0154(A, PI:~D'Odorico), and 
286.A-5025(A, PI:~Venemans).
%

\subsection{Data Reduction}

%
The data reduction was performed with the open source software \textsc{PypeIt} \citep[][]{Prochaska2019pypeit1, Prochaska2019pypeit2, Prochaska2020pypeit3} following standard procedures for near--IR spectroscopy:
\begin{enumerate}
    \item single exposures are corrected for instrumental signatures (e.g., flat--field);
    \item the wavelength solution for each exposure is obtained by comparing the sky--spectrum with the location, in vacuum, of \oh{} \citep[from][]{Rousselot2000} and water lines\footnote{\url{https://hitran.org/}};
    \item cosmic rays are removed with the \textsc{L. A. Cosmic} algorithm \citep[][]{vanDokkum2001};
    \item sky subtraction is performed on the 2D images both by differencing the AB exposures and by iteratively removing residuals with a 2D b--spline fit that follows the curvature of the spectral features \citep[e.g.,][]{Kelson2003};
    \item the quasar's trace is automatically detected and optimally extracted \citep[][]{Horne1986};
    \item single 1D spectra are flux calibrated using \xs{} standard stars\footnote{\url{https://www.eso.org/sci/facilities/paranal/instruments/xshooter/tools/specphot_list.html}} \citep[][]{Moehler2014};
    \item all 1D flux--calibrated spectra of each object are combined, after re--scaling them to the same flux level;
    \item telluric absorption features are removed from the stacked spectrum using telluric grids created from the \textsc{Line--By--Line Radiative Transfer Model} \citep[LBLRTM4;][]{Clough2005, Clough2014} and a principal component analysis model of the quasar emission \citep[][]{Davies2018};    
    \item absolute flux calibration is reached by re--scaling the spectrum to the published $J$-- (or $K$--)band magnitude of the objects (see \autoref{tab:sample}).
\end{enumerate}
The reduced spectra have a median signal--to--noise per pixel in the $J$--band of $\textrm{S/N}\sim6$.
%

\subsection{Fit of the spectra}\label{sec:fit}

%
To characterize the \civ{} ($\lambda_\textrm{rest} = 1549.06$\,\AA{}) and \mgii{} ($\lambda_\textrm{rest} = 2798.75$\,\AA{})  broad emission line properties of our quasars, we followed the standard procedure of fitting the spectra of high--redshift quasars with a global continuum plus a model of the emission lines \citep[e.g.,][, \citetalias{Schindler2020}]{Derosa2014, Mazzucchelli2017, Shen2019, Onoue2020, Yang2021}.
The fitting procedure was handled using the \textsc{Python} package \textsc{Sculptor} \citep[]{Schindler2022} after binning each spectrum by a factor of 4 in wavelength.
%

\subsubsection{The quasar's continuum}
%
The global continuum is constructed considering the following components:
\begin{itemize}
    \item accretion disk related emission is reproduced with a power--law normalized at 2500\,\AA{} with slope $\alpha_{\lambda}$;
    \item the Balmer continuum at $\lrest<3646$\,\AA\ is modeled by assuming the presence of gas clouds with uniform electron temperature ($T_e= 1.5\times10^4$\,K) which are partially optically thick \citep[e.g.,][]{Dietrich2003} and by fixing the normalization to be 30\% of the power law flux at the Balmer edge \citep[see, e.g.,][ for further details]{Onoue2020};
    \item the iron pseudo--continuum beneath the \mgii{} broad emission line is modeled with a template derived from the narrow--line Seyfert 1 galaxy I~Zwicky-1 \citep[][]{Vestergaard2001} with redshift and FWHM anchored to the fit of the \mgii{} line (see below for details).
\end{itemize}
The continuum emission of five quasars in our sample (P009$-$10\footnote{
See also \citet{Farina2019} for peculiarities on the optical part of the spectrum of P009$-$10.
}, J0046$-$2837, P065$-$26, P065$-$19, and J2100$-$1715) show at wavelengths $\lrest\lesssim2000$\,\AA{} an apparent drop with respect to the fitted power law plus Balmer continuum emission.
This is likely due to absorbing material located in front of the accretion disk \citep[e.g.,][]{Gallerani2010}.
For these objects, the fit was performed separately for spectral regions around the \civ{} and around the \mgii{} emission lines.
%

\subsubsection{Quasar broad line emission}
%
The ultraviolet \civ{} and \mgii{} broad emission lines were fitted to derive single--epoch black hole masses and Eddington ratios for our sample of quasars (see \autoref{sec:BlackHoleMasses} and \autoref{sec:EddingtonRatios}).
\paragraph{\civ}
The broad \civ{} line is modeled with two Gaussian components in the rest--frame wavelength range $\lrest=1470-1600\,{\textrm{\AA}}$.
If the resultant two components fit appears indistinguishable from a single Gaussian due to the signal--to--noise ratio of a spectrum and/or to the symmetry of the line, the model of the line is reduced to a single Gaussian.
BAL and other absorption features present close to the line are identified and excluded from the fit. %
\paragraph{\mgii}
The \mgii{} emission is modeled with a single Gaussian profile in the wavelength range $2700\,\textrm{\AA{}}< \lambda_\textrm{rest} < 2900\,\textrm{\AA{}}$. 
The fits to the \mgii{} line and to the iron pseudo--continuum are connected. This means that, for each iteration, the redshifts and the FWHMs of the \mgii{} line are used as proxies for the corresponding values in the iron emission fit, until convergence is reached.
As discussed in \citetalias{Schindler2020}, the adoption of an iron template that takes the iron contribution beneath the broad \mgii{} line into account \citep[e.g.,][]{Tsuzuki2006} causes a \textit{narrowing} of the measured line FWHM by $\sim20$\% with respect to fit which use the \citet{Vestergaard2001} model \citep[][, see also \citealt{Trakhtenbrot2012} for the use of a different iron template]{Woo2018, Onoue2020}.
To estimate the black hole masses from the \mgii{} line, however, we will adopt the estimators of \citet{Vestergaard2009} and \citet[][, see \autoref{sec:BlackHoleMasses}]{Shen2011}, which are based on the \citet{Vestergaard2001} iron template.
To maximize consistency, the \mgii{} line properties reported in this paper will thus be determined employing the \citet{Vestergaard2001} model.
%

\section{Bolometric Luminosities, Black--Hole Masses, and Eddington Ratios}\label{sec:SMBH}

\begin{figure*}[p]
    \centering
    \includegraphics[width=0.97\textwidth]{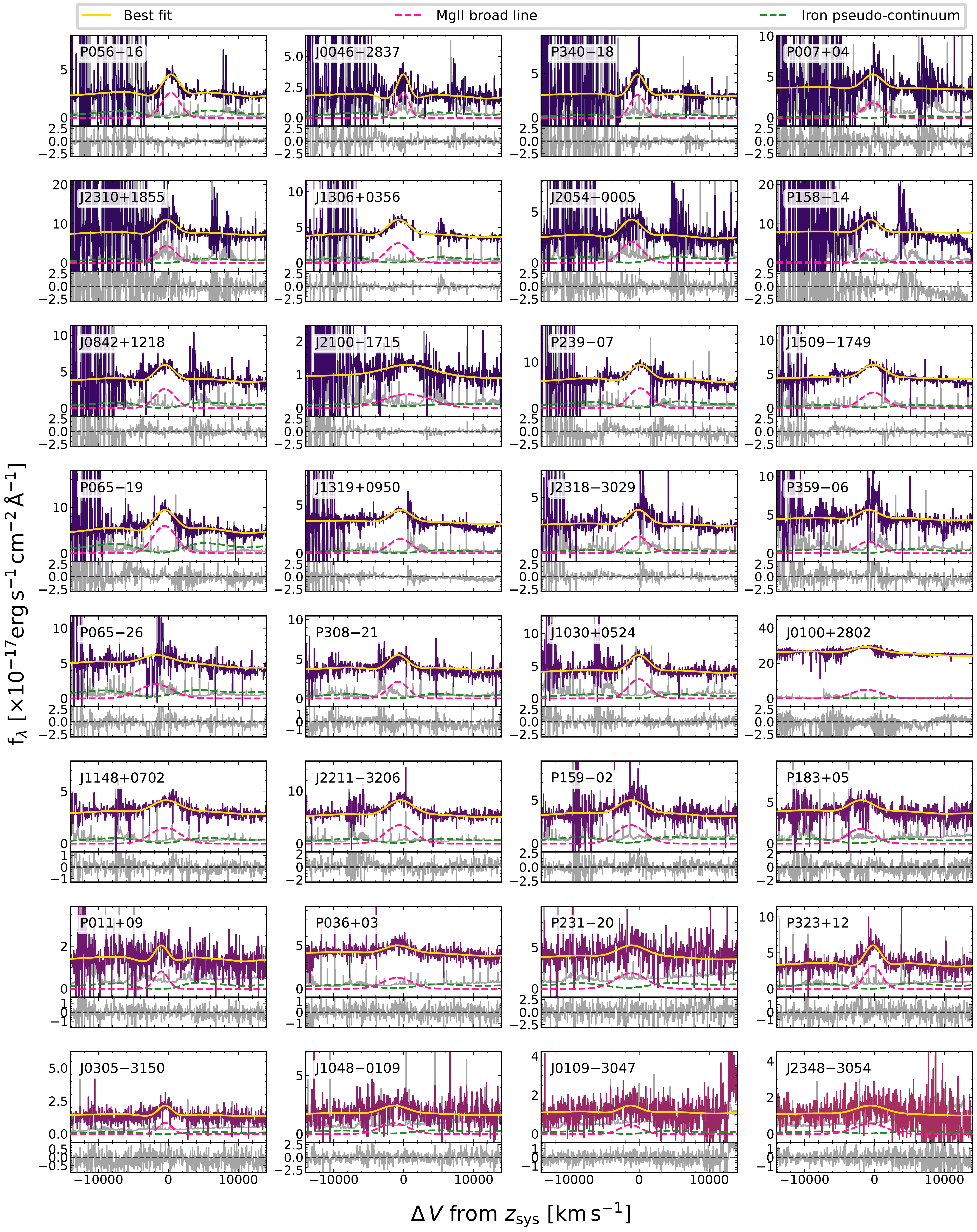}
    \caption{
    Atlas of the \mgii{} emission for the 32 quasars in our sample for which the line was successfully fitted.
    Quasars are ordered by redshift, from low redshift on the top left to the highest redshift on the bottom right.
    In the \textit{Top Panel}, we show the spectrum (in different shades of purple) and the corresponding 1--$\sigma$ error for each object.
    The fits to the \mgii{} broad line and the iron pseudo--continuum are shown in pink and green dashed lines, respectively.
    The combined best--fit of the continuum and the broad line emission (see \autoref{sec:fit}) is plotted as a solid yellow line.
    Residuals are displayed in gray in the \textit{Bottom Panel}.
    In each panel, the spectrum is displayed in a $\pm14000\,\kms{}$ range around the expected position of the \mgii{} line at the systemic redshift ($z_\textrm{sys}$).
    }
    \label{fig:fitmgii}
\end{figure*}

\begin{figure*}[p]
    \centering
    \includegraphics[width=0.97\textwidth]{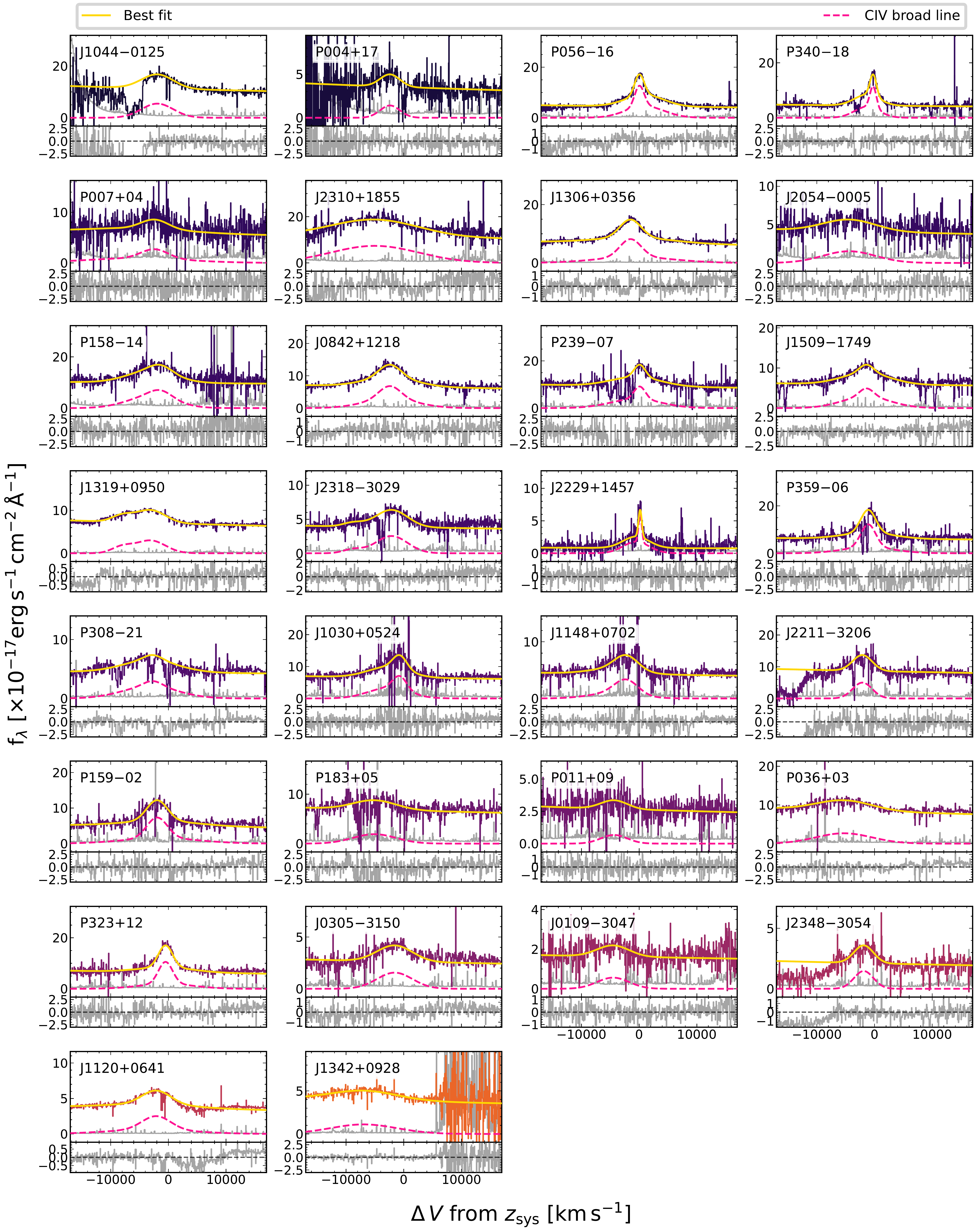}
    \caption{
    Same as \autoref{fig:fitmgii}, but showing the \civ{} emission for the sub--sample of 34 quasars for which the line was successfully fitted.
    Each panel plots a region $\pm17000\,\kms$ around the systemic redshift reported in \autoref{tab:sample}.
    }
    \label{fig:fitciv}
\end{figure*}

We successfully fitted the \mgii{} broad emission line of 32 quasars and the \civ{} line of 34 quasars in our sample (see \autoref{fig:fitmgii} and \autoref{fig:fitciv}, respectively).
The broad line emission has not been modeled in case it was redshifted in proximity of strong telluric absorption bands and/or in spectral regions with relatively low signal--to--noise ratios.
The derived continuum luminosities, line redshifts, integrated line luminosities, and FWHMs are listed in \autoref{tab:fitmgii} for \mgii{} and in \autoref{tab:fitciv} for \civ{}.
The quoted values are 50$^\textrm{th}$ (median), 16$^\textrm{th}$ and 84$^\textrm{th}$ percentiles obtained by re--fitting each spectrum 1,000 times after adding a Gaussian perturbation generated from the error vector to the observed flux.
These quantities will be used to derive the physical properties of the central SMBHs.

{\begin{deluxetable*}{lCCCCCCC}
\tabletypesize{\footnotesize} 
\tablecaption{Properties of the \mgii{} broad line emission and of the continuum at $\lambda_\textrm{rest}=3000$\,\AA{} \label{tab:fitmgii}}
\tablehead{
\colhead{ID}                                                              &
\colhead{$3000\textrm{{\AA{}}}\,f_{\lambda,3000\textrm{\AA{}}}$}          &
\colhead{$3000\textrm{{\AA{}}}\,L_{\lambda,3000\textrm{\AA{}}}$}          &
\colhead{$\lambda_\textrm{MgII}$}                                         &
\colhead{$z_\textrm{MgII}$}                                               &
\colhead{$\textrm{FWHM}_\textrm{MgII}$}                                   &
\colhead{$f_\textrm{MgII}$}                                               &
\colhead{$L_\textrm{MgII}$}                                               \\
\colhead{}                        &
\colhead{($10^{-13}\,\ergscm{}$)} &
\colhead{($10^{46}\,\ergs{}$)}    &
\colhead{(\AA{})}                 &
\colhead{}                        &
\colhead{($\kms{}$)}              &
\colhead{($10^{-16}\,\ergscm{}$)} &
\colhead{($10^{44}\,\ergs{}$)}    
}
\startdata
P004$+$17    & 0.325^{+0.010}_{-0.011} & \phn1.21^{+0.04}_{-0.04} & \nodata                       & \nodata                    & \nodata                          & \nodata               & \nodata               \\
P007$+$04    & 0.980^{+0.033}_{-0.039} & \phn3.91^{+0.13}_{-0.16} & 19576.6^{   +14.9}_{   -13.1} & 5.9946^{+0.0053}_{-0.0047} & 3203^{      +1925}_{   \phn-591} & \phn6.0^{+0.9}_{-0.9} & \phn2.4^{+0.4}_{-0.3} \\
P009$-$10    & 0.940^{+0.010}_{-0.010} & \phn3.75^{+0.04}_{-0.04} & \nodata                       & \nodata                    & \nodata                          & \nodata               & \nodata               \\
P011$+$09    & 0.336^{+0.007}_{-0.007} & \phn1.60^{+0.03}_{-0.03} & 20835.5^{\phn+8.0}_{   -14.1} & 6.4444^{+0.0029}_{-0.0050} & 2238^{   \phn+744}_{   \phn-373} & \phn1.8^{+0.4}_{-0.2} & \phn0.9^{+0.2}_{-0.1} \\
J0046$-$2837 & 0.390^{+0.017}_{-0.017} & \phn1.55^{+0.07}_{-0.07} & 19566.2^{\phn+4.5}_{\phn-4.6} & 5.9910^{+0.0016}_{-0.0016} & 1974^{   \phn+134}_{   \phn-133} & \phn4.0^{+0.4}_{-0.4} & \phn1.6^{+0.2}_{-0.2} \\
J0100$+$2802 & 6.608^{+0.007}_{-0.007} &    29.78^{+0.03}_{-0.03} & 20425.9^{\phn+1.7}_{\phn-1.7} & 6.2981^{+0.0006}_{-0.0006} & 4962^{\phn\phn+65}_{\phn\phn-62} &    24.5^{+0.3}_{-0.3} &    11.0^{+0.1}_{-0.1} \\
J0109$-$3047 & 0.281^{+0.007}_{-0.007} & \phn1.49^{+0.04}_{-0.04} & 21710.3^{   +26.6}_{   -27.2} & 6.7570^{+0.0095}_{-0.0097} & 3530^{   \phn+590}_{   \phn-620} & \phn1.7^{+0.3}_{-0.3} & \phn0.9^{+0.2}_{-0.2} \\
P036$+$03    & 0.996^{+0.006}_{-0.006} & \phn4.85^{+0.03}_{-0.03} & 21039.7^{\phn+8.1}_{\phn-9.4} & 6.5174^{+0.0029}_{-0.0034} & 4840^{   \phn+364}_{   \phn-341} & \phn6.1^{+0.3}_{-0.3} & \phn3.0^{+0.2}_{-0.1} \\
J0305$-$3150 & 0.366^{+0.004}_{-0.005} & \phn1.83^{+0.02}_{-0.02} & 21277.7^{\phn+6.4}_{\phn-7.1} & 6.6024^{+0.0023}_{-0.0025} & 2347^{   \phn+239}_{   \phn-280} & \phn2.0^{+0.2}_{-0.2} & \phn1.0^{+0.1}_{-0.1} \\
P056$-$16    & 0.523^{+0.006}_{-0.006} & \phn2.06^{+0.02}_{-0.02} & 19522.6^{\phn+2.2}_{\phn-2.3} & 5.9754^{+0.0008}_{-0.0008} & 3049^{   \phn+121}_{   \phn-118} & \phn7.8^{+0.2}_{-0.2} & \phn3.1^{+0.1}_{-0.1} \\
P065$-$26    & 0.984^{+0.021}_{-0.020} & \phn4.21^{+0.09}_{-0.09} & 19993.5^{\phn+7.9}_{\phn-8.8} & 6.1436^{+0.0028}_{-0.0032} & 5558^{   \phn+352}_{   \phn-303} &    11.3^{+0.7}_{-0.8} & \phn4.8^{+0.3}_{-0.3} \\
P065$-$19    & 0.889^{+0.018}_{-0.018} & \phn3.72^{+0.07}_{-0.08} & 19905.3^{\phn+2.1}_{\phn-2.0} & 6.1121^{+0.0008}_{-0.0007} & 3861^{   \phn+109}_{   \phn-103} &    22.9^{+0.6}_{-0.6} & \phn9.6^{+0.3}_{-0.3} \\
J0842$+$1218 & 0.917^{+0.009}_{-0.009} & \phn3.76^{+0.04}_{-0.04} & 19768.3^{\phn+3.8}_{\phn-4.1} & 6.0632^{+0.0014}_{-0.0015} & 3422^{   \phn+153}_{   \phn-135} & \phn9.0^{+0.4}_{-0.4} & \phn3.7^{+0.2}_{-0.2} \\
J1030$+$0524 & 1.041^{+0.016}_{-0.016} & \phn4.65^{+0.07}_{-0.07} & 20440.7^{\phn+4.9}_{\phn-4.4} & 6.3034^{+0.0017}_{-0.0016} & 3511^{   \phn+209}_{   \phn-192} &    10.4^{+0.5}_{-0.5} & \phn4.7^{+0.2}_{-0.2} \\
P158$-$14    & 2.219^{+0.051}_{-0.049} & \phn9.08^{+0.21}_{-0.20} & 19745.6^{\phn+6.1}_{\phn-7.7} & 6.0550^{+0.0022}_{-0.0027} & 2794^{   \phn+208}_{   \phn-194} & \phn9.7^{+0.9}_{-1.0} & \phn4.0^{+0.4}_{-0.4} \\
P159$-$02    & 0.769^{+0.014}_{-0.013} & \phn3.53^{+0.06}_{-0.06} & 20573.2^{\phn+7.0}_{\phn-7.2} & 6.3507^{+0.0025}_{-0.0026} & 4798^{   \phn+286}_{   \phn-252} &    10.2^{+0.4}_{-0.5} & \phn4.7^{+0.2}_{-0.2} \\
J1044$-$0125 & 1.384^{+0.024}_{-0.023} & \phn5.07^{+0.09}_{-0.08} & \nodata                       & \nodata                    & \nodata                          & \nodata               & \nodata               \\
J1048$-$0109 & 0.401^{+0.012}_{-0.011} & \phn2.05^{+0.06}_{-0.06} & 21382.9^{   +24.1}_{   -20.5} & 6.6400^{+0.0086}_{-0.0073} & 4703^{   \phn+620}_{   \phn-635} & \phn4.4^{+0.6}_{-0.6} & \phn2.2^{+0.3}_{-0.3} \\
J1120$+$0641 & 0.565^{+0.002}_{-0.002} & \phn3.31^{+0.01}_{-0.01} & \nodata                       & \nodata                    & \nodata                          & \nodata               & \nodata               \\
J1148$+$0702 & 0.721^{+0.009}_{-0.008} & \phn3.24^{+0.04}_{-0.04} & 20499.0^{\phn+5.2}_{\phn-6.0} & 6.3242^{+0.0019}_{-0.0022} & 4895^{   \phn+165}_{   \phn-154} & \phn7.4^{+0.3}_{-0.3} & \phn3.3^{+0.1}_{-0.1} \\
P183$+$05    & 0.928^{+0.012}_{-0.012} & \phn4.36^{+0.06}_{-0.06} & 20682.5^{   +11.4}_{   -10.5} & 6.3898^{+0.0041}_{-0.0037} & 4459^{   \phn+311}_{   \phn-334} & \phn7.9^{+0.6}_{-0.5} & \phn3.7^{+0.3}_{-0.2} \\
J1306$+$0356 & 0.902^{+0.006}_{-0.006} & \phn3.64^{+0.02}_{-0.02} & 19633.8^{\phn+1.7}_{\phn-1.7} & 6.0151^{+0.0006}_{-0.0006} & 3895^{\phn\phn+69}_{\phn\phn-58} &    10.8^{+0.2}_{-0.2} & \phn4.3^{+0.1}_{-0.1} \\
J1319$+$0950 & 0.818^{+0.003}_{-0.003} & \phn3.43^{+0.01}_{-0.01} & 19936.2^{\phn+2.9}_{\phn-3.1} & 6.1231^{+0.0010}_{-0.0011} & 3747^{   \phn+112}_{   \phn-113} & \phn5.5^{+0.2}_{-0.2} & \phn2.3^{+0.1}_{-0.1} \\
J1342$+$0928 & 0.558^{+0.004}_{-0.001} & \phn3.77^{+0.03}_{-0.01} & \nodata                       & \nodata                    & \nodata                          & \nodata               & \nodata               \\
J1509$-$1749 & 1.166^{+0.010}_{-0.011} & \phn4.87^{+0.04}_{-0.05} & 19922.9^{\phn+3.3}_{\phn-3.4} & 6.1184^{+0.0012}_{-0.0012} & 4067^{   \phn+156}_{   \phn-153} & \phn9.3^{+0.4}_{-0.4} & \phn3.9^{+0.2}_{-0.1} \\
P231$-$20    & 0.877^{+0.021}_{-0.021} & \phn4.34^{+0.10}_{-0.10} & 21152.8^{   +15.0}_{   -17.8} & 6.5578^{+0.0054}_{-0.0063} & 5196^{   \phn+588}_{   \phn-493} & \phn9.7^{+0.9}_{-0.9} & \phn4.8^{+0.5}_{-0.4} \\
P239$-$07    & 1.419^{+0.018}_{-0.020} & \phn5.90^{+0.07}_{-0.08} & 19906.3^{\phn+2.8}_{\phn-2.7} & 6.1124^{+0.0010}_{-0.0010} & 3611^{   \phn+135}_{   \phn-150} &    15.4^{+0.6}_{-0.6} & \phn6.4^{+0.2}_{-0.3} \\
P308$-$21    & 0.975^{+0.009}_{-0.009} & \phn4.25^{+0.04}_{-0.04} & 20195.9^{   +16.4}_{\phn-9.5} & 6.2159^{+0.0059}_{-0.0034} & 3355^{   \phn+203}_{   \phn-361} & \phn7.0^{+0.6}_{-1.0} & \phn3.1^{+0.3}_{-0.4} \\
J2054$-$0005 & 0.594^{+0.018}_{-0.015} & \phn2.40^{+0.07}_{-0.06} & 19628.7^{\phn+7.4}_{\phn-7.3} & 6.0133^{+0.0026}_{-0.0026} & 3633^{   \phn+215}_{   \phn-230} & \phn7.5^{+0.5}_{-0.5} & \phn3.1^{+0.2}_{-0.2} \\
J2100$-$1715 & 0.227^{+0.005}_{-0.005} & \phn0.93^{+0.02}_{-0.02} & 19851.9^{   +18.6}_{   -22.3} & 6.0930^{+0.0066}_{-0.0080} & 8012^{   \phn+814}_{      -1466} & \phn3.3^{+0.3}_{-0.5} & \phn1.4^{+0.1}_{-0.2} \\
P323$+$12    & 0.786^{+0.008}_{-0.008} & \phn3.89^{+0.04}_{-0.04} & 21219.3^{\phn+4.4}_{\phn-4.6} & 6.5816^{+0.0016}_{-0.0016} & 2803^{   \phn+174}_{   \phn-184} & \phn8.8^{+0.4}_{-0.4} & \phn4.4^{+0.2}_{-0.2} \\
J2211$-$3206 & 1.285^{+0.025}_{-0.026} & \phn5.82^{+0.11}_{-0.12} & 20505.1^{\phn+3.8}_{\phn-4.1} & 6.3264^{+0.0014}_{-0.0014} & 4666^{   \phn+196}_{   \phn-156} &    16.4^{+0.7}_{-0.6} & \phn7.4^{+0.3}_{-0.3} \\
J2229$+$1457 & 0.131^{+0.008}_{-0.007} & \phn0.55^{+0.03}_{-0.03} & \nodata                       & \nodata                    & \nodata                          & \nodata               & \nodata               \\
P340$-$18    & 0.647^{+0.008}_{-0.007} & \phn2.58^{+0.03}_{-0.03} & 19578.8^{\phn+3.2}_{\phn-3.2} & 5.9954^{+0.0012}_{-0.0012} & 2468^{   \phn+190}_{   \phn-144} & \phn6.3^{+0.3}_{-0.3} & \phn2.5^{+0.1}_{-0.1} \\
J2310$+$1855 & 1.905^{+0.023}_{-0.021} & \phn7.60^{+0.09}_{-0.09} & 19580.6^{\phn+6.6}_{\phn-6.8} & 5.9961^{+0.0024}_{-0.0024} & 3312^{   \phn+238}_{   \phn-213} &    14.3^{+0.9}_{-1.0} & \phn5.7^{+0.4}_{-0.4} \\
J2318$-$3029 & 0.669^{+0.009}_{-0.010} & \phn2.82^{+0.04}_{-0.04} & 19987.2^{\phn+5.6}_{\phn-5.2} & 6.1413^{+0.0020}_{-0.0018} & 3460^{   \phn+177}_{   \phn-160} & \phn5.2^{+0.3}_{-0.3} & \phn2.2^{+0.1}_{-0.1} \\
J2348$-$3054 & 0.265^{+0.007}_{-0.007} & \phn1.46^{+0.04}_{-0.04} & 22063.2^{   +23.0}_{   -26.7} & 6.8831^{+0.0082}_{-0.0095} & 5182^{   \phn+843}_{   \phn-811} & \phn3.0^{+0.4}_{-0.4} & \phn1.7^{+0.2}_{-0.2} \\
P359$-$06    & 1.170^{+0.013}_{-0.014} & \phn4.98^{+0.06}_{-0.06} & 20013.2^{   +12.0}_{   -14.7} & 6.1507^{+0.0043}_{-0.0053} & 3825^{   \phn+427}_{   \phn-370} & \phn5.9^{+0.7}_{-0.6} & \phn2.5^{+0.3}_{-0.2} \\
\enddata
\end{deluxetable*}
}

{\begin{deluxetable*}{lCCCCCCC}
\tabletypesize{\footnotesize} 
\tablecaption{Properties of the \civ{} broad line emission and of the continuum at $\lambda_\textrm{rest}=1350$\,\AA{} \label{tab:fitciv}}
\tablehead{
\colhead{ID}                                                              &
\colhead{$1350\textrm{{\AA{}}}\,f_{\lambda,1350\textrm{\AA{}}}$}          &
\colhead{$1350\textrm{{\AA{}}}\,L_{\lambda,1350\textrm{\AA{}}}$}          &
\colhead{$\lambda_\textrm{CIV}$}                                          &
\colhead{$z_\textrm{CIV}$}                                                &
\colhead{$\textrm{FWHM}_\textrm{CIV}$}                                    &
\colhead{$f_\textrm{CIV}$}                                                &
\colhead{$L_\textrm{CIV}$}                                                \\
\colhead{}                        &
\colhead{($10^{-13}\,\ergscm{}$)} &
\colhead{($10^{46}\,\ergs{}$)}    &
\colhead{(\AA{})}                 &
\colhead{}                        &
\colhead{($\kms{}$)}              &
\colhead{($10^{-16}\,\ergscm{}$)} &
\colhead{($10^{44}\,\ergs{}$)}    
}
\startdata
P004$+$17    & 0.623^{+0.032}_{-0.029} & \phn2.31^{+0.12}_{-0.11} & 10474.9^{\phn+6.9}_{\phn-6.7} & 5.7620^{+0.0045}_{-0.0043} &     \phn4071^{   \phn+451}_{   \phn-462} & \phn3.1^{+0.7}_{-0.6} & \phn1.2^{+0.2}_{-0.2} \\
P007$+$04    & 0.911^{+0.053}_{-0.052} & \phn3.63^{+0.21}_{-0.21} & 10757.3^{   +10.6}_{   -11.5} & 5.9442^{+0.0068}_{-0.0074} &     \phn7278^{      +1332}_{      -1090} &    16.0^{+6.6}_{-5.2} & \phn6.4^{+2.6}_{-2.1} \\
P009$-$10    & 0.482^{+0.028}_{-0.025} & \phn1.92^{+0.11}_{-0.10} & 10646.1^{   +13.7}_{   -17.1} & 5.8725^{+0.0088}_{-0.0111} &        15746^{      +2315}_{      -2274} & \phn9.4^{+2.0}_{-1.9} & \phn3.8^{+0.8}_{-0.7} \\
P011$+$09    & 0.437^{+0.011}_{-0.011} & \phn2.07^{+0.05}_{-0.05} & 11403.1^{   +14.3}_{   -13.8} & 6.3611^{+0.0092}_{-0.0089} &     \phn5378^{   \phn+994}_{   \phn-807} & \phn2.0^{+0.4}_{-0.4} & \phn0.9^{+0.2}_{-0.2} \\
J0046$-$2837 & 0.210^{+0.061}_{-0.058} & \phn0.83^{+0.24}_{-0.23} & \nodata                       & \nodata                    &  \nodata                                 & \nodata               & \nodata               \\
J0100$+$2802 & 8.389^{+0.010}_{-0.010} &    37.81^{+0.05}_{-0.05} & \nodata                       & \nodata                    &  \nodata                                 & \nodata               & \nodata               \\
J0109$-$3047 & 0.248^{+0.008}_{-0.008} & \phn1.32^{+0.04}_{-0.04} & 11885.4^{   +12.0}_{   -11.6} & 6.6725^{+0.0078}_{-0.0075} &     \phn6636^{   \phn+799}_{   \phn-798} & \phn2.0^{+0.3}_{-0.3} & \phn1.1^{+0.2}_{-0.2} \\
P036$+$03    & 1.396^{+0.009}_{-0.009} & \phn6.79^{+0.04}_{-0.04} & 11473.9^{\phn+4.0}_{\phn-4.0} & 6.4068^{+0.0026}_{-0.0026} &        11640^{   \phn+557}_{   \phn-496} &    16.8^{+0.7}_{-0.7} & \phn8.2^{+0.4}_{-0.3} \\
J0305$-$3150 & 0.425^{+0.006}_{-0.006} & \phn2.12^{+0.03}_{-0.03} & 11732.5^{\phn+3.5}_{\phn-3.6} & 6.5737^{+0.0022}_{-0.0023} &     \phn7277^{   \phn+301}_{   \phn-282} & \phn6.2^{+0.3}_{-0.3} & \phn3.1^{+0.1}_{-0.1} \\
P056$-$16    & 0.762^{+0.013}_{-0.014} & \phn3.00^{+0.05}_{-0.05} & 10794.7^{\phn+0.5}_{\phn-0.5} & 5.9684^{+0.0003}_{-0.0003} &     \phn2642^{\phn\phn+57}_{\phn\phn-50} &    28.4^{+0.5}_{-0.6} &    11.2^{+0.2}_{-0.2} \\
P065$-$26    & 1.097^{+0.018}_{-0.017} & \phn4.70^{+0.08}_{-0.07} & 10920.2^{\phn+6.2}_{\phn-5.9} & 6.0494^{+0.0040}_{-0.0038} &     \phn7766^{   \phn+268}_{   \phn-283} &    11.6^{+0.7}_{-0.7} & \phn5.0^{+0.3}_{-0.3} \\
P065$-$19    & 0.499^{+0.021}_{-0.021} & \phn2.09^{+0.09}_{-0.09} & 10953.2^{\phn+3.2}_{\phn-3.6} & 6.0707^{+0.0021}_{-0.0023} &     \phn5638^{   \phn+245}_{   \phn-215} &    35.1^{+1.6}_{-1.5} &    14.7^{+0.7}_{-0.6} \\
J0842$+$1218 & 1.067^{+0.014}_{-0.014} & \phn4.38^{+0.06}_{-0.06} & 10872.3^{\phn+2.0}_{\phn-1.8} & 6.0185^{+0.0013}_{-0.0011} &     \phn6027^{   \phn+135}_{   \phn-137} &    26.4^{+1.1}_{-1.0} &    10.8^{+0.4}_{-0.4} \\
J1030$+$0524 & 1.034^{+0.019}_{-0.020} & \phn4.62^{+0.09}_{-0.09} & 11284.7^{   +14.0}_{   -15.7} & 6.2847^{+0.0091}_{-0.0101} &     \phn4733^{   \phn+517}_{   \phn-679} &    21.6^{+1.2}_{-1.2} & \phn9.6^{+0.6}_{-0.5} \\
P158$-$14    & 1.443^{+0.045}_{-0.042} & \phn5.90^{+0.18}_{-0.17} & 10886.4^{   +21.6}_{   -15.5} & 6.0276^{+0.0140}_{-0.0100} &     \phn7703^{   \phn+369}_{   \phn-339} &    32.0^{+5.2}_{-2.8} &    13.1^{+2.1}_{-1.2} \\
P159$-$02    & 0.768^{+0.016}_{-0.018} & \phn3.53^{+0.07}_{-0.08} & 11359.3^{\phn+2.5}_{\phn-2.3} & 6.3329^{+0.0016}_{-0.0015} &     \phn4921^{   \phn+210}_{   \phn-183} &    26.6^{+1.5}_{-1.4} &    12.2^{+0.7}_{-0.6} \\
J1044$-$0125 & 1.872^{+0.067}_{-0.068} & \phn6.85^{+0.25}_{-0.25} & 10443.2^{   +20.1}_{   -30.2} & 5.7415^{+0.0130}_{-0.0195} &     \phn6478^{      +1363}_{      -1090} &    19.5^{+6.7}_{-4.8} & \phn7.1^{+2.4}_{-1.8} \\
J1048$-$0109 & 0.553^{+0.019}_{-0.020} & \phn2.82^{+0.10}_{-0.10} & \nodata                       & \nodata                    &  \nodata                                 & \nodata               & \nodata               \\
J1120$+$0641 & 0.566^{+0.002}_{-0.003} & \phn3.31^{+0.01}_{-0.02} & 12435.2^{\phn+1.3}_{\phn-1.1} & 7.0274^{+0.0009}_{-0.0007} &     \phn6952^{\phn\phn+91}_{\phn\phn-86} &    11.9^{+0.3}_{-0.3} & \phn7.0^{+0.2}_{-0.2} \\
J1148$+$0702 & 0.666^{+0.010}_{-0.011} & \phn3.01^{+0.05}_{-0.05} & 11267.3^{    +9.0}_{\phn-8.9} & 6.2734^{+0.0058}_{-0.0057} &     \phn5734^{   \phn+295}_{   \phn-295} &    11.8^{+1.0}_{-0.7} & \phn5.3^{+0.4}_{-0.3} \\
P183$+$05    & 1.097^{+0.017}_{-0.019} & \phn5.15^{+0.08}_{-0.09} & 11328.2^{   +11.5}_{   -10.5} & 6.3128^{+0.0074}_{-0.0067} &     \phn8927^{   \phn+768}_{   \phn-649} & \phn9.2^{+1.0}_{-0.9} & \phn4.3^{+0.5}_{-0.4} \\
J1306$+$0356 & 1.097^{+0.011}_{-0.010} & \phn4.43^{+0.04}_{-0.04} & 10840.5^{\phn+0.7}_{\phn-0.7} & 5.9979^{+0.0005}_{-0.0004} &     \phn5236^{\phn\phn+83}_{\phn\phn-99} &    32.1^{+0.9}_{-1.0} &    13.0^{+0.4}_{-0.4} \\
J1319$+$0950 & 1.175^{+0.005}_{-0.005} & \phn4.93^{+0.02}_{-0.02} & 10932.8^{\phn+2.5}_{\phn-2.8} & 6.0575^{+0.0016}_{-0.0018} &     \phn8933^{   \phn+118}_{   \phn-110} &    13.3^{+0.2}_{-0.2} & \phn5.6^{+0.1}_{-0.1} \\
J1342$+$0928 & 0.614^{+0.003}_{-0.002} & \phn4.15^{+0.02}_{-0.01} & 12921.3^{\phn+4.4}_{\phn-4.1} & 7.3412^{+0.0029}_{-0.0026} &        13969^{   \phn+263}_{   \phn-334} & \phn8.3^{+0.2}_{-0.3} & \phn5.6^{+0.1}_{-0.2} \\
J1509$-$1749 & 0.890^{+0.011}_{-0.012} & \phn3.72^{+0.04}_{-0.05} & 10981.3^{\phn+1.8}_{\phn-1.9} & 6.0888^{+0.0012}_{-0.0012} &     \phn5537^{   \phn+183}_{   \phn-175} &    18.7^{+0.9}_{-0.7} & \phn7.8^{+0.4}_{-0.3} \\
P231$-$20    & 1.284^{+0.036}_{-0.035} & \phn6.35^{+0.18}_{-0.17} & \nodata                       & \nodata                    &  \nodata                                 & \nodata               & \nodata               \\
P239$-$07    & 1.481^{+0.023}_{-0.024} & \phn6.15^{+0.09}_{-0.10} & 11016.1^{\phn+6.3}_{   -16.6} & 6.1113^{+0.0041}_{-0.0107} &     \phn3633^{   \phn+827}_{   \phn-481} &    28.8^{+3.8}_{-2.1} &    12.0^{+1.6}_{-0.9} \\
P308$-$21    & 0.648^{+0.006}_{-0.006} & \phn2.82^{+0.03}_{-0.03} & 11103.4^{\phn+7.4}_{\phn-6.9} & 6.1677^{+0.0048}_{-0.0045} &     \phn8035^{   \phn+749}_{   \phn-861} &    14.7^{+0.6}_{-0.6} & \phn6.4^{+0.3}_{-0.2} \\
J2054$-$0005 & 0.657^{+0.025}_{-0.030} & \phn2.66^{+0.10}_{-0.12} & 10744.1^{   +11.3}_{   -10.6} & 5.9357^{+0.0073}_{-0.0068} &        10795^{      +2049}_{      -1669} & \phn8.9^{+2.2}_{-1.6} & \phn3.6^{+0.9}_{-0.6} \\
J2100$-$1715 & 0.126^{+0.010}_{-0.008} & \phn0.52^{+0.04}_{-0.03} & 10937.1^{   +12.4}_{   -15.8} & 6.0603^{+0.0080}_{-0.0102} &     \phn7433^{      +2324}_{   \phn-999} & \phn1.6^{+0.5}_{-0.3} & \phn0.7^{+0.2}_{-0.1} \\
P323$+$12    & 1.084^{+0.012}_{-0.012} & \phn5.37^{+0.06}_{-0.06} & 11734.0^{\phn+0.8}_{\phn-1.0} & 6.5747^{+0.0005}_{-0.0007} &     \phn3286^{\phn\phn+93}_{\phn\phn-83} &    25.4^{+0.8}_{-0.8} &    12.6^{+0.4}_{-0.4} \\
J2211$-$3206 & 1.386^{+0.048}_{-0.042} & \phn6.28^{+0.22}_{-0.19} & 11289.0^{\phn+3.8}_{\phn-3.3} & 6.2875^{+0.0024}_{-0.0021} &     \phn3996^{   \phn+250}_{   \phn-246} &    10.9^{+1.1}_{-1.0} & \phn4.9^{+0.5}_{-0.5} \\
J2229$+$1457 & 0.127^{+0.010}_{-0.010} & \phn0.53^{+0.04}_{-0.04} & 11084.8^{\phn+0.5}_{\phn-0.6} & 6.1556^{+0.0003}_{-0.0004} &  \phn\phn886^{\phn\phn+51}_{\phn\phn-49} & \phn6.8^{+0.4}_{-0.3} & \phn2.9^{+0.2}_{-0.1} \\
P340$-$18    & 0.712^{+0.013}_{-0.014} & \phn2.84^{+0.05}_{-0.06} & 10834.8^{\phn+0.4}_{\phn-0.4} & 5.9943^{+0.0002}_{-0.0003} &     \phn1767^{\phn\phn+44}_{\phn\phn-42} &    16.9^{+0.4}_{-0.4} & \phn6.7^{+0.2}_{-0.2} \\
J2310$+$1855 & 1.749^{+0.040}_{-0.041} & \phn6.98^{+0.16}_{-0.16} & 10665.4^{\phn+4.0}_{\phn-3.9} & 5.8849^{+0.0026}_{-0.0025} &        18297^{   \phn+651}_{   \phn-708} &    72.9^{+4.2}_{-4.0} &    29.1^{+1.7}_{-1.6} \\
J2318$-$3029 & 0.592^{+0.010}_{-0.011} & \phn2.49^{+0.04}_{-0.04} & 10990.9^{\phn+3.8}_{\phn-4.0} & 6.0950^{+0.0024}_{-0.0026} &     \phn6733^{   \phn+397}_{   \phn-339} &    10.3^{+0.6}_{-0.6} & \phn4.3^{+0.3}_{-0.2} \\
J2348$-$3054 & 0.350^{+0.010}_{-0.011} & \phn1.93^{+0.06}_{-0.06} & 12162.0^{\phn+4.4}_{\phn-5.0} & 6.8510^{+0.0028}_{-0.0032} &     \phn3982^{   \phn+272}_{   \phn-256} & \phn3.2^{+0.3}_{-0.3} & \phn1.8^{+0.1}_{-0.1} \\
P359$-$06    & 0.929^{+0.015}_{-0.015} & \phn3.95^{+0.06}_{-0.06} & 11069.3^{\phn+1.4}_{\phn-1.0} & 6.1456^{+0.0009}_{-0.0007} &     \phn3520^{   \phn+123}_{   \phn-117} &    30.6^{+1.2}_{-1.3} &    13.0^{+0.5}_{-0.5} \\
\enddata
\end{deluxetable*}}

\subsection{Bolometric Luminosities}\label{sec:BolometricLuminosities}

%
We estimate the bolometric luminosities ($\Lbol$) from the monochromatic luminosities at 3000\,\AA{} following the relation:
\begin{equation}\label{eq:lbol}
L_\textrm{bol} = 5.15 \times 3000\,\textrm{\AA}\,L_{\lambda,3000\textrm{\AA}}
\end{equation}
originally derived from a compilation of 259 bright SDSS quasars by \citet{Richards2006} and then updated by \citet{Shen2011}.
This conversion is commonly used in high--redshift quasars studies \citep[e.g.,][]{DeRosa2011, Mazzucchelli2017, Shen2019, Eilers2020, Onoue2020, WangFeige2021XRay}.
Given the $\sim0.3$\,dex uncertainties in the determination of $\Lbol$ \citep[e.g.,][]{Richards2006, Shen2011}, the choice of a different conversion has little impact on the general results of this paper.
For instance, if we use the correction provided by \citet{Runnoe2012}, the bolometric luminosity of sources in our sample would be, on average, $\sim20$\% higher, while using the one from \citet{Trakhtenbrot2012} would result in bolometric luminosities that are $\sim20$\% lower.
Bolometric luminosities for all quasars in our sample are listed in \autoref{tab:mbh}.
%

\subsection{Single--Epoch Virial Black Hole Masses}\label{sec:BlackHoleMasses}

%
Under the assumption that the clouds in the BLR orbit around a SMBH purely following gravitational dynamics, the mass of the central black hole can be estimated from the virial theorem:
$\mbh{}=G^{-1}\,R_\mathrm{BLR}\,v_\mathrm{BLR}^2$, 
where $R_\mathrm{BLR}$ is the radius of the BLR, and $v_\mathrm{BLR}$ is the cloud velocity.
Both $R_\mathrm{BLR}$ and $v_\mathrm{BLR}$ can be directly derived from a quasar spectrum by considering that:
\textit{(i)} broad emission--line reverberation--mapping of local AGNs \citep[e.g.,][]{Peterson2004} has revealed a tight correlation between the $R_\mathrm{BLR}$ and the AGN luminosity \citep[e.g.,][]{Kaspi2005, Bentz2006, Bentz2009}; and
\textit{(ii)} $v_\mathrm{BLR}$ and the FWHM of a broad emission--line are related through a deprojection factor $f$ of the order unity \citep[e.g.,][]{Gravity2018, Gravity2020}.
Thus, despite some potential biases (see e.g., \citealt{Shen2012} and \citealt{Shen2013}), the so--called single--epoch virial black hole masses can be calculated as: 
\begin{equation}\label{eq:mbh}
    \frac{\mbh{}}{\msun} = 10^{a}
             \left(\frac{\mathrm{FWHM}}{10^3\,\kms}\right)^2
             \left(\frac{\lrest\,L_{\lambda,\lrest}}{10^{44}\,\ergs{}}\right)^{b},
\end{equation}
where $L_{\lambda,\lrest}$ is the monochromatic continuum luminosity of the quasar at the rest frame wavelength $\lrest$ and the coefficients $a$ and $b$ are empirically derived either directly against reverberation--mapping masses or by cross--correlating measurements from different broad lines.
In \autoref{tab:mbh} we report black hole masses obtained from the following estimators: 
($a=6.86$, $b=0.5$, $x=3000$\,\AA{}) from \citet{Vestergaard2009} and 
($a=6.74$, $b=0.62$, $x=3000$\,\AA{}) from \citet{Shen2011} for the \mgii{} broad line and 
($a=6.66$, $b=0.53$, $x=1350$\,\AA{}) from \citet{Vestergaard2006} for the \civ{} broad line.
We stress that individual black hole masses inferred with this approach are afflicted by systematic errors of $\sim0.3-0.5$\,dex \citep[e.g.][]{Vestergaard2009, Shen2012}.
These are much larger than the statistical uncertainties associated with the fitting of the broad lines, which are reported in \autoref{tab:mbh}. 
%

\subsubsection{Which black hole mass estimator?}\label{sec:whichbh}

\begin{figure}[tb]
    \centering
    \includegraphics[width=0.98\columnwidth]{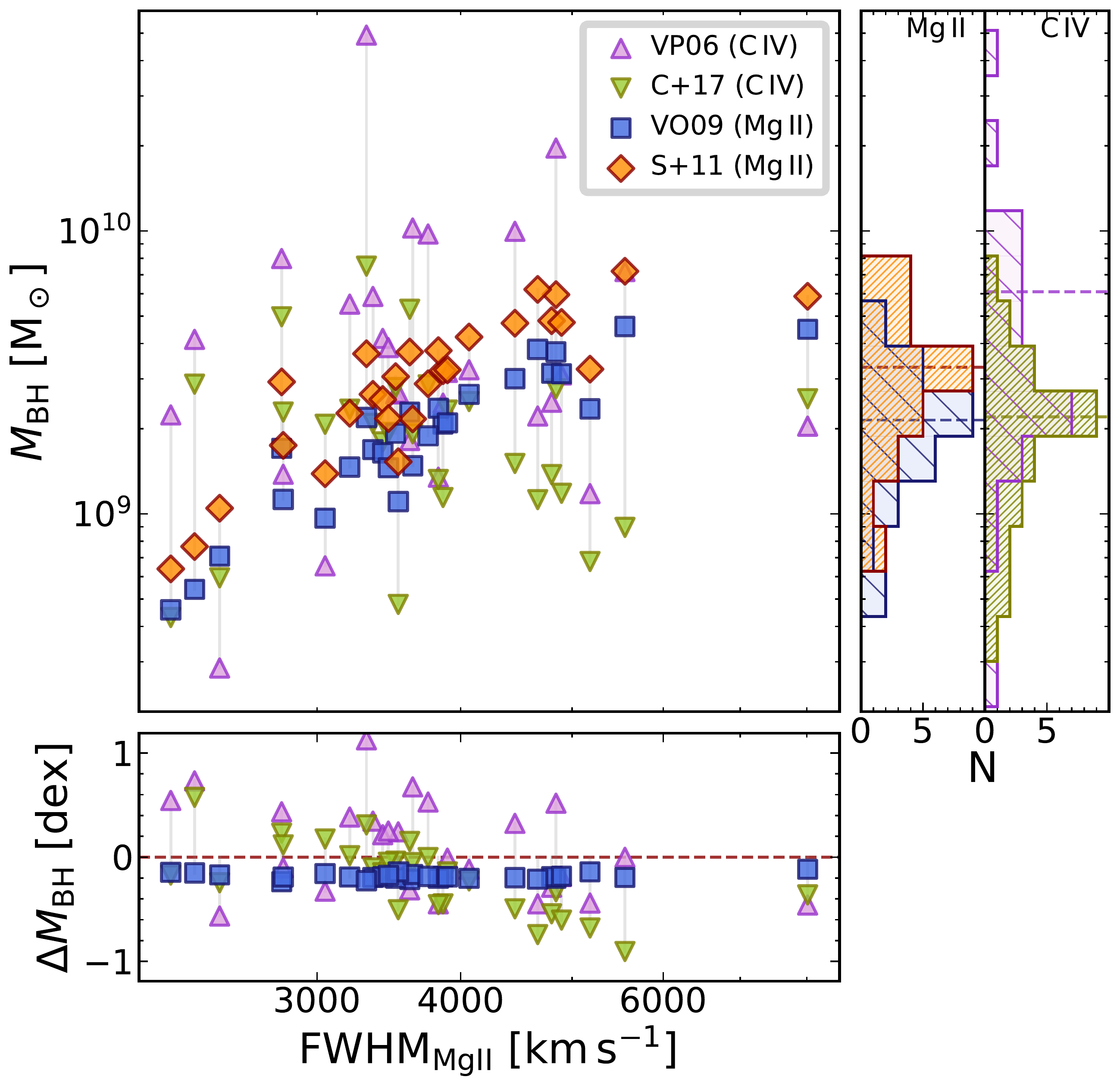}
    \caption{
    Black hole mass versus FWHM of the \mgii{} line for the 28 quasars in our sample for which both the \mgii{} and \civ{} lines were successfully fitted (see \autoref{sec:fit}).
    Pink triangles, green triangles, blue squares, and orange diamonds mark masses obtained with the \citet{Vestergaard2006}, \citet{Coatman2017}, \citet{Vestergaard2009}, and \citet{Shen2011} estimators, respectively (see \autoref{tab:mbh}).
    Pale gray lines connect different mass estimates of the same object.
    The histograms on the right illustrate how the different recipes affect the mass distribution (the color scheme is the same as in the main panel).
    Horizontal dashed lines locate the corresponding averages.
    The average (median) masses of this sub--sample are $3.3\times$, $2.1\times$, $6.1\times$, and $2.2\times10^{9}\,\msun$ ($3.1\times$, $2.0\times$, $3.1\times$, and $2.0\times10^{9}\,\msun$) if calculated with the \citet{Shen2011}, \citet{Vestergaard2009}, \citet{Vestergaard2006}, and \citet{Coatman2017} estimators, respectively.
    The scatter plot on the bottom displays mass differences with respect to the \citeauthor{Shen2011} estimator (see \autoref{sec:BlackHoleMasses}).
    }
    \label{fig:mgii_vs_civ}
\end{figure}

%
Given their extensive use in reverberation mapping investigations \citep[e.g.,][]{Kaspi2000, Bentz2006, Bentz2009, Bentz2013, Derosa2018}, the Balmer \ha{} and \hb{} lines are typically considered the most reliable indicators for black hole mass estimates.
At $z\gsim4$, however, these lines move out of the $K$--band and are, thus, virtually inaccessible from the ground.
Hence, for high--redshift quasar studies, the \mgii{} and \civ{} lines are commonly used as surrogates.
In the last two decades, the number of AGNs with measured time lags between continuum and \mgii{} and \civ{} line flux fluctuations has been drastically increased \citep[e.g.,][]{Peterson2004, Kaspi2007, Woo2008, Derosa2015, Sun2015, Lira2018, Kaspi2021}.
Nevertheless, the derived radius--luminosity relations are still poorly constrained and/or show significantly larger scatter with respect to the ones based on \hb{} \citep[e.g.,][, and references therein]{Grier2019, Homayouni2020}.
A strong correlation between the width of the \mgii{} and the \hb{} lines has been observed \citep[e.g.,][]{Trakhtenbrot2012, Shen2012}, suggesting that estimators relying on this low--ionization line deliver trustworthy black hole masses for high--redshift quasars.
On the other hand, the frequent occurrence of prominent blue--shifts and asymmetric line shapes, and the poor correlation with the width of the \hb{} (or \mgii{}) line imply that the high--ionization \civ{} line is more affected by non--virial components and, thus, generates more biased mass estimates \citep[e.g.,][]{Baskin2005, Marziani2012, Richards2011, Trakhtenbrot2012, Shen2012, Shen2013, Coatman2016, Coatman2017, Marziani2019, Zuo2020, WangShu2020}.
In addition, the frequent presence of strong absorption system makes the reconstruction of the intrinsic \civ{} line profile arduous.
This is particularly relevant at $z \gtrsim 6.5$, when bright quasars experience a sharp increase in the mean \civ{} blueshift \citep[e.g.,][; \citetalias{Schindler2020}]{Derosa2014, Mazzucchelli2017, Meyer2019, Reed2019, Shen2019, Yang2021} and an increase in the frequency of BAL features \citep[e.g.,][]{Yang2021, Bischetti2022}.
To reduce the influence of the non--virial contributions on the \civ{} emission, we also estimate black hole masses employing the empirical \civ{}--blueshift dependent correction to the \citet{Vestergaard2006} relation provided by \citet[][, but see \citealt{Mejia2018}, for limitations of this approach]{Coatman2017}\footnote{
For the majority of our sample the systemic redshift of the quasar host galaxies are measured from the \ciimu{} line \citepalias[see \autoref{tab:sample} and][]{Schindler2020} while \citeauthor{Coatman2017} used systemic redshifts derived from the the \ha{} (and \hb{}) Balmer lines.
Fortunately, redshifts determined from these Hydrogen lines are typically similar to the systemic ones \citep[e.g.,][]{Bonning2007}.
The difference between these two approaches has only a marginal effect on the derived black hole masses.
}.
\civ{} based black hole masses and Eddington ratios calculated using the \citeauthor{Coatman2017} correction are listed in \autoref{tab:mbh}).
The quasars P056$-$16, P239$-$07, and J2229$+$1457 present a redshift of the \civ{} line with respect to the systemic redshift \citepalias[see][]{Schindler2020}.
For these, the \citeauthor{Coatman2017} correction has not been applied.
Henceforth we will consider as \textit{fiducial} the black hole masses derived from the \mgii{} line using the \citet{Shen2011} estimator.
This is selected mainly to remain consistent with previous studies of large samples of intermediate--redshift quasars \citep[in particular the one from SDSS, e.g.,][]{Shen2011} that we will compare our results with.
For objects that only have a successful \civ{} line fit (see \autoref{sec:fit}), we will instead adopt the masses calculated using the \citet{Vestergaard2006} estimator.
These choices will have an impact on our result.
To quantify this, in \autoref{fig:mgii_vs_civ} we compare black hole mass estimates for the sub--sample of 28~quasars for which we have both the \mgii{} and the \civ{} line fits.
In general, due to different zero--points, the \mgii{}--based estimator from \citet[][]{Vestergaard2009} predicts masses that are systematically offset by $\sim0.2$\,dex with respect to that of \citet{Shen2011}.
Consequently, if we had used the \citeauthor{Vestergaard2009} recipe as our fiducial one, all Eddington ratios (see \autoref{sec:EddingtonRatios}) would be a factor $\sim1.5\times$ higher.
Masses derived from the \civ{} line with the \citet{Vestergaard2006} estimator are consistent within $\sim0.4$\,dex with those calculated using the \citet{Shen2011} calibration.
The \citet{Coatman2017} correction consistently reduces the scatter between the \citet{Vestergaard2006} and the \citet{Shen2011} black hole masses for quasars with $\textrm{FWHM}_\textrm{MgII}\lesssim4000\,\kms$.
Yet, it systematically under--estimates masses for quasars with a broader \mgii{} line.
This causes the masses derived with the \citet{Coatman2017} correction to be, typically, $\sim0.2$\,dex lower than the \citet{Shen2011} ones for the entire sub--sample, and $\sim0.5$\,dex lower if we limit the comparison to quasars with $\textrm{FWHM}_\textrm{MgII}>4000\,\kms$.
To avoid the introduction of biases in the masses of the most massive objects, we therefore opt not to use the \citeauthor{Coatman2017} correction for the measured \civ{} FWHMs.

\subsection{Eddington Ratios}\label{sec:EddingtonRatios}

%
The so--called Eddington ratio ($\Eratio{}$) measures the ratio of the bolometric luminosity of a quasar (see \autoref{sec:BolometricLuminosities}) to its Eddington luminosity ($\Ledd{}$): i.e., the theoretical maximum luminosity that can be emitted when radiation pressure and gravity are in equilibrium in a spherical geometry \citep[][]{Eddington1926}.
While the exact limit depends on the chemical composition of the gas surrounding the SMBH, the Eddington luminosity that is typically reported assumes hydro--static equilibrium of pure ionized hydrogen:
\begin{equation}\label{eq:Ledd}
\begin{aligned}
    \Ledd{} & = \frac{4 \pi G \mbh{} m_\mathrm{p} c}{\sigma_\mathrm{T}} \\ 
    & = 1.257\times10^{38}\,\ergs{}\left(\frac{\mbh{}}{\msun{}}\right),
\end{aligned}
\end{equation}
where, $G$ is the gravitational constant, $m_\mathrm{p}$ is the mass of a proton, $c$ is the speed of light, and $\sigma_\mathrm{T}$ is the Thomson scattering cross--section. 
Eddington ratios ($\Eratio{}=\Lbol{}/\Ledd{}$) calculated for different mass estimators (see \autoref{sec:BlackHoleMasses}) are reported in \autoref{tab:mbh}.
{\begin{deluxetable*}{lCCCCCCCCC}
\tabletypesize{\footnotesize} 
\tablecaption{Properties of the quasars observed with \xs\label{tab:mbh}}
\tablehead{
\colhead{ID}                            &
\colhead{$L_\textrm{bol}$}              &
\colhead{$\mbh^\textrm{VO09}$}          &
\colhead{$\Eratio^\textrm{VO09}$}       &
\colhead{$\mbh^\textrm{S+11}$}          &
\colhead{$\Eratio^\textrm{S+11}$}       &
\colhead{$\mbh^\textrm{VP06}$}          &
\colhead{$\Eratio^\textrm{VP06}$}       &
\colhead{$\mbh^\textrm{C+17}$}          &
\colhead{$\Eratio^\textrm{C+17}$}       \\
\colhead{}                      &
\colhead{($10^{46}\,\ergs{}$)}  &
\colhead{($10^9\,\msun{}$)}     &
\colhead{}                      &
\colhead{($10^9\,\msun{}$)}     &
\colhead{}                      &
\colhead{($10^9\,\msun{}$)}     &
\colhead{}                      &
\colhead{($10^9\,\msun{}$)}     &
\colhead{}            
}
\startdata
 P004$+$17    & \phn\phn6.21^{+0.20}_{-0.21} & \nodata              & \nodata              & \nodata                  & \nodata              & \phn1.35^{+0.28}_{-0.27} & 0.36^{+0.08}_{-0.06} & 0.59^{+0.12}_{-0.11}  &  0.84^{+0.17}_{-0.13} \\
 P007$+$04    &    \phn20.13^{+0.67}_{-0.81} & 1.46^{+2.31}_{-0.49} & 1.08^{+0.57}_{-0.66} & \phn2.27^{+3.58}_{-0.75} & 0.70^{+0.36}_{-0.42} & \phn5.51^{+2.01}_{-1.48} & 0.29^{+0.11}_{-0.08} & 2.34^{+0.90}_{-0.63}  &  0.68^{+0.26}_{-0.20} \\
 P009$-$10    &    \phn19.32^{+0.21}_{-0.21} & \nodata              & \nodata              & \nodata                  & \nodata              &    18.44^{+5.32}_{-4.68} & 0.36^{+0.10}_{-0.08} & 2.35^{+0.61}_{-0.54}  &  2.81^{+0.80}_{-0.65} \\
 P011$+$09    & \phn\phn8.22^{+0.16}_{-0.16} & 0.46^{+0.36}_{-0.14} & 1.42^{+0.64}_{-0.62} & \phn0.64^{+0.49}_{-0.19} & 1.02^{+0.46}_{-0.45} & \phn2.24^{+0.87}_{-0.62} & 0.29^{+0.11}_{-0.08} & 0.43^{+0.14}_{-0.10}  &  1.52^{+0.46}_{-0.36} \\
 J0046$-$2837 & \phn\phn7.98^{+0.34}_{-0.34} & 0.35^{+0.04}_{-0.04} & 1.80^{+0.30}_{-0.24} & \phn0.49^{+0.06}_{-0.06} & 1.30^{+0.21}_{-0.17} & \nodata                  & \nodata              & \nodata               &  \nodata              \\
 J0100$+$2802 &       153.39^{+0.16}_{-0.15} & 9.73^{+0.26}_{-0.24} & 1.25^{+0.03}_{-0.03} &    19.28^{+0.51}_{-0.47} & 0.63^{+0.02}_{-0.02} & \nodata                  & \nodata              & \nodata               &  \nodata              \\
 J0109$-$3047 & \phn\phn7.69^{+0.20}_{-0.20} & 1.11^{+0.40}_{-0.36} & 0.55^{+0.26}_{-0.14} & \phn1.53^{+0.56}_{-0.49} & 0.40^{+0.19}_{-0.10} & \phn2.68^{+0.64}_{-0.58} & 0.23^{+0.06}_{-0.04} & 0.48^{+0.11}_{-0.09}  &  1.27^{+0.27}_{-0.22} \\
 P036$+$03    &    \phn24.96^{+0.14}_{-0.15} & 3.74^{+0.59}_{-0.51} & 0.53^{+0.08}_{-0.07} & \phn5.96^{+0.93}_{-0.81} & 0.33^{+0.05}_{-0.04} &    19.64^{+1.86}_{-1.63} & 0.10^{+0.01}_{-0.01} & 2.77^{+0.30}_{-0.24}  &  0.72^{+0.07}_{-0.07} \\
 J0305$-$3150 & \phn\phn9.42^{+0.10}_{-0.12} & 0.54^{+0.11}_{-0.12} & 1.39^{+0.40}_{-0.25} & \phn0.77^{+0.16}_{-0.18} & 0.98^{+0.29}_{-0.17} & \phn4.14^{+0.34}_{-0.30} & 0.18^{+0.01}_{-0.01} & 2.87^{+0.35}_{-0.27}  &  0.26^{+0.03}_{-0.03} \\
 P056$-$16    &    \phn10.59^{+0.11}_{-0.11} & 0.96^{+0.08}_{-0.07} & 0.87^{+0.08}_{-0.07} & \phn1.39^{+0.11}_{-0.10} & 0.61^{+0.05}_{-0.05} & \phn0.66^{+0.03}_{-0.02} & 1.28^{+0.05}_{-0.05} & \nodata               &  \nodata              \\
 P065$-$26    &    \phn21.69^{+0.46}_{-0.45} & 4.60^{+0.57}_{-0.48} & 0.38^{+0.05}_{-0.05} & \phn7.20^{+0.88}_{-0.75} & 0.24^{+0.03}_{-0.03} & \phn7.19^{+0.47}_{-0.49} & 1.23^{+0.14}_{-0.12} & 0.90^{+0.06}_{-0.05}  &  9.85^{+1.15}_{-1.05} \\
 P065$-$19    &    \phn19.14^{+0.38}_{-0.39} & 2.08^{+0.12}_{-0.10} & 0.73^{+0.04}_{-0.04} & \phn3.21^{+0.19}_{-0.16} & 0.47^{+0.03}_{-0.03} & \phn2.46^{+0.21}_{-0.17} & 2.66^{+0.65}_{-0.52} & 1.14^{+0.05}_{-0.05}  &  5.77^{+1.27}_{-1.08} \\
 J0842$+$1218 &    \phn19.37^{+0.20}_{-0.20} & 1.64^{+0.15}_{-0.12} & 0.94^{+0.08}_{-0.08} & \phn2.54^{+0.23}_{-0.18} & 0.61^{+0.05}_{-0.05} & \phn4.17^{+0.18}_{-0.18} & 0.37^{+0.02}_{-0.02} & 1.80^{+0.08}_{-0.07}  &  0.85^{+0.04}_{-0.04} \\
 J1030$+$0524 &    \phn23.97^{+0.37}_{-0.36} & 1.93^{+0.23}_{-0.20} & 0.99^{+0.12}_{-0.11} & \phn3.06^{+0.37}_{-0.31} & 0.62^{+0.08}_{-0.07} & \phn2.65^{+0.60}_{-0.71} & 0.71^{+0.26}_{-0.13} & 2.79^{+1.87}_{-0.44}  &  0.68^{+0.12}_{-0.27} \\
 P158$-$14    &    \phn46.75^{+1.08}_{-1.04} & 1.70^{+0.25}_{-0.22} & 2.18^{+0.34}_{-0.31} & \phn2.93^{+0.43}_{-0.38} & 1.27^{+0.20}_{-0.18} & \phn7.99^{+0.70}_{-0.65} & 0.46^{+0.04}_{-0.04} & 4.97^{+2.83}_{-1.14}  &  0.75^{+0.23}_{-0.28} \\
 P159$-$02    &    \phn18.21^{+0.32}_{-0.32} & 3.14^{+0.38}_{-0.31} & 0.46^{+0.05}_{-0.05} & \phn4.82^{+0.58}_{-0.48} & 0.30^{+0.03}_{-0.03} & \phn2.48^{+0.20}_{-0.17} & 0.57^{+0.05}_{-0.04} & 1.38^{+0.12}_{-0.10}  &  1.03^{+0.08}_{-0.08} \\
 J1044$-$0125 &    \phn26.09^{+0.45}_{-0.43} & \nodata              & \nodata              & \nodata                  & \nodata              & \phn6.12^{+2.73}_{-1.85} & 0.34^{+0.14}_{-0.10} & 3.37^{+0.52}_{-0.47}  &  0.61^{+0.10}_{-0.08} \\
 J1048$-$0109 &    \phn10.55^{+0.30}_{-0.30} & 2.29^{+0.64}_{-0.57} & 0.37^{+0.13}_{-0.08} & \phn3.29^{+0.93}_{-0.82} & 0.26^{+0.09}_{-0.06} & \nodata                  & \nodata              & \nodata               &  \nodata              \\
 J1120$+$0641 &    \phn17.05^{+0.07}_{-0.07} & \nodata              & \nodata              & \nodata                  & \nodata              & \phn4.79^{+0.12}_{-0.11} & 0.28^{+0.01}_{-0.01} & 2.40^{+0.06}_{-0.05}  &  0.57^{+0.01}_{-0.01} \\
 J1148$+$0702 &    \phn16.71^{+0.22}_{-0.19} & 3.13^{+0.21}_{-0.19} & 0.42^{+0.03}_{-0.03} & \phn4.75^{+0.31}_{-0.28} & 0.28^{+0.02}_{-0.02} & \phn3.10^{+0.32}_{-0.31} & 0.43^{+0.05}_{-0.04} & 1.18^{+0.11}_{-0.09}  &  1.11^{+0.09}_{-0.10} \\
 P183$+$05    &    \phn22.43^{+0.30}_{-0.30} & 3.00^{+0.43}_{-0.43} & 0.59^{+0.10}_{-0.08} & \phn4.72^{+0.68}_{-0.67} & 0.38^{+0.06}_{-0.05} & \phn9.97^{+1.71}_{-1.36} & 0.18^{+0.03}_{-0.02} & 1.51^{+0.37}_{-0.25}  &  1.18^{+0.23}_{-0.23} \\
 J1306$+$0356 &    \phn18.74^{+0.12}_{-0.12} & 2.10^{+0.07}_{-0.06} & 0.71^{+0.02}_{-0.02} & \phn3.23^{+0.11}_{-0.10} & 0.46^{+0.01}_{-0.02} & \phn3.16^{+0.10}_{-0.11} & 0.46^{+0.02}_{-0.01} & 2.33^{+0.06}_{-0.06}  &  0.63^{+0.02}_{-0.02} \\
 J1319$+$0950 &    \phn17.67^{+0.06}_{-0.06} & 1.89^{+0.11}_{-0.11} & 0.75^{+0.05}_{-0.04} & \phn2.88^{+0.17}_{-0.17} & 0.49^{+0.03}_{-0.03} & \phn9.76^{+0.24}_{-0.23} & 0.14^{+0.00}_{-0.00} & 2.86^{+0.11}_{-0.10}  &  0.49^{+0.02}_{-0.02} \\
 J1342$+$0928 &    \phn19.43^{+0.13}_{-0.01} & \nodata              & \nodata              & \nodata                  & \nodata              &    21.78^{+0.80}_{-1.00} & 0.07^{+0.00}_{-0.00} & 1.97^{+0.09}_{-0.08}  &  0.79^{+0.03}_{-0.03} \\
 J1509$-$1749 &    \phn25.09^{+0.22}_{-0.24} & 2.65^{+0.21}_{-0.19} & 0.76^{+0.06}_{-0.06} & \phn4.22^{+0.33}_{-0.30} & 0.47^{+0.04}_{-0.03} & \phn3.23^{+0.21}_{-0.20} & 0.61^{+0.04}_{-0.04} & 2.49^{+0.18}_{-0.16}  &  0.80^{+0.06}_{-0.06} \\
 P231$-$20    &    \phn22.34^{+0.54}_{-0.52} & 4.06^{+1.00}_{-0.74} & 0.43^{+0.10}_{-0.08} & \phn6.40^{+1.56}_{-1.18} & 0.28^{+0.06}_{-0.05} & \nodata                  & \nodata              & \nodata               &  \nodata              \\
 P239$-$07    &    \phn30.37^{+0.38}_{-0.43} & 2.29^{+0.17}_{-0.18} & 1.05^{+0.10}_{-0.08} & \phn3.73^{+0.28}_{-0.29} & 0.65^{+0.06}_{-0.05} & \phn1.82^{+0.91}_{-0.45} & 1.30^{+0.42}_{-0.43} & \nodata               &  \nodata              \\
 P308$-$21    &    \phn21.89^{+0.19}_{-0.19} & 1.69^{+0.20}_{-0.35} & 1.03^{+0.27}_{-0.11} & \phn2.65^{+0.32}_{-0.56} & 0.66^{+0.17}_{-0.07} & \phn5.87^{+1.14}_{-1.19} & 0.29^{+0.07}_{-0.05} & 2.09^{+0.52}_{-0.49}  &  0.82^{+0.25}_{-0.16} \\
 J2054$-$0005 &    \phn12.37^{+0.37}_{-0.31} & 1.48^{+0.18}_{-0.17} & 0.66^{+0.10}_{-0.07} & \phn2.17^{+0.27}_{-0.25} & 0.45^{+0.06}_{-0.05} &    10.24^{+3.95}_{-2.79} & 0.10^{+0.04}_{-0.03} & 1.92^{+0.83}_{-0.52}  &  0.51^{+0.19}_{-0.15} \\
 J2100$-$1715 & \phn\phn4.80^{+0.10}_{-0.11} & 4.49^{+0.92}_{-1.51} & 0.08^{+0.04}_{-0.01} & \phn5.88^{+1.18}_{-1.96} & 0.06^{+0.03}_{-0.01} & \phn2.04^{+1.44}_{-0.46} & 1.00^{+0.42}_{-0.40} & 2.55^{+1.33}_{-0.71}  &  0.82^{+0.45}_{-0.30} \\
 P323$+$12    &    \phn20.04^{+0.20}_{-0.21} & 1.13^{+0.14}_{-0.14} & 1.42^{+0.21}_{-0.16} & \phn1.75^{+0.22}_{-0.22} & 0.92^{+0.13}_{-0.11} & \phn1.38^{+0.08}_{-0.07} & 1.14^{+0.06}_{-0.06} & 2.29^{+0.10}_{-0.10}  &  0.69^{+0.03}_{-0.03} \\
 J2211$-$3206 &    \phn29.97^{+0.58}_{-0.61} & 3.82^{+0.30}_{-0.25} & 0.63^{+0.05}_{-0.05} & \phn6.21^{+0.48}_{-0.41} & 0.38^{+0.03}_{-0.03} & \phn2.22^{+0.25}_{-0.24} & 1.05^{+0.11}_{-0.09} & 1.12^{+0.10}_{-0.10}  &  2.10^{+0.17}_{-0.16} \\
 J2229$+$1457 & \phn\phn2.86^{+0.17}_{-0.15} & \nodata              & \nodata              & \nodata                  & \nodata              & \phn0.03^{+0.00}_{-0.00} & 7.71^{+1.10}_{-0.96} & \nodata               &  \nodata              \\
 P340$-$18    &    \phn13.28^{+0.16}_{-0.15} & 0.71^{+0.11}_{-0.08} & 1.49^{+0.19}_{-0.20} & \phn1.05^{+0.16}_{-0.12} & 1.01^{+0.13}_{-0.14} & \phn0.29^{+0.01}_{-0.01} & 3.71^{+0.19}_{-0.18} & 0.59^{+0.03}_{-0.03}  &  1.78^{+0.08}_{-0.09} \\
 J2310$+$1855 &    \phn39.13^{+0.47}_{-0.44} & 2.19^{+0.33}_{-0.27} & 1.42^{+0.20}_{-0.19} & \phn3.68^{+0.55}_{-0.46} & 0.85^{+0.12}_{-0.11} &    49.23^{+3.13}_{-3.27} & 0.06^{+0.00}_{-0.00} & 7.51^{+0.47}_{-0.52}  &  0.41^{+0.03}_{-0.02} \\
 J2318$-$3029 &    \phn14.53^{+0.20}_{-0.21} & 1.45^{+0.15}_{-0.13} & 0.79^{+0.08}_{-0.08} & \phn2.17^{+0.23}_{-0.19} & 0.53^{+0.05}_{-0.05} & \phn3.87^{+0.44}_{-0.37} & 0.30^{+0.03}_{-0.03} & 1.94^{+0.23}_{-0.19}  &  0.59^{+0.06}_{-0.06} \\
 J2348$-$3054 & \phn\phn7.51^{+0.20}_{-0.19} & 2.35^{+0.85}_{-0.67} & 0.25^{+0.10}_{-0.07} & \phn3.25^{+1.17}_{-0.93} & 0.18^{+0.08}_{-0.05} & \phn1.18^{+0.15}_{-0.14} & 0.50^{+0.06}_{-0.05} & 0.68^{+0.08}_{-0.07}  &  0.87^{+0.10}_{-0.08} \\
 P359$-$06    &    \phn25.65^{+0.29}_{-0.31} & 2.36^{+0.56}_{-0.44} & 0.86^{+0.20}_{-0.17} & \phn3.78^{+0.89}_{-0.70} & 0.54^{+0.12}_{-0.10} & \phn1.35^{+0.09}_{-0.09} & 1.49^{+0.10}_{-0.09} & 1.32^{+0.10}_{-0.09}  &  1.52^{+0.12}_{-0.11} \\
\enddata 
\end{deluxetable*}
}
%

\section{Results and Discussion}\label{sec:results}

\begin{figure}[tb]
    \centering
    \includegraphics[width=0.98\columnwidth]{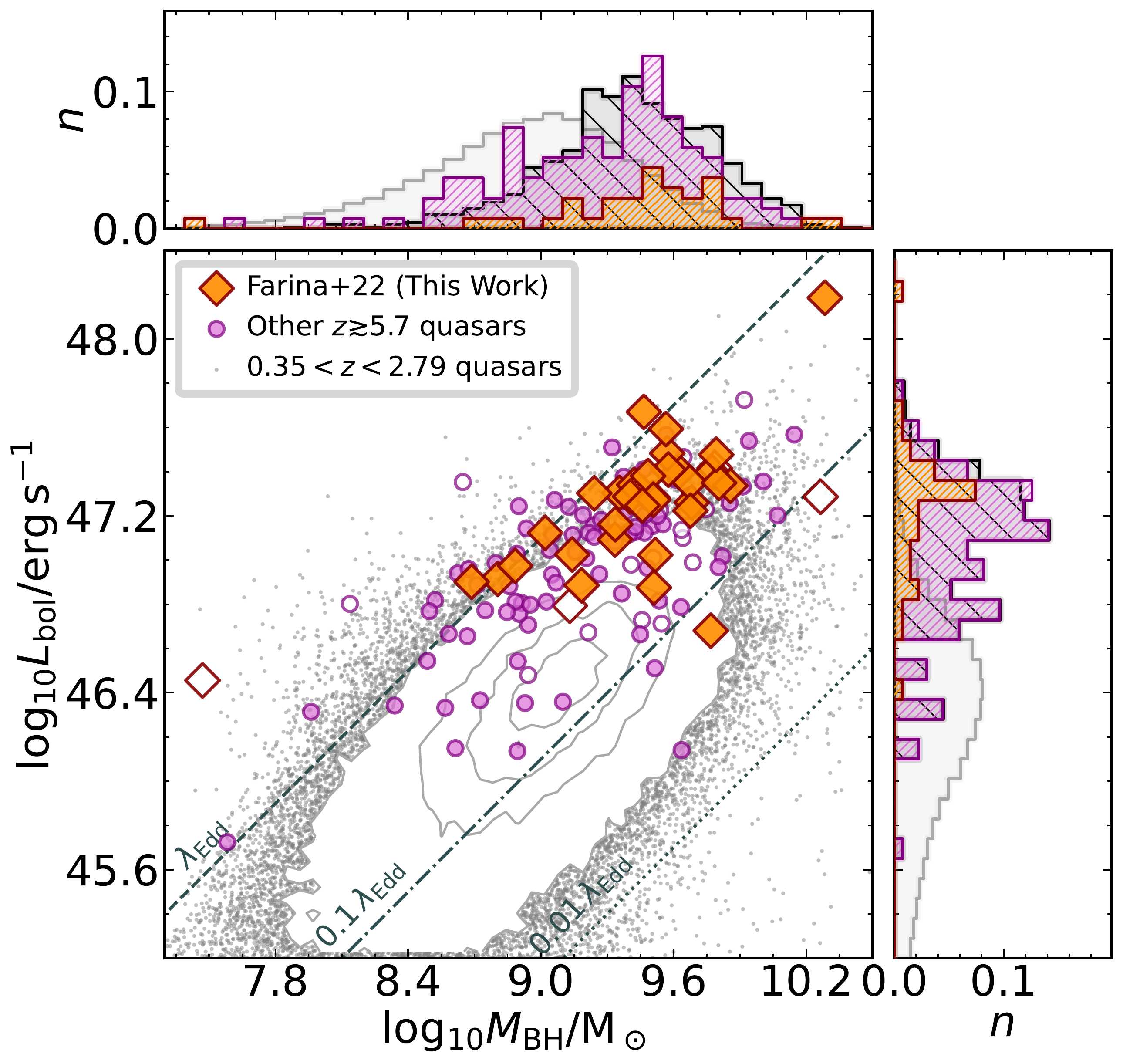}
    \caption{
    Distribution of our \textit{\mbox{\xs{}/ALMA} sample} (orange diamonds) and from other $z \gtrsim 5.7$ studies (violet circles) in the black hole mass versus bolometric luminosity plane.
    Filled symbols indicate that the \mgii{} line was used to derive the black hole mass, while empty symbols indicate the use of the \civ{} line.
    Gray contours and gray points indicate the location of SDSS DR7 $0.35 < z < 2.79$ quasars for which either \mgii{} or \civ{} lines were fitted by \citet{Shen2011}.
    Diagonal gray lines mark lines of constant Eddington ratio, with dotted, dash-dotted and dashed lines indicating $0.01\times$, $0.1\times$, and $1\times$ the Eddington limit, respectively. 
    Histograms in the right and top panels are normalized to the total number of objects and follow the same color scheme as the main panel.
    The top and right panel also shows the distribution of $0.6 \lesssim z \lesssim 2.3$  luminosity--matched quasars as black hatched histograms. 
    }
    \label{fig:mbhlbol}
\end{figure}

%
To summarize our findings: our sample of \NQSO{} bright (with $\Lbol{} \gtrsim 2.9\times10^{46}\,\ergs$) $z > 5.7$ quasars have black hole masses ranging from $\sim3.0\times10^7\,\msun{}$ to $\sim2.2\times10^{10}\,\msun{}$ (with an average of $\sim4.6\times10^9\,\msun{}$ and a median of $\sim3.2\times10^9\,\msun{}$) and typically have accretion rates between 0.1 and 1 Eddington (see \autoref{fig:mbhlbol}, orange diamonds).
Our results are complemented by all the $z \gtrsim 5.7$ quasars for which deep optical--to--near--IR spectroscopic observations are available in the literature (see \autoref{fig:sample}).
This literature sample consists of \NQSOLIT{} quasars from \citet{Willott2010BH}, \citet{DeRosa2011}, \citet{Mazzucchelli2017}, \citet{Chehade2018}, \citet{Onoue2019}, \citet{Reed2019}, \citet{Pons2019}, \citet{Shen2019}, \citet{Matsuoka2019}, \citet{Andika2020}, \citet{Eilers2021}, \citet{WangFeige2020z7, WangFeige2021, WangFeige2021XRay}, \citet{Banados2021}, and \citet[][, pink circles in \autoref{fig:mbhlbol}]{Yang2020z75, Yang2021}.
All these measurements have been homogenized to the same cosmological parameter and to the same bolometric luminosity and black hole mass estimators used in this paper (see \autoref{sec:SMBH} for details).
Quasars duplicated among these samples are removed and only the most recent measurement is taken into account.
The \mgii{} emission line of the $z \gtrsim 7.1$ quasars in our sample is redshifted into the order 11 of the \xs{} near--IR arm.
At these wavelengths, the efficiency of the instrument is hampered by the rise of the thermal background.
Instead of the values derived from the \civ{} line listed in \autoref{tab:mbh}, we will recalculate black hole masses and Eddington ratios of the two quasars J1120$+$0641 ($z = 7.08$) and J1342$+$0928 ($z=7.54$) using the measurements of the \mgii{}--line (based on \textit{Gemini}/GNIRS spectra) reported in \citet{Yang2021}.
In the following sections, we will refer to the \NQSO{} quasars studied here as the \textit{\xs{}/ALMA sample}, while to the \NQSO{}$+$\NQSOLIT{}$=$135 $z\gtrsim5.7$ quasars with black hole mass estimates as the \textit{full sample}.
In all the figures, quasars part of the \textit{\xs{}/ALMA sample} will be plot as orange diamonds (filled in case the black hole mass is derived from the \mgii{} line and empty if from the \civ{} line), while measurements from the literature will be shown as pink circles (with the same convention for black hole mass estimators).
Histograms will follow the same color scheme and the \textit{\xs{}/ALMA sample} and the literature sample will be stacked one on the top of the other to represent the distribution of the \textit{full sample}.
Both their normalization will be set by the relevant number of objects in the \textit{full sample}.
Box plots above the histograms will indicate the median and the 25$^\textrm{th}$ and 75$^\textrm{th}$ percentiles, while the whiskers will mark the 16$^\textrm{th}$ and 84$^\textrm{th}$ ones.
In \autoref{fig:mbhlbol} we also show the location of lower redshift ($0.35 < z < 2.79$) SDSS DR7 quasars that have either \mgii{} or \civ{} black hole mass estimates by \citet[][, gray contours and points]{Shen2011}.
The entirety of the full sample of $z \gtrsim 5.7$ quasars, spanning $\sim2.5$ orders of magnitude in $\Lbol{}$, appears to inhabit the same locus of similarly luminous objects at lower redshifts.
To quantify this, we constructed a bolometric--luminosity matched control sample of $0.56 < z < 2.30$ quasars.
Given that we calculate bolometric luminosities following \autoref{eq:lbol}, this choice corresponds to matching the monochromatic luminosities at 3000\,\AA{} of high-- and low--redshift quasars.
The redshift range was selected to ensure a good spectral coverage of the \mgii{} and of the \civ{} broad emission lines, to avoid potential contamination from host galaxies \citep[e.g.,][]{Shen2011}, and to reduce the impact of biases introduced due to the low efficiency in the SDSS color selection of $2.7 \lesssim z \lesssim 3.5$ quasars \citep[e.g.,][]{Richards2006, Richards2006LF, Worseck2011}.
Specifically, starting from the SDSS DR7 sample, we first removed all sources for which the \mgii{} and the \civ{} lines were too narrow or poorly measured, i.e., for both lines we required: \texttt{FWHM\_LINE}\(>\)2\(\times\)\texttt{FWHM\_LINE\_ERR}, \texttt{EW\_LINE}\(>\)2\(\times\)\texttt{EW\_LINE\_ERR}, and \texttt{FWHM\_LINE}\(>\)1000\,$\kms{}$ \citep[see e.g., ][]{Richards2006}.
For each of the $z \gtrsim 5.7$ quasars, we then selected 10 radio--quiet sources with bolometric luminosities closest to the luminosity of the quasar. 
In addition, consistency in the black hole mass estimators is accomplished by making sure that each high--redshift quasar and its 10 low--redshift analogues have masses calculated with the same estimator (black histograms in \autoref{fig:mbhlbol}).
A two---dimensional \citeauthor{Kolmogorov1933}--\citeauthor{Smirnov1948} test \citep[][]{Fasano1987} comparing the distribution in the black hole masses versus bolometric luminosity plane of the \textit{full sample} and the luminosity matched control sample returns a $p-\textrm{value}\sim3.6\times10^{-2}$.
Thus, the difference between the two is not significant enough (at the $\sim{}2.1\sigma$ level) to imply that they are drawn from distinct parent populations.
However, while the distribution of bolometric luminosities of the two samples are consistent by construction, the control sample appears to have, on average, slightly larger black hole masses with respect to the population of $z \gtrsim 5.7$ quasars (top panel of \autoref{fig:mbhlbol}).
A two sample \citeauthor{Kolmogorov1933}--\citeauthor{Smirnov1948} test performed with bootstrap resampling over the black hole masses distributions of the \textit{full sample} and of the luminosity matched control sample results in $p-\textrm{value}\sim1.9\times10^{-3}$, implying that the two distribution are different at the $\sim{}3.1\,\sigma$ level ($\sim{}3.2\,\sigma$ if only black hole masses derived from the \mgii{} are considered).
In the following sections we will quantify this difference in terms of Eddington ratios.
Note that we excluded from these comparisons the ultra--luminous quasar J0100$+$2802.
While the known number of such bright quasars is increasing \citep[e.g.,][]{Schindler2017, Schindler2018, Schindler2019, Calderone2019}, they are still under--represented in the SDSS sample.
Its inclusion would otherwise bias the control sample towards lower luminosities.
%


\subsection{The Eddington Ratio Distribution}\label{sec:eratio}

%
Our analysis shows that the bright quasars in our \textit{\mbox{\xs{}/ALMA} sample} are accreting at a moderately high Eddington ratio with a median of $\Eratio=0.48^{+0.06}_{-0.02}$ (with errors derived from bootstrap resampling of the distributions) with 16$^\textrm{th}$ and 84$^\textrm{th}$ percentiles equal to $p_\mathrm{16th}=0.29$ and $p_\mathrm{84th}=0.94$.
More generally, $z \gtrsim 5.7$ quasars in the \textit{full sample} have Eddington ratios within the fairly narrow range with $p_\mathrm{16th}=0.23$ and $p_\mathrm{84th}=0.97$ (with a median of $\Eratio=0.47^{+0.05}_{-0.01}$).
For comparison, \citet{Shen2019} report a median Eddington ratios of $\sim0.31$.
While \citet{Mazzucchelli2019} and \citet{Yang2021}, using the \citet{Vestergaard2009} black hole mass estimator, report median values of $\sim0.31$ and $\sim0.85$, respectively.
Intriguingly, despite the large systematics in the black hole mass and bolometric luminosity estimates (of the order of $0.3-0.5$\,dex, see \autoref{sec:SMBH}), only three quasars appear to have $\Eratio > 2.0$.
All three have masses derived from the \civ{} line which are intrinsically less accurate\footnote{
A detailed discussion on J2229$+$1457, the quasar in the \textit{\xs{}/ALMA sample} with the highest Eddington ration ($\Eratio\sim7.7$), is presented in \autoref{app:j2229}.
} (see \autoref{sec:BlackHoleMasses}).
Therefore, there is no clear evidence for a significant population of $5.7 \lesssim z \lesssim 7.6$ quasars accreting at super--Eddington rates.
However, one should keep mind that accretion disc models predict that, at luminosities approaching Eddington, advective energy transport dominates over radiative cooling, causing the Eddington ratio to \textit{saturate} at $\Eratio\gtrsim3$ \citep[][, see also \citealt{Jiang2019}]{Abramowicz1988, Watarai2000, Sadowski2009, Sadowski2011, Madau2014, Kubota2019}.
\begin{figure}[tb]
    \centering
    \includegraphics[width=0.98\columnwidth]{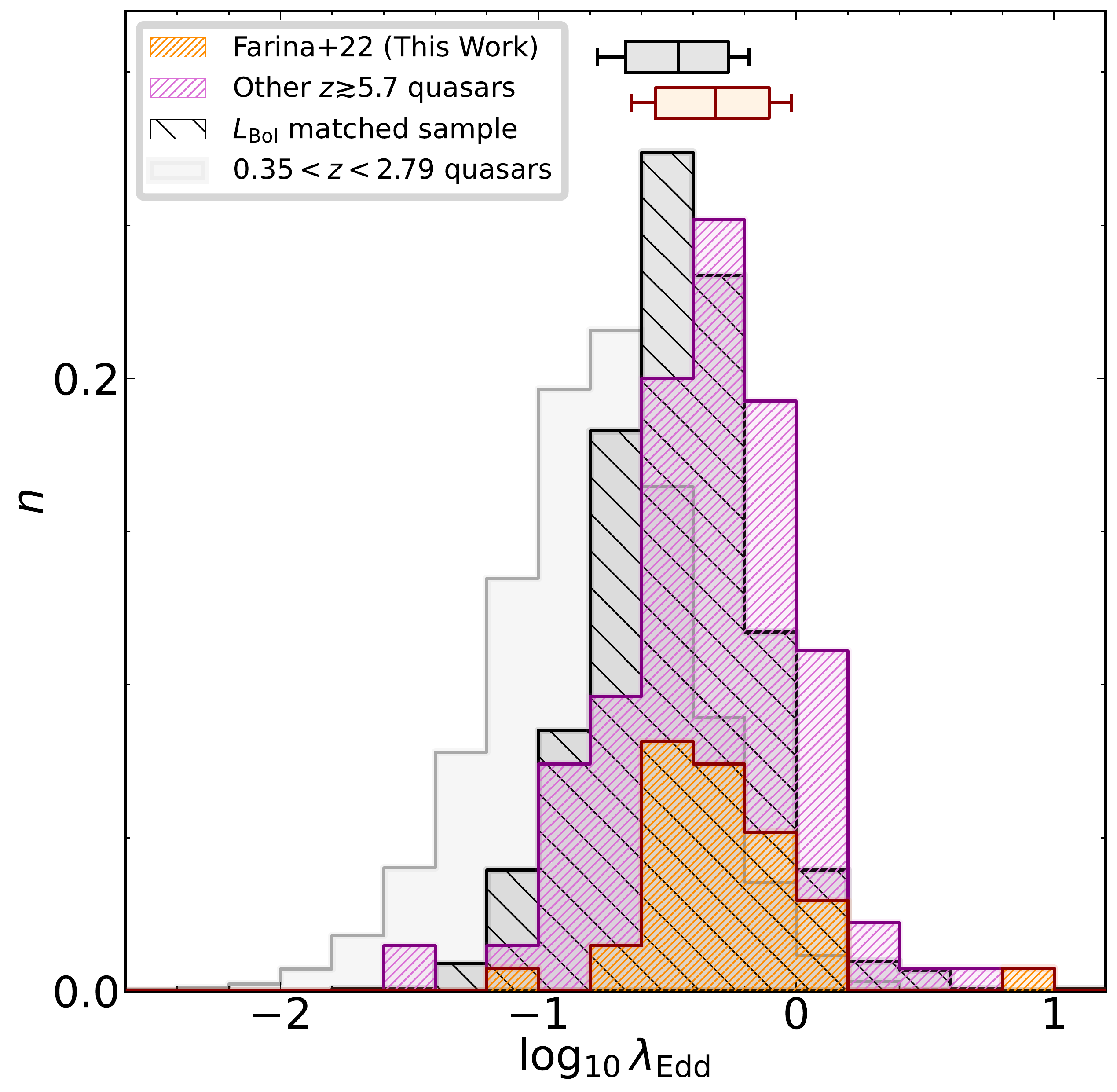}
    \caption{
    Logarithmic Eddington ratio distributions for the full sample of $z \gtrsim 5.7$ quasars (orange for our \textit{\xs{} sample} and violet for other studies) and for the bolometric luminosity--matched sample described in \autoref{sec:results} (black hatched histogram).
    The high and low redshift distributions have similar dispersions ($\sim 0.3$) but $z \gtrsim 5.7$ quasars have a median Eddington ratio (top red rectangle) that is higher than their lower redshift counterparts (top black rectangle).
    For the sake of comparison, we also show the sample of $0.35 < z < 2.79$ SDSS quasars as a pale gray histogram.
    }
    \label{fig:eratios}
\end{figure}
These findings are summarized in \autoref{fig:eratios}, where we compare the Eddington ratio distribution of our \textit{full sample} to the one of the bolometric luminosity--matched low--redshift quasars (excluding the ultra--luminous quasar J0100$+$2802, see \autoref{sec:results}).
In logarithmic space, both distributions are well fitted by a Gaussian function with a logarithmic dispersion of $\sim 0.3$ (consistent with values reported by \citealt{Mazzucchelli2017} and \citealt{Shen2019}).
However, the median Eddington ratio of the control sample ($\Eratio=0.35^{+0.01}_{-0.01}$ with $p_\mathrm{16th}=0.17$ and $p_\mathrm{84th}=0.66$) is lower than the high--redshift value. 
We test the significance of this difference by performing a \citeauthor{Kolmogorov1933}--\citeauthor{Smirnov1948} test with bootstrap resampling over the two distributions.
The derived $p$--value is $5.2\times10^{-6}$ ($8.1\times10^{-8}$, if we consider only masses derived from the \mgii{} line), this allows us to reject the null hypothesis that the two samples are drawn from the same distribution at the $\sim4.6\sigma$ ($\sim5.4\sigma$) level.
This difference could be driven by the intrinsic flux limit of high--redshift quasar searches \citep[e.g.,][]{Malmquist1922, Malmquist1925}.
Which can translate into selecting more massive black holes at fixed galaxy mass \citep[e.g.,][]{Lauer2007} and into a bias towards objects with higher bolometric luminosities and, thus, higher Eddington ratios\footnote{This bias is, however, mitigated by the fact that both $\Lbol$ and $\mbh$ are linearly proportional to the monochromatic luminosity of the quasar's $L_{\lambda,3000\textrm{\AA}}$ (see \autoref{eq:lbol} and \autoref{eq:mbh}).}.
Given that quasars in the \textit{full sample} come from different surveys with different depths [for instance the 5--$\sigma$ $z$--band magnitude limit of the \textit{Subaru} High--$z$ Exploration of Low--Luminosity Quasars survey is $z_{5\sigma}\sim25.1$\,mag \citep{Matsuoka2016}, while that of \textit{Pan--STARRS1} is  $z_{5\sigma}\sim22.1$\,mag \citep{Banados2016}], one can naively expect that brighter quasars are impacted by this bias more.
\autoref{fig:eratios_lbol} compares the logarithmic distributions of Eddington ratios of our full sample of high--redshift quasars split into three bins of bolometric luminosity, each containing the same number of objects.
The median Eddington ratios of high--redshift quasars in these bins are: 
$\Eratio=0.47^{+0.13}_{-0.11}$ for the low--luminosity bin (\(\Lbol < 10^{46.94}\,\ergs{}\)),
$0.45^{+0.08}_{-0.03}$ for the intermediate--luminosity bin (\(10^{46.94}\,\ergs{} \le \Lbol < 10^{47.22}\,\ergs{}\)), and 
$0.51^{+0.03}_{-0.05}$ 
for the high--luminosity bin (\(\Lbol \ge 10^{47.22}\,\ergs{}\)).
The distribution of $\log_{10}\Eratio{}$ for low--luminosity quasars is slightly wider ($\sigma\sim0.45$) than for brighter quasars ($\sigma\sim0.29$ and $\sim0.26$ for intermediate and bright objects, respectively).
But the width of these distributions is broadly consistent with their corresponding low--redshift luminosity--matched samples ($\sigma\sim0.31$, $\sigma\sim0.25$, and $\sigma\sim0.28$, for low--, intermediate--, and high--luminosity quasars, respectively).
The broader distribution observed for in the high--redshift low--luminosity bin could be, partially, associated to intrinsically lower signal--to--noise ratio of the spectra of faint quasars \citep[e.g.,][]{Denney2009}.
It is however beyond the scope of the current study to fully settle this issue.
In general, the median values of the Eddington ratios of the control samples ($\Eratio=0.26^{+0.01}_{-0.01}$, $0.37^{+0.01}_{-0.02}$, and $0.44^{+0.02}_{-0.02}$ for the faint, intermediate, and high luminosity--matched control samples, respectively) are consistently lower than the high--redshift ratios, independent of the luminosity bin.
With the largest difference observed in the low luminosity bin.
This suggests that selection biases should not play a major role, and that $z \sim 6$ quasars at a given luminosity are actually accreting faster than their lower--redshift counterparts or that the a non--virial component contributes to the broadening of the BLR.
\begin{figure}[tb]
    \centering
    \includegraphics[width=0.98\columnwidth]{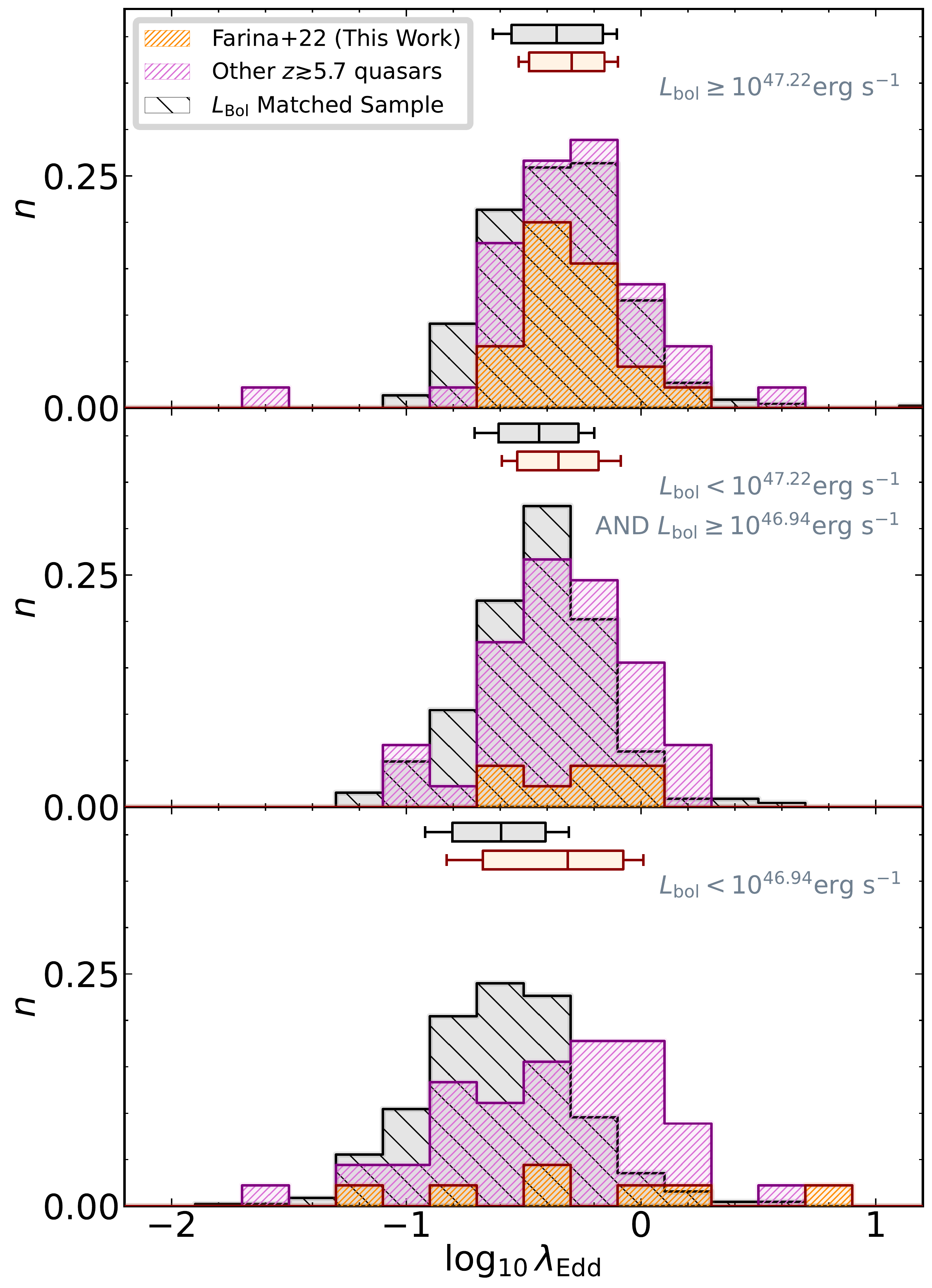}
    \caption{
    Eddington ratio distribution for the bright (top), intermediate (middle), and faint (bottom) samples of $z \gsim 5.7$ quasars in this study.
    For comparison, the Eddington ratios of the luminosity--matched samples are also shown.
    The rectangles on the top of each panel mark the median values of the different distributions and show that high--redshift quasars have higher median accretion rates than the lower redshift ones for all luminosity ranges.
    The color code is the same as in \autoref{fig:eratios}.
    }
    \label{fig:eratios_lbol}
\end{figure}
%

\subsection{Is there an evolution of intrinsic quasar properties at \texorpdfstring{$z > 6.5$}{redshift greater than 6.5}?}\label{sec:eratioredshift}

\begin{figure}[tb]
    \centering
    \includegraphics[width=0.98\columnwidth]{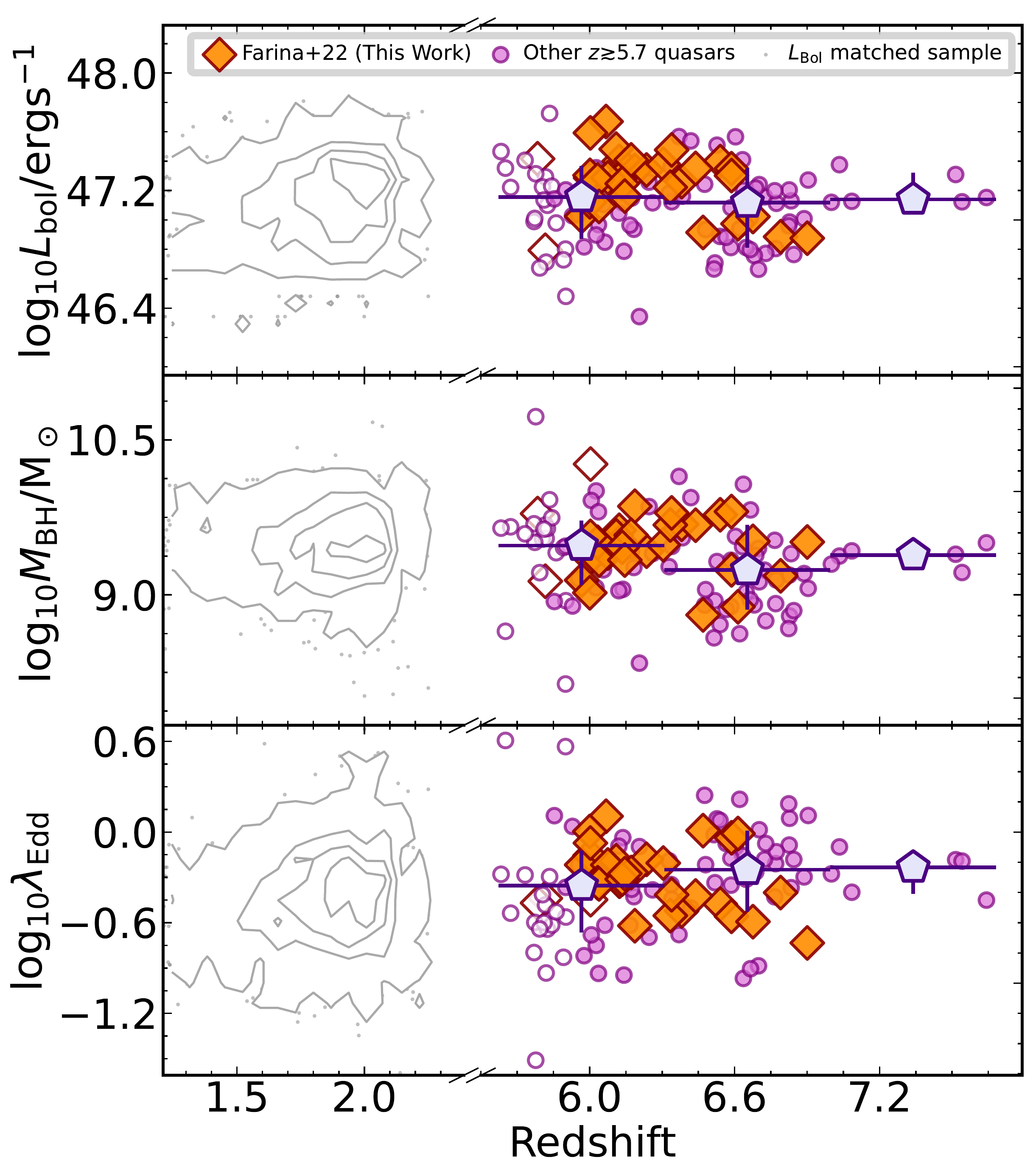}
    \caption{
    Redshift distribution of (from top to bottom) bolometric luminosities, black hole masses, and Eddington ratios for bright (M$_{1450} < -25.2$\,mag) $z \gtrsim 5.7$ quasars (color coded as in \autoref{fig:mbhlbol}).
    Blue pentagons mark median values in three redshift bins.
    Error--bars on the $y$--axis represent the 16$^\textrm{th}$ and the 84$^\textrm{th}$ percentiles of the distribution, while on the $x$--axis they indicate the size of the redshift bin.
    For comparison, the distribution of the luminosity--matched sample is also shown by gray contours.
    }
    \label{fig:eratios_redshift}
\end{figure}

%
The X--ray emission of luminous quasars is thought to originate from inverse Compton scattering of UV/optical photons produced by the accretion disk on a compact, hot corona surrounding the inner parts of the disk  \citep[e.g.,][]{Galeev1979, Haardt1991, Svensson1994}.
Recent \textit{XMM--Newton}, \textit{Chandra}, \textit{Swift}, and \textit{eRosita} investigations of $z\sim6$ quasars are providing insights into black hole accretion physics, complementing the UV emission line observations \citep[e.g.,][]{Ai2017, Nanni2017, Nanni2018, Connor2019, Connor2020, Connor2021J0100, Pons2020, Medvedev2020, Medvedev2021, Khorunzhev2021}.
Particular interest has been given to the intrinsic photon index of the $\sim0.5-10$\,keV power--law continuum ($\Gamma$), because of its correlation with the Eddington ratio \citep[e.g.,][]{Shemmer2008, Risaliti2009, Fanali2013}.
Studies of $z > 6.5$ quasars \citep[e.g.,][]{Moretti2014, Page2014, Banados2018XRay, Vito2019, Connor2021} revealed an average photon index of \(\Gamma=2.32^{+0.31}_{-0.30}\) \citep[][]{WangFeige2021XRay}.
This is steeper than what is reported for bright quasars in the redshift range $1.5\lesssim z \lesssim 4.5$ \citep[\(\Gamma=1.92^{+0.09}_{-0.08}\),][]{Just2007} and up to $z\sim6$ \citep[\(\Gamma=1.92^{+0.28}_{-0.27}\),][]{Nanni2017}.
This steepening suggests that $z > 6.5$ quasars should be accreting faster \citep[by a factor $\sim20$, if we take the face values of the steepening and the relation between hard X--ray spectral slope and the Eddington ratio derived by][ or by \citealt{Brightman2013}]{Risaliti2009} than lower--redshift analogues \citep[][]{WangFeige2021XRay}.
In \autoref{fig:eratios_redshift}, we test if this evolutionary trend also appears in our sample of bright quasars.
To ensure a uniform luminosity selection in the entire redshift range $5.7\lesssim z \lesssim 7.6$, we consider only quasars brighter than M$_{1450}=-25.2$\,mag \citep[i.e., the characteristic magnitude of $z\sim6$ quasars, e.g.,][, see \autoref{fig:sample}]{Jiang2016}, and we exclude the hyper--luminous quasar J0100$+$2802.
The median properties are calculated in three, uniformly spaced, redshift bins centered at $z=6.0$, $6.7$, and $7.3$ containing 60, 50, and 6 quasars, respectively.
The median bolometric luminosities in these three bins all fall in the small range  $\Lbol\sim10^{47.1}-10^{47.2}\,\ergs{}$, confirming that we are probing similarly luminous quasars at different redshifts.
The median Eddington ratios in the three redshift bins ($\Eratio = 
 0.44_{-0.06}^{+0.04}$,  
$0.56_{-0.07}^{+0.09}$, and, 
$0.59_{-0.12}^{+0.06}$, at  
$z=6.0$, $6.7$, and $7.3$, respectively) are marginally increased above $z \sim 6.5$ and are significantly higher than the median value in the $z \sim 2$ luminosity--matched sample ($\Eratio=0.37_{-0.01}^{+0.01}$).
Despite the different black hole mass estimators and fitting procedures used by \citet{Shen2019} and \citet{Yang2021}, our findings are qualitatively compatible with their results. 
Using a sample of 50 quasars at $5.6 < z < 6.5$, \citeauthor{Shen2019} report Eddington ratios consistent with those of their luminosity matched low redshift sample. 
While \citealt{Yang2021} show an increase in accretion rates in a sample of 37 quasars at $6.3 < z < 7.7$ with respect to $0.4 < z < 2.1$ quasars with similar luminosity.
It is important to note that other properties of the luminous quasar population also evolve above $z \sim 6.5$.
For instance, the number density of $\MMCDL{} < -26.0\,\mathrm{mag}$ objects drops twice as fast from $z \sim 6$ to $z \sim 6.7$ than across the range $3 \lesssim z \lesssim 5$ \citep[][]{WangFeige2019LF}. 
The average blueshift of the broad \civ{} emission line relative to low--ionization lines rises significantly above $z \sim 6$ \citep[][; \citetalias{Schindler2020}]{Meyer2019, Yang2021}.
The fraction of $z > 6.5$ broad absorption line quasars increases to $\sim 24\%$ compared to $\sim 15\%$ at lower redshifts \citep[][]{Yang2021, Bischetti2022}. 
Lastly, the fraction of $z \sim 6$ quasars showing weak emission lines (i.e., those with \civ{} rest frame equivalent widths of $W_{\mathrm{r}}^{\mathrm{CIV}}<10$\,\AA{}) appears to be higher than in bright SDSS quasars \citep[e.g.,][]{Banados2014, Shen2019}. 
These general evolutionary trends, however, are only marginally reflected in the changes in the accretion rate of reionization--era quasars, that merely increase by $\sim0.1$\,dex from $z \sim 6$ to $z \sim 7$.
This suggests that the steep $\Gamma$ reported by \citet[][]{WangFeige2021XRay}, from the joint analysis of the X--ray spectra of six $z\gtrsim6.5$ quasars, can not be explained by higher Eddington ratios alone\footnote{Note that all quasars in \citet{WangFeige2021XRay} are part of our \textit{full sample}.}.
Indeed, while the link between the X--ray spectral properties of quasars and the underlying accretion physics is now accepted, the details of these processes are still not well understood \citep[e.g.,][]{Trakhtenbrot2017Xray}.
A larger sample of $z > 6.5$ quasars observed in X--ray will be necessary to pin--down the different actors contributing to the observed evolution of $\Gamma$ at these high--redshifts.
%

\subsection{Constraints on the formation of first massive black holes}\label{sec:bhseed}

\begin{figure*}[tb]
    \centering
    \includegraphics[width=1.98\columnwidth]{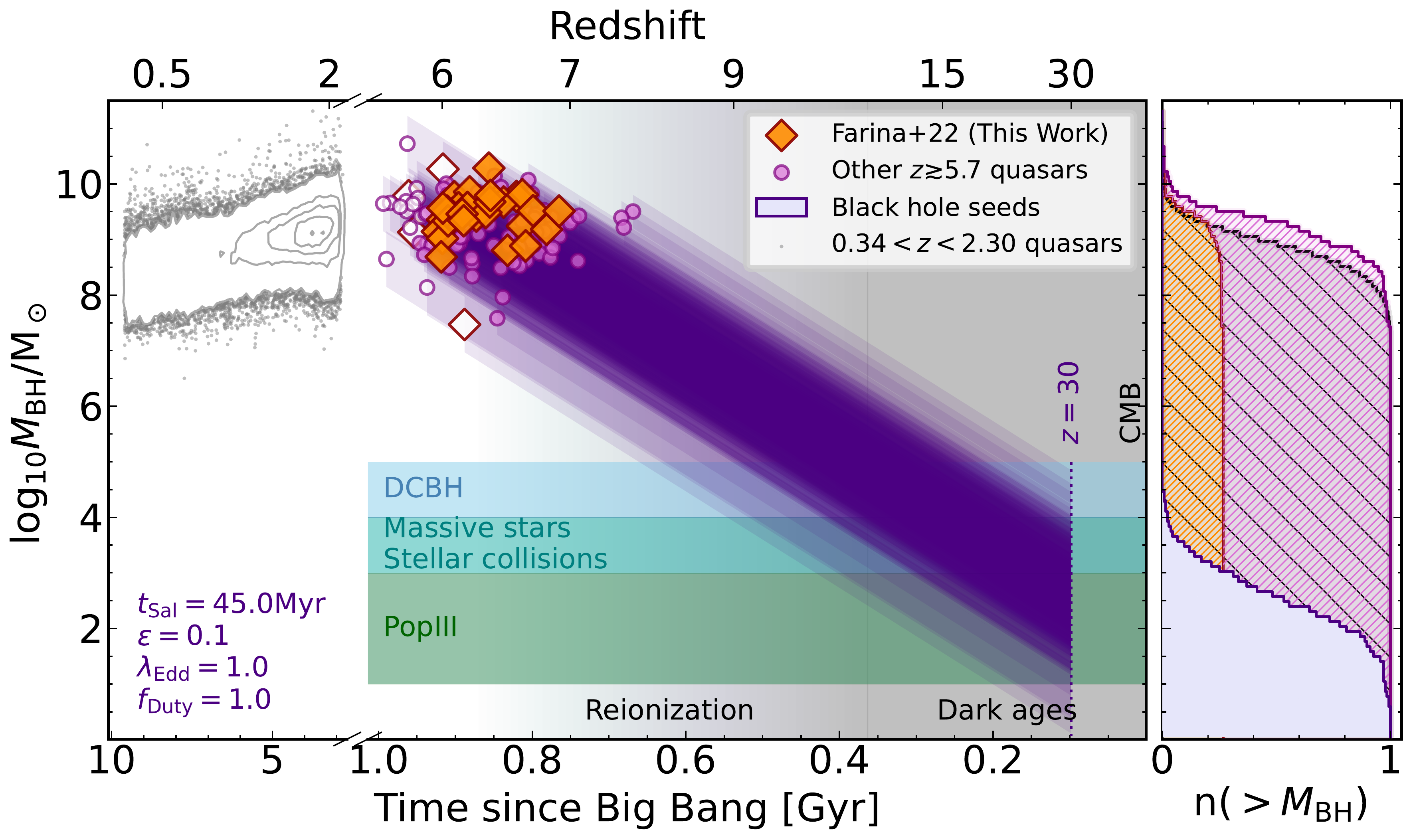}
    \caption{
    Evolutionary tracks of black hole seeds (purple lines) to reach the observed quasar masses at $z \sim 6$ (orange diamonds and pink circles).
    These are calculated using \autoref{eq:massevolution} with parameters: $f_\mathrm{duty} = 1$, $\epsilon = 0.1$, $\Eratio = 1$, and $t_\mathrm{Sal} = 45\,\mathrm{Myr}$.
    Mass ranges for different black hole seed formation scenarios are indicated by shaded regions.
    For the sake of comparison, gray contours and points indicate the location of $1.3 < z < 2.3$ SDSS DR7 quasars \citep[][]{Shen2011}.
    In the right panel, we show the cumulative distributions of observed black holes (orange and pink for the $z \sim 6$ quasars, and black for the $0.34 < z < 2.30$ quasars) and expectations for the considered evolutionary pathway starting at $z = 30$ (purple).
    }
    \label{fig:bh_seeds}
\end{figure*}

%
The fiducial time scale of SMBH growth is the so--called Salpeter time scale \citep[\(t_\mathrm{Sal}\),][]{Salpeter1964}.
This is calculated by considering that the bolometric luminosity of a quasar's accretion disk can be related to the total mass inflow rate ($\dot M$) and, thus, the growth rate of the black hole (${\dot M}_\mathrm{BH}$), following:
\begin{equation}\label{eq:accretiondisk}
    \Lbol = \epsilon \dot M c^2 = \frac{\epsilon}{1-\eta}{\dot M}_\mathrm{BH} c^2,
\end{equation}
where $\eta$ is the accretion \textit{inefficiency} (i.e., the fraction of the total accreted mass/energy which is lost during the process) and $\epsilon$ is the radiative efficiency of black hole accretion.
In the following, we will consider the case of radiatively--efficient accretion events, so we will assume $\epsilon = \eta$.
Combining \autoref{eq:lbol} with \autoref{eq:accretiondisk} and taking the quasar's duty cycle ($f_\mathrm{duty}$) into account, one obtains:
\begin{equation}\label{eq:massevolution}
\frac{\mbh\left(t-t_0\right)}{\mbh\left(t_0\right)}=\exp{\left[{f_\mathrm{duty}\,(1-\epsilon)\,\Eratio\,\frac{t-t_0}{t_\mathrm{Sal}}}\right]},
\end{equation}
where the Salpeter time is defined as:
\begin{equation}
t_\mathrm{Sal}=\epsilon\,\frac{\sigma_\mathrm{T} c}{4 \pi G m_\mathrm{p}}\approx\epsilon\,450\,\mathrm{Myr}.
\end{equation}
This value is calculated in the idealized case of pure hydrogen accretion.
However, we can consider a more realistic case where the accreting gas has primordial abundances of hydrogen ($X = 0.75$) and helium \citep[$Y = 0.25$, see, e.g.,][, and references therein]{Cyburt2016}.
Given that a helium atom has a mass of $\sim 4 \times m_\mathrm{p}$ and 2 free electrons, a quasar's Eddington luminosity will be a factor $1/\left(1-Y/2\right)\approx1.14$ higher and thus, the Salpeter time will be correspondingly shortened to $t_\mathrm{Sal} \approx \epsilon\,395\,\mathrm{Myr}$ \citep[see also,][]{Worseck2021}.
As noted by \citet{Mayer2019}, considering a plasma with higher metallicity has little impact on the mass growth time scale.
The radiative efficiency is a crucial quantity to determine the early growth of SMBHs.
For individual quasars, $\epsilon$ will cover a range of values depending on different parameters such as black hole spin \citep[e.g.,][]{Reynolds2014, Reynolds2021, Capellupo2016} or optical depth of the accretion disk \citep[e.g.,][]{Narayan1995, Sadowski2014}.
For instance, \citet[][]{Trakhtenbrot2017} compared the expected emission of a thin accretion disk with the observed luminosities and black hole masses of 20 $z\sim6$ quasars and obtained radiative efficiencies in the range $0.03 \lesssim \epsilon \lesssim 0.3$.
Following the bulk of the literature on $z \gtrsim 6$ quasars, here we assume a \textit{universal} and constant value of $\epsilon = 0.1$.
This is consistent with $\epsilon = 0.06$ calculated for a radiatively--efficient, \citeauthor{Shakura1973} thin--disk accretion model \citep[with an upper limit of $\epsilon \lesssim 0.3$ for a maximally rotating Kerr black hole, e.g.,][]{Thorne1974} and with constraints derived from the comparison between the total mass in SMBHs observed in the local Universe and the integrated emission of all quasars over cosmic times \citep[the so--called \citeauthor{Soltan1982} argument, e.g.,][]{Shankar2009}.
\autoref{fig:bh_seeds} shows the growth path of black hole seeds assuming maximal accretion ($\Eratio = 1$) for the entire history of the Universe ($f_\mathrm{duty} = 1$).
Under these conditions, the black hole seeds would have to have masses of $4 \msun{} \lesssim M_\mathrm{seed} \lesssim 2.7 \times 10^{4}\msun{}$ at $z = 30$ to grow to the observed mass distribution at $z \gtrsim 6$ \citep[see, e.g., ][ for a discussion on the redshifts of black hole seeds formation]{Sassano2021}.
In the following, we will briefly summarize the different scenarios that have been proposed to explain the formation of seeds with such masses.
For a more complete overview on the topic, we direct the interested reader to the several excellent reviews on this topic from \citet{Volonteri2010}, \citet{Latif2016}, \citet{Inayoshi2020}, \citet{Greene2020}, and \citet{Trakhtenbrot2020}.
The \textit{light--seeds} (with masses of up to a few hundred solar masses) scenario assumes that black hole progenitors are remnants of Population III (Pop\,III) stars.
These first stars are believed to form at $z \gtrsim 20$, when the limited cooling ability of the primordial gas leads to inefficient fragmentation, resulting in a top--heavy initial mass function with masses in the range $10 - 1000\,\msun{}$ \citep[][]{Abel2002, Karlsson2013, Hirano2014}.
While Pop\,III stars with masses $140 - 260\,\msun{}$ will explode as pair--instability supernovae and leave no remnants, those with masses $40 - 140\,\msun{}$ and $> 260\,\msun{}$ will collapse into black holes containing at least half of the progenitor mass \citep[e.g.,][]{Fryer2001, Heger2003}.
These \textit{light--seeds} are expected to form in the highest peaks of the primordial density field and, thus, will cluster in the cores of the most massive halos at high--redshift \citep[e.g.,][]{Madau2001, Volonteri2003}.
However, as shown by \autoref{fig:bh_seeds}, it is unattainable for these Pop\,III remnants could grow to $\gtrsim 10^{9}\,\msun{}$ by $z \gtrsim 6.5$.
Even considering that, for their entire life (i.e., $f_\mathrm{duty} = 1$), they accrete gas at a rate close to the Eddington limit and with low values of $\epsilon \lesssim 0.1$ \citep[][]{Haiman2001, Tanaka2009}.
Such conditions are unlikely to be met in the early Universe. 
Strong feedback from the progenitor stars, from supernovae, and from the accretion onto the black hole seeds are all expected to act within the shallow gravitational potential wells of the first galaxies, reducing the available gas for accretion \citep[e.g.,][]{Mori2002, Johnson2007, Pelupessy2007}.
These requirements can be relaxed with short episodes of moderately super--Eddington accretion \citep[e.g.,][]{Madau2014, Lupi2016, Pacucci2017, Jiang2019}. 
Or, if the gas can efficiently lose angular momentum, even with hyper--Eddington accretion \citep[e.g.,][]{Volonteri2005, Tanaka2009, Inayoshi2016}. 
Such accretion events would accelerate the growth in mass of, initially, small black hole seeds by orders of magnitude within a few million years.
We stress that while our data shows Eddington ratios of $\Eratio \lesssim 1$ for all quasars with black hole masses derived from the \mgii{} line (see \autoref{sec:EddingtonRatios}), it does not rule out such a scenario.
It does, however, put some constraints on it. 
If the first quasars do violate the Eddington limit, our results imply that this should only happen mildly, and/or at earlier cosmic times, and/or for a short period of time, and/or in an highly obscured phase \citep[e.g.,][]{Kelly2013}.
Alternatively, the necessity of a rapid and continuous accretion could be mitigated if the progenitors of the first quasars are born directly as more massive seeds. 
One scenario to form such massive seeds is through stellar collisions. It has been proposed that the first episodes of efficient star formation are able to foster the formation of very compact nuclear star clusters in metal--poor proto--galaxies.
In these environments, direct collisions can occur on timescales shorter than the lifetime of massive stars \citep[][]{PortegiesZwart2002, PortegiesZwart2004} and can lead to the formation of a super--massive star that could potentially leave a black hole remnant with a mass in the range $\sim 10^2 - 10^4\,\msun{}$ \citep[][]{Devecchi2009}.
Finally, it has also been suggested that even more massive seeds can form at $z \lesssim 15$ through the direct monolithic collapse of low angular momentum massive gas clouds, resulting in super--massive stars that will promptly end their life in black holes with masses around $\sim 10^4 - 10^5\,\msun{}$ \citep[][]{Oh2002, Koushiappas2004, Begelman2006, Lodato2006, Mayer2010, Mayer2019}.
Several studies have investigated the (rare) physical conditions necessary for the formation of these so--called Direct Collapse Black Holes \citep[DCBHs, e.g.,][]{Visbal2014, Habouzit2016, Woods2019}.
For instance, it has been proposed that fragmentation (which would deter the formation of massive seeds) can be prevented by the dissociation of H$_2$ molecules (the main cooling channel in the young Universe) by a strong Lyman--Werner radiation field \citep[e.g.,][]{Bromm2003, Latif2015, Luo2020}. 
The necessary radiation could originate from nearby highly star--forming galaxies \citep[e.g.,][]{Regan2017} or from heated gas and shocks induced by gas--rich major mergers \citep[e.g.,][]{Mayer2010}, which could simultaneously facilitate the black hole growth, triggering \(\gtrsim 1,000 \msunyr{}\) gas inflows \citep[e.g.,][]{Mayer2019}.
It is important to note that the limits on the mass of the black hole seeds set by our sample are strongly influenced by the choice of the physical parameters used in \autoref{eq:massevolution}.
For example, the tension between the short formation time and the high mass of the quasar black holes can be eased if we consider that the presence of primordial helium can shorten the Salpeter time to $t_\mathrm{Sal} \approx \epsilon\,395\,\mathrm{Myr}$ and that the radiative efficiency calculated for a \citeauthor{Shakura1973} thin disk as $\epsilon = 0.06$\footnote{
Even lower values of $\epsilon \sim 0.008 - 0.010$ (below the range allowed by the thin-disk model) have been derived by \citet{Davies2019} from the impact of the ionizing flux of two $z > 7$ quasars on the surrounding neutral IGM.
} 
(see Top Panel of \autoref{fig:bh_seeds_fast_slow} in \autoref{app:test_seeds}).
On the other hand, if we can also take into account that the median observed Eddington ratio of $z > 5.7$ quasars is $\Eratio{} \sim 0.47$ (see \autoref{sec:eratio}) and that clustering measurements suggest a duty cycle of $f_\mathrm{duty} = 0.03-0.60$ for bright $z \sim 4$ quasars \citep[e.g.,][]{Shen2007, White2008}, implying $f_\mathrm{duty} \sim 0.9$ at $z \sim 6$ \citep[e.g.,][]{Shankar2010}.
Using these values, \autoref{eq:massevolution} predicts that black hole seeds need to have masses as high as $\sim 10^{5}\,\msun{}$ at $z = 30$ (see bottom panel of \autoref{fig:bh_seeds_fast_slow} in \autoref{app:test_seeds}).
%

\subsection{The earliest super--massive black holes and their hosts}\label{sec:hosts}

%
For all quasars in our \textit{\xs{}/ALMA sample}, sub--mm ALMA data has been collected \citep[see][ for details]{Venemans2017ALMAz7, Venemans2019, Decarli2018, Decarli2019Merger, Banados2019Pisco, Eilers2020, Venemans2020, Neeleman2021, Walter2022}.
These observations revealed bright \ciimu{} line emission in 34 out of \NQSO{} quasars, with luminosities typically in the range of $10^9 - 10^{10}\,\lsun{}$. The data also revealed the presence of large quantities of dust with masses of $10^7 - 10^9\,\msun{}$ (as derived from the FIR luminosities ranging from $\sim 0.3 \times 10^{11}\,\lsun{}$ to $\sim 13 \times 10^{13}\,\lsun{}$\footnote{The FIR luminosities of $z \sim 6$ quasars are typically calculated assuming that the dust spectral energy distribution is described by a modified black body with temperature $T_\textrm{dust} = 47\,\mathrm{K}$ and an emissivity index $\beta = 1.6$ \citep[e.g.,][]{Beelen2006, Leipski2014, Venemans2016}}, e.g., \citealt{Venemans2018}).
In addition, using $<0\farcs25$ (corresponding to $< 1.4$\,kpc at $z = 6$) ALMA imaging of the \ciimu{} line emission of quasar host galaxies, \citet{Neeleman2021} inferred dynamical masses ($\mdyn{}$) of the order $\sim1-10\times10^{10}\,\msun{}$. 
The authors also reported that $\sim 30\%$ of hosts show a disturbed \ciimu{} morphology, possibly due to a recent merger or interaction with close companions \citep[see also][]{Decarli2017, Decarli2019Merger, Decarli2019J1030, Willott2017, Banados2019Pisco, Mazzucchelli2019, Neeleman2019, Venemans2019, Meyer2022Overdesity}.
In what follows, we examine these properties in light of our comprehensive analysis of the proprieties of the black holes in the first quasars.
%

\subsubsection{Merger signatures and quasar activity}

\begin{figure}[tb]
    \centering
    \includegraphics[width=0.98\columnwidth]{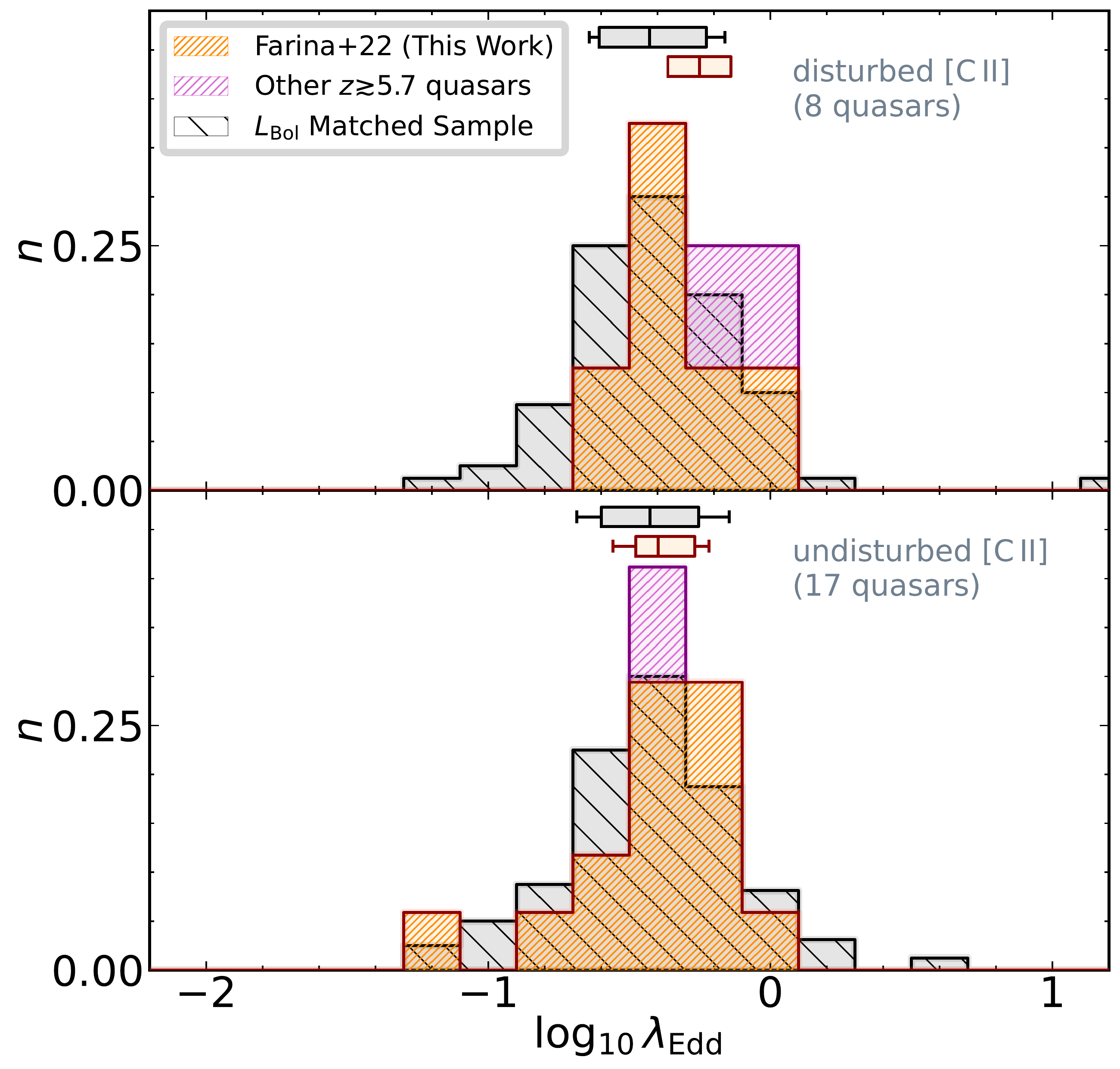}
    \caption{
    Logarithmic Eddington ratio distributions of the quasars observed at $<0\farcs25$ resolution with ALMA by \citet{Neeleman2021}.
    Host galaxies that show disturbed and undisturbed \ciimu{} morphologies are shown in the upper and lower panels, respectively.
    For the sake of comparison, the Eddington ratios of the luminosity--matched samples are also shown in black.
    However, note that no information on the hosts is taken into account for the selection of the $z\sim0.3-2.3$ quasars.
    The box plots at the top of each panel mark the median values of the various distributions.
    High redshift quasars residing in galaxies with a disturbed morphology have a median Eddington ratio higher than those in undisturbed hosts.
    However, the two distribution do not differ enough to suggest differences between their parent populations.
    The color code is the same as in \autoref{fig:eratios}.
    }
    \label{fig:eratios_host}
\end{figure}

%
If gravitational interactions between gas--rich galaxies are able to induce strong gas inflows \citep[e.g.,][]{Barnes1991, Barnes1996} that trigger quasar activity \citep[e.g.,][]{Dimatteo2005, Hopkins2010}, then host galaxies with a disturbed morphology should naively have, on average, higher accretion rates.
Indeed, merger signatures are expected to be detectable for $\lesssim2.4\times(1+z)^{-2}$\,Gyr \citep[][]{Snyder2017, Lupi2022}, a time much longer than current constraints on $z \sim 6$ quasar lifetimes \citep[\(t_\mathrm{Q} \sim 10^{6}\)\,yr, based on the measurements of their proximity zones, e.g.,][; see also \citealt{Khrykin2019, Worseck2021}]{Eilers2017, Eilers2021, Davies2020, Morey2021}. 
Even when considering that the gas perturbed by the merger event may take several dynamical times ($t_\mathrm{dyn}\sim4\times10^{6}$\,yr for the typical host in our sample) to funnel down to the central regions of a galaxy \citep[e.g.,][]{Solanes2018}, the ensuing peak of quasar activity can be delayed by $\gtrsim100$\,Myr \citep[e.g.,][]{Dimatteo2005}.
\autoref{fig:eratios_host} shows the logarithmic distribution of the Eddington ratios divided into disturbed and undisturbed morphology of the host galaxy, as classified by \citet{Neeleman2021}.
Galaxies with a disturbed morphology harbor black holes accrete at a median Eddington ratio of $\Eratio{}=0.56^{+0.09}_{-0.10}$. 
At first glance, this appears to be higher than the remaining sample (with a median Eddington ratio of  $\Eratio{}=0.40 ^{+0.07}_{-0.02}$).
However, a \citeauthor{Student}'s t--test on the two samples returns a $p-\textrm{value}\sim0.15$.
Thus, the two median values are not statistically different ($\sim1.4\,\sigma$).
This suggests that the growth of a $z \sim 6$ black hole is not strongly affected by merger events, or that the peak of black hole activity may occur during an obscured phase \citep[e.g.,][]{Dimatteo2012, Dubois2013, Trebitsch2019, Davies2019}.
It is worth noting that these findings echo results from deep HST/WFC3 observations of the hosts of bright ($\Lbol{} \gtrsim 4 \times 10^{46}\,\ergs{}$) and massive ($\mbh{} \gtrsim 3 \times 10^{8}\,\msun{}$) quasars at $z \sim 2$.
Indeed, these investigations also revealed merger features in $\sim 25\%$ of the cases and no clear link between accretion rates and disturbed stellar morphologies \citep[e.g.,][]{Mechtley2016, Marian2019}.
However, the importance of merger in triggering the black hole activity will become more relevant at later cosmic times \citep[$z<0.2$, e.g.,][]{Marian2020}.
%

\subsubsection{The black hole--to--galaxy mass ratio}

\begin{figure}[tb]
    \centering
    \includegraphics[width=0.98\columnwidth]{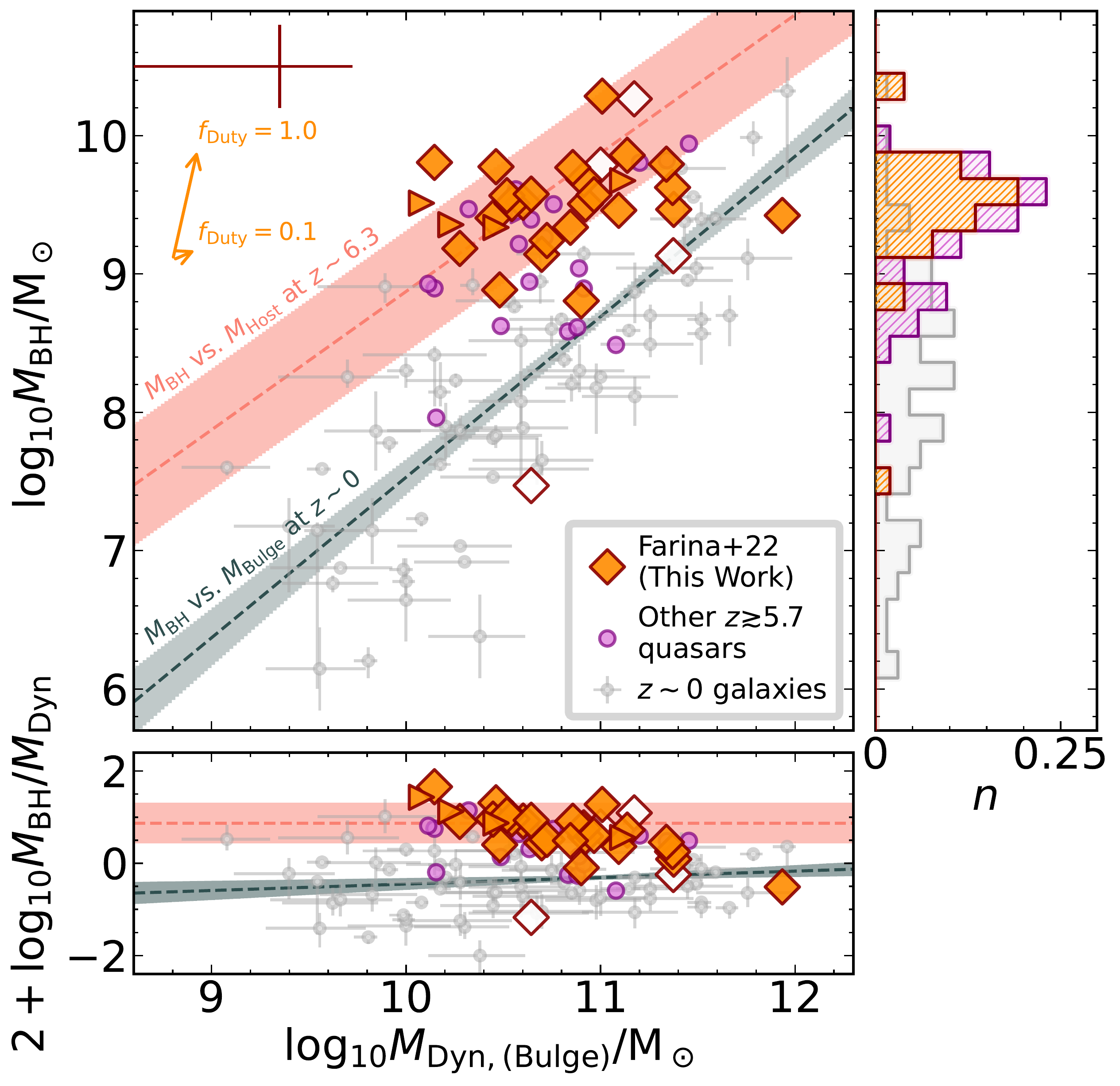}
    \caption{
    Relation between the dynamical mass of the host galaxies and the mass of the central black holes of the $z \gtrsim 5.7$ quasars for the \textit{\xs{}/ALMA sample} \citep[orange diamonds, with right pointing orange triangles indicating lower limits, see also][]{Neeleman2021} and for other studies (violet circles). 
    The typical error (including systematics) is shown as a cross in the top left corner.
    For comparison, the best fit of the relation between $\mbh{}$ and $M_\mathrm{Bulge}$ derived by \citet{Kormendy2013} is shown as a gray dashed line.
    The extrapolation of the $\mbh$/$M_\mathrm{Host}$ ratio evolution derived by \citet{Decarli2010Evolution} for $z \sim 6.3$ (the median redshift of the quasar sample) is shown as an orange dashed line.
    Light gray points are data for 66 local galaxies derived by \citet{Savorgnan2016}.
    The bottom panel highlights the differences with respect to the local relation by rescaling the $y$--axis by a factor of 100.
    Histograms in the right panel show the black hole mass distributions for local galaxies (in gray) and for high--redshift quasars (in orange and pink).
    }
    \label{fig:MbhMdyn}
\end{figure}

%
Despite similarities in accretion and host galaxy properties across redshifts, it has been noted since the first resolved detection of the CO emission of a $z > 6$ quasar host galaxy \citep[SDSS J1148$+$5251 at $z = 6.42$, ][]{Walter2003, Walter2004} that, when compared to local measurements of the $\mbh$--$M_\mathrm{Bulge}$ relation \citep[e.g.,][]{Haring2004, Kormendy2013, Graham2016, deNicola2019}, the black holes of the first quasars are over--massive relative to their hosts.
This was recently confirmed by \citet{Neeleman2021} with a large sample of bright quasars, using the black hole masses estimated in this work.
In \autoref{fig:MbhMdyn}, we further reinforce these findings by complementing their results with all available estimates of dynamical masses of $z \gtrsim 5.7$ hosts from the literature \citep[][]{Willott2015, Willott2017, Izumi2018, Izumi2019, Izumi2021, Pensabene2020}.
For unresolved observations \citep[][]{Willott2013, Mazzucchelli2017, Decarli2018, Yang2019MultiLine, Eilers2020, Andika2020, WangFeige2021, Khusanova2022}, we additionally include estimates of the host galaxy masses by using the relations between $\mdyn{}$ and FWHM and size of the \ciimu{} emitting region (where published) provided by \citet{Neeleman2021}.
The median $\mbh{}$/$\mdyn{}$ ratio of the 52 quasars with available \ciimu{} data is $\sim0.038$. This is a factor of $\gtrsim10\times$ larger than the median $\mbh{}$/$M_\mathrm{Bulge}$ ratio of the 66 local galaxies with precise black hole and bulge masses derived by \citet{Savorgnan2016}.
The $\mbh{}$/$\mdyn{}$ ratios of $z \gtrsim 5.7$ quasars are broadly consistent with the $z \sim 6.3$ (the median redshifts of the considered quasars) extrapolation of the $\mbh{}$--$M_\mathrm{Host}$ evolution presented by \citet[][, see also \citealt{Peng2006, Merloni2010}]{Decarli2010Evolution}.
Note that these comparisons are based on the working assumptions that dynamical masses and bulge (or stellar) masses are comparable \citep[but see, e.g.,][]{Lupi2019} and that biases due to the sample selection function are negligible \citep[but see, e.g.,][]{Lauer2007}.
While the direct detection of the stellar light of high redshift quasars with \textit{JWST} will quantify the consequences of the first assumption \citep[e.g.,][]{Marshall2021}, we can already (partially) estimate the impact of the latter.
If we split the sample into two bins of bolometric luminosity, both containing the same number of objects, we observe that quasars with $\Lbol{} > 1.4 \times 10^{47}\,\ergs{}$ have a median $\mbh{}$--$\mdyn{}$ ratio of $\sim0.056$.
Which is a factor of $\sim2\times$ higher than the median value for the remainder of the sample ($\mbh{}/\mdyn{}\sim0.025$).
While this indicates that fainter quasars also lie, in average, above the local $\mbh{}$--$M_\mathrm{Bulge}$ relation, the lower mass ratio observed in the fainter quasar bin suggests that selection biases may indeed play a major role in shaping the observed evolution of the relation \citep[e.g.,][; and \citealt{Marshall2020Sim} for results from cosmological hydrodynamical simulations]{Willott2017, Izumi2021}.
%

\subsection{The first black holes and their gas reservoirs}\label{sec:lya}

\begin{figure}[tb]
    \centering
    \includegraphics[width=0.98\columnwidth]{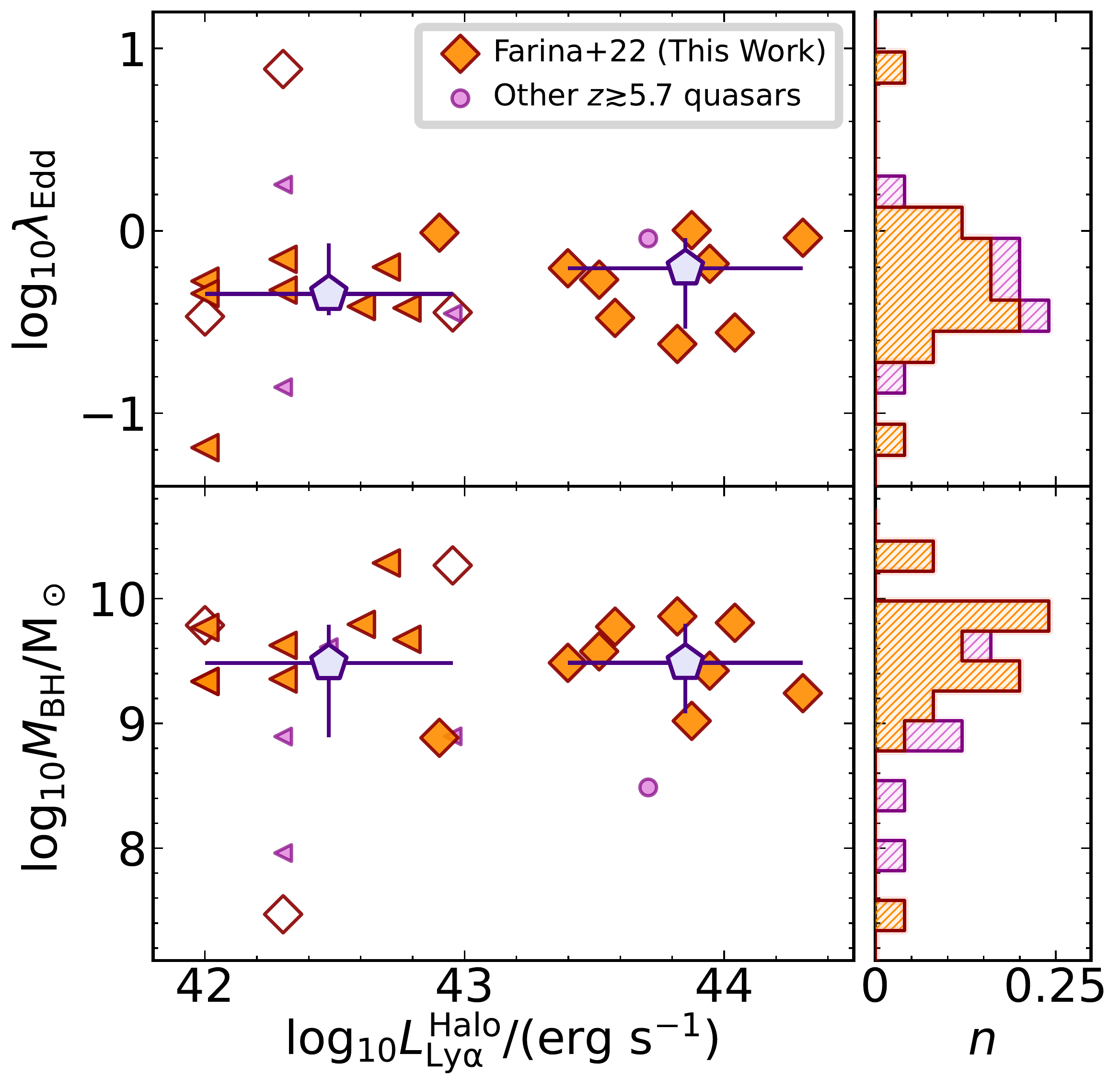}
    \caption{
    Black hole masses (bottom) and Eddington ratios (top) of quasars from the \textit{\xs{}/ALMA sample} (20 targets) and from the literature (5 targets) as a function of total $\lya{}$ luminosity of the extended halo.
    The color code of the points is the same as in \autoref{fig:mbhlbol} and left pointing triangles indicate upper limits on the \lya{} halo luminosities.
    Median values calculated after splitting the sample at an halo luminosity of $10^{43}\,\ergs{}$ are shown as blue pentagons.
    The 16$^\mathrm{th}$ and 84$^\mathrm{th}$ percentiles of the distribution are shown as error bars along the $y$--axis.
    Error bars along the $x$--axis indicate the luminosity range covered by each bin.
    Histograms in the right panel show the distribution of black hole masses and Eddington ratios considered.
    }
    \label{fig:haloluminosity_mbh}
\end{figure}

%
Our \textit{\xs{}/ALMA sample} largely overlaps (20 out of \NQSO{}~targets) with the \textit{REQUIEM} survey, a \textit{VLT}/MUSE survey of high--redshift quasars presented by \citet[][, see also \citealt{Farina2017} and \citealt{Drake2019}]{Farina2019}.
This study probed the extended \lya{} emission surrounding $z \sim 6$ quasars on scales of several kilo--parsecs.
While these halos are routinely detected around $z\gtrsim2$ quasars \citep[e.g.,][]{Borisova2016, Arrigoni2019, Farina2019, Cai2019, Drake2020, Fossati2021}, the physical mechanisms that power their observed emission remain uncertain. 
The currently favored scenarios include a combination of emission from collisionally excited gas and recombination radiation produced after quasar photo--ionization, and/or scattering of \lya{} photons from the broad line region \citep[e.g.,][]{Hennawi2013, Arrigoni2017, Farina2017, Cantalupo2017, Cantalupo2019, Costa2022}.
Nonetheless, all these mechanisms require the presence of large amounts of cool ($T < 10^{4}$\,K) gas.
Thus, the detection of \lya{} nebulae around $z \gtrsim 6$ quasars is an indication of the presence of large (possibly $\gtrsim10^{10}\,\msun{}$) gas reservoirs that are able to simultaneously sustain the intense star--formation rates of the hosts and feed the black hole\footnote{
We remind the reader that the minimum timescale for cold gas in the halo to reach the black hole is the free fall time, estimated to be $\gtrsim 50$\,Myr.
}.
Bearing in mind that obscuration, variations in the quasar activity, the presence of additional sources of ionizing radiation, and the specific density distribution of the emitting material can each influence the intensity of \lya{} emission, we will assume, for the sake of simplicity, that the brightest \lya{} nebulae trace the most massive gas reservoirs.
Given that inflow of cold gas is expected to be the main driver of the early and rapid growth of the first quasars \citep[e.g.,][]{Dimatteo2012, Dubois2013}, one can expect that the most massive and highly accreting quasars are surrounded by the most luminous \lya{} nebulae.
In \autoref{fig:haloluminosity_mbh}, we plot $\mbh{}$ and $\Eratio{}$ as a function of the luminosity of the \lya{} halo ($L_{\lya{}}^\mathrm{Halo}$).
If we split the sample in two bins at a halo luminosity of $L_{\lya{}}^\mathrm{Halo}=10^{43}\,\ergs{}$ \citep[a limit set by the typical depth of the \textit{VLT}/MUSE observations in ][]{Farina2019}, we observe no major difference in the median black hole masses ($\mbh{}=3.1^{+1.3}_{-0.9}\times10^9\,\msun$ and $3.1^{+0.7}_{-1.3}\times10^9\,\msun$, for faint and bright halos, respectively). We see a mild (but low significance, i.e., $\sim0.7\sigma$ obtained from a \citeauthor{Student}'s t--test on the two samples) increase in the median Eddington ratio with halo luminosity ($\Eratio{}=0.45^{+0.08}_{-0.07}$ for faint halos and $0.62^{+0.28}_{-0.08}$ for bright halos).
While the absence of correlation between the cool gas reservoirs and black hole properties might seem surprising, we stress that both the \textit{REQUIEM} and the \textit{\xs{}/ALMA} surveys preferentially target UV--bright quasars (see \autoref{sec:sample}).
Thus, we have biased our study to a narrow Eddington ratio and black hole mass range, with only three quasars that have $\mbh{}<5\times10^{8}\,\msun{}$.
A sample of faint quasars with reliable black hole mass estimates and a sensitive investigation of the extended emission would be necessary to further test for the presence of correlations.
%

\subsubsection{Are the first black holes and their dark matter halos linked?}

\begin{figure}[tb]
    \centering
    \includegraphics[width=0.98\columnwidth]{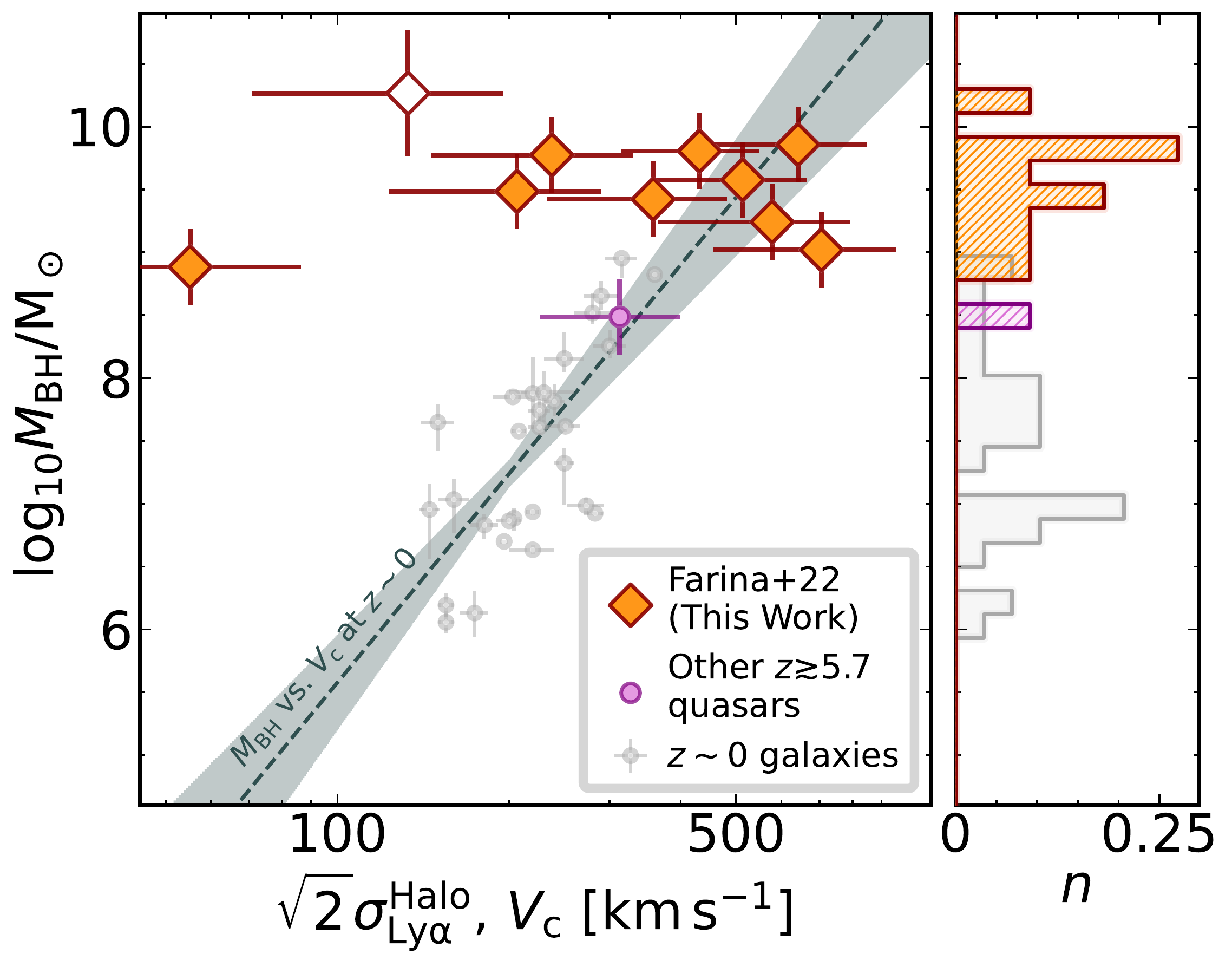}
    \caption{
    Distribution of $z \gtrsim 5.7$ quasars in the circular velocity vs. black hole mass plane.
    Circular velocities are derived from the 1D velocity dispersion of the \lya{} halos considering $V_\mathrm{c}=\sqrt{2}\sigma^\mathrm{Halo}_\mathrm{\lya{}}$.  
    Black hole masses and circular velocities for local galaxies are taken from the review of \citet[][, gray points]{Kormendy2013}.
    The gray dashed line is the best fit to the $z \sim 0$ data following the procedure in \citet[][]{Volonteri2011}.
    Histograms in the right panel show the distribution of black hole masses of the high--redshift and local samples.
    The color code is the same as in \autoref{fig:MbhMdyn}.
    }
    \label{fig:mbhvcirc}
\end{figure}

%
The presence of a correlation between $\mbh{}$ and the mass of the dark matter halo host has been a matter of discussion since it has been first proposed by \citet{Ferrarese2002}.
Given the smaller scatter of this relation, it has been argued that it is more fundamental than relations with  other host galaxy properties \citep[e.g.,][]{Baes2003, Marasco2021}.
However, the level of the scatter including different galaxy populations and the strength of the correlation has been put into question by several studies \citep[e.g.,][]{Kormendy2011, Kormendy2013, Sabra2015}.
\citet{Volonteri2011} proposed that the relation between super--massive black holes and dark matter halos should be tighter at early cosmic times, when the growth is expected to be dominated by the merger history of the host galaxy (see \autoref{sec:bhseed}).
Dynamical measurements of the mass of dark matter halos ($M_\mathrm{h}$) of single galaxies commonly rely on globular cluster kinematics and/or spatially resolved \hi{} rotation curves. Thus, testing this prediction at high--redshift is observationally challenging \citep[e.g.,][]{Sun2013}.
And currently there are no direct measurements of the circular velocity ($V_\mathrm{c}$) of the halo of $z \gtrsim 6$ quasars \citep[for indirect estimates see, e.g.,][]{Shimasaku2019}.
However, we can obtain a coarse estimate of $V_\mathrm{c}$ under the assumption that the extended \lya{} emission traces gas in gravitational motion within the dark matter halo.
Using this, we can derive the circular velocity from the kinematic information of the \lya{} extended emission, considering $V_\mathrm{c}=\sqrt{2}\,\sigma_\mathrm{\lya{}}^\mathrm{Halo}$
\citep[where \(\sigma_\mathrm{\lya{}}^\mathrm{Halo}\) is the measured 1D root--mean--square velocity dispersion of the halo, e.g.,][]{Tormen1997}.
This is supported by simulations showing that in massive high--redshift halos, the dark matter component dominates the kinematics at scales larger than $\sim10$\,kpc \citep[][]{Costa2015}. 
It is also supported by observations of the quiescent kinematics of the extended \lya{} emission \citep[e.g.,][]{Arrigoni2019, Farina2019, Lau2022, Drake2022} and, more in general, of the gas residing in the circumgalactic medium of quasar \citep[e.g.,][]{Prochaska2009, Lau2018}.
With large velocity dispersions ($>200\,\kms$), potentially associated with outflows, observed only in the closest regions to the quasar hosts \citep[e.g.,][]{Guo2020}.
In \autoref{fig:mbhvcirc}, we compare our rough estimates of the velocity dispersion with those of local galaxies, as presented in \citet{Kormendy2013}.
Despite the large scatter, most of the quasars sit on the high velocity--high mass extrapolation of the best fit of the local relation \citep[see][ for details]{Volonteri2011}, especially when bearing in mind that our sample is highly biased towards massive objects.
In general, the observed velocities are consistent with gravitational motions within a dark matter halo with masses $M_\mathrm{DM}\gtrsim10^{12.5}\,\msun$.
The first quasars are indeed expected to reside in such massive halos.
Indeed, the number density of $\MMCDL{}<-26$\,mag quasars (the typical quasar part of the \textit{REQUIEM} survey), $\phi(\MMCDL{}<-26\,\mathrm{mag})\sim7\times10^{-10}$\,cMpc$^{-3}$ \citep[using the estimates of][]{Matsuoka2018LF}, matches the integral of the $z \sim 6.3$ halo mass function from \citet{Behroozi2013} at a mass of $M_\mathrm{DM}\sim10^{12.8}\,\msun$, if we assume an high duty cycle of $f_\mathrm{duty}=0.9$ \citep[see \autoref{sec:bhseed} and, e.g.,][]{Shankar2010}.
The tentative results are, thus, that the \lya{} emitting gas in the circum--galactic medium is bound by gravitation and that local relation holds out to high--redshifts.
The on--set of this relation at early cosmic time supports a scenario where the rapid formation of the first quasars happens in the rarest peaks of the initial distribution of density fluctuations, meaning that the quasars are hosted by the most massive dark matter halos at that time \citep[e.g.,][]{Volonteri2012}.
Although these halos may evolve at later times into less extreme systems \citep[e.g.,][]{Angulo2012, Fanidakis2013, Tenneti2018}.
There are, however, several potential systematics associated with our approach.
For one, resonant scattering could lead to a strong modification of the spectral shape of the \lya{} line \citep[e.g.,][]{Dijkstra2017, Costa2022}, making the determination of the true gas kinematics less reliable. 
This would imply that measurements of the circular velocity from the \lya{} extended emission are an overestimate of the true kinematics \citep[up to a factor $\sim1.5\times$, e.g.,][]{Costa2022}, and thus that the black holes in $z \sim 6$ quasars are over--massive with respect to their dark matter halos.
Detection of non--resonant line emission from the circum--galactic medium \citep[e.g.,][]{Cantalupo2019, Guo2020} with JWST will be able to mitigate these systematics.
In addition, the presence of companion galaxies at small separations from the quasar hosts may perturb the gas.
This may be the case for the largest outlier in \autoref{fig:mbhvcirc}: the quasar J0305--3150 at $z \sim 6.6$ with a peculiarly narrow extended emission \citep[][]{Farina2019}. 
The presence of a close--by companion located in the same direction of the halo may suggest that the \lya{} line is not tracing the kinematics of the halo, but rather the gas perturbed by the close interaction \citep[][]{Farina2017}.
%

\section{SUMMARY AND CONCLUSIONS}\label{sec:conclusion}

%
We present black hole masses and Eddington ratios obtained from deep (typically two hours per target) near--IR \xs{} spectroscopy of \NQSO{} bright quasars at $5.78 < z < 7.54$.
Our data are complemented by ALMA observations of the \ciimu{} emission (providing information on the host galaxies) and \textit{VLT}/MUSE sensitive searches of the extended \lya{} emission (providing insight into the cool gas reservoirs).
This is the second paper in a series aimed at inferring the general properties of bright quasars during the epoch of reionization.
In \citetalias{Schindler2020}, we investigated the properties of the quasar broad line regions, showing an increase in the blue--shift of the \civ{} emission lines with respect to low redshift quasar samples. 
Here, we calculate accurate black hole masses and Eddington ratios derived primarily from the \mgii{} broad emission line.
After including data from the literature, we compiled a sample of \NQSO{} + \NQSOLIT{} quasars at $5.7 \lesssim z \lesssim 7.6$.
The main results from the study of this large sample of bright quasars are:
\paragraph{(i)} 
High--redshift quasars are powered by massive ($\mbh{}\gtrsim10^8\,\msun{}$), strongly accreting black holes (with a median Eddington ratio of $\Eratio{}=0.48$).
While $z \gtrsim 5.7$ quasars sit in a similar locus in the bolometric luminosity vs. black hole mass plane as a luminosity--matched sample of $z\lesssim2.3$ quasars, their accretion rates are, on average, $\sim0.2$\,dex larger.
The data also tentatively suggests a mild evolution of the median accretion rate from $z \sim 6$ to $z \sim 7$.
\paragraph{(ii)} 
Only three quasars show $\Eratio{} \gg 1$. However, all of them have black hole masses derived from the \civ{} broad emission line.
We interpret this as caused by the larger intrinsic uncertainties associated with this line, instead of owing to underlying physical processes.
Thus, we have no strong evidence of a significant population of quasars accreting at super--Eddington.
\paragraph{(iii)}
When compared with local studies, the black holes appear to be over--massive with respect to their host galaxy masses derived from the \ciimu{} emission line.
There is no evidence for changing accretion behavior with a more or less perturbed morphology, for example caused by a merger of the host galaxy.
\paragraph{(iv)}
The extent of the reservoirs of cool ($T \sim 10^{4}$\,K) gas within the halo of the quasars does not correlate with the accretion rate of the $10^{9}\,\msun{}\lesssim\mbh{}\lesssim10^{10}\,\msun{}$ central black holes.\newline
Given these findings we can speculate on the fate of the first quasars.
The first SMBHs appear to already be as massive as the most massive quasars at $z\sim0.3-2.3$, more than 2\,Gyr later (see \autoref{fig:mbhlbol}).
It seems that the rapid black hole growth required at early times (shown in \autoref{fig:bh_seeds}) is strongly reduced for the remainder of the Universe's history.
Instead, the host galaxy needs to continue forming stars to grow its mass and move towards the local $\mbh{}$--$\mdyn{}$ relation (see \autoref{fig:MbhMdyn}).
Such stellar mass growth can be sustained by the cool gas reservoirs discussed in \autoref{sec:lya} \citep[e.g.,][]{Walter2020}.

In \autoref{fig:MbhMdyn}, we illustrate the evolution of the typical quasar in our sample. 
It has a black hole mass of $3\times10^{9}\,\msun{}$, an accretion rate of $\Eratio=0.48$, and a host galaxy mass of $7\times10^{10}\,\msun$ with a dust--obscured star formation rate of $\sim220\,\msunyr{}$. 
This quasar would move across the $\mbh{}$ -- $\mdyn{}$ plane as indicated by the larger orange arrow, if we assume that its growth continues uninterrupted ($f_\mathrm{duty}=1.0$) for 100\,Myr (roughly corresponding to the depletion timescale for star formation).
Under these conditions the super--massive black hole will overgrow its host galaxy even further, moving the quasar away from the local relation.
Given that other accretion properties of the quasar population appear to be constant with redshift, this implies that the duty cycle of quasars needs to drastically evolve.
For instance, if the duty cycle is reduced to $f_\mathrm{duty}\lesssim0.1$ (indicated by the shorter orange arrow), the relative growth in mass of the black hole would be smaller than that of the host, pushing the quasar in the direction of the local relation \citep[e.g.,][]{Trakhtenbrot2011}.
This study showcases the critical role of combining data from different wavelengths and instruments to derive constraints on the physical properties of the first SMBHs.
Systematic uncertainties described in this work could be conceivably reduced in the near future.
\textit{JWST} \citep[][]{Gardner2006} will provide precise black hole mass and Eddington ratio estimates via rest--frame optical investigation of the first quasars.
Complementary, \textit{Euclid} is expected to push quasar searches to fainter objects and higher redshifts \citep[e.g.,][]{Euclid2019}, reducing the impact of selection biases.
%


\section*{Acknowledgments}

EPF is grateful to V.\ Springel and M.\ Strauss for the hospitality at MPA and at Princeton University while writing this manuscript.
It is a pleasure to thank M.\ Strauss, J.\ Green, and M.\ Neeleman for discussion, comments, and suggestions that helped improving this paper. 
EPF, FW, MO, RAM and SEIB acknowledge funding through the ERC Advanced Grant 740246 (\textit{Cosmic Gas}).
ACE, TAG, and FW acknowledge support by NASA through the NASA Hubble Fellowship grants ($\#$HF2-51434, $\#$HF2-51480, and \#HF2-51448, respectively) awarded by the Space Telescope Science Institute, which is operated by the Association of Universities for Research in Astronomy, Inc., for NASA, under contract NAS5-26555. 
JFH acknowledges support from the National Science Foundation under Grant No. 1816006. 
JTS, JFH acknowledge funding through the ERC European Research Council (ERC) under the European Union's Horizon 2020 research and innovation program (grant agreement No 885301).
BT acknowledges support from the Israel Science Foundation (grant number 1849/19) and from the European Research Council (ERC) under the European Union's Horizon 2020 research and innovation program (grant agreement number 950533).
ABD acknowledges support from the UK Science and Technology Facilities Council (STFC) under grant ST/V000624/1.
EPF is supported by the international Gemini Observatory, a program of NSF's NOIRLab, which is managed by the Association of Universities for Research in Astronomy (AURA) under a cooperative agreement with the National Science Foundation, on behalf of the Gemini partnership of Argentina, Brazil, Canada, Chile, the Republic of Korea, and the United States of America.
Based on observations made with ESO Telescopes at the La Silla or Paranal Observatories under programme ID(s):
60.A-9418(A),
084.A-0360(A),
084.A-0390(A),
085.A-0299(A),
086.A-0162(A),
087.A-0890(A),
088.A-0897(A),
089.A-0814(A),
091.C-0934(B),
093.A-0707(A),
096.A-0095(A),
096.A-0418(A),
096.A-0418(B),
097.B-1070(A),
098.B-0537(A),
0100.A-0625(A),
0100.A-0898(A),
0101.B-0272(A),
0102.A-0154(A), and
286.A-5025(A).

\software{
    \textsc{Astropy} \citep{Astropy2013, Astropy2018},
    \textsc{IPython} \citep{Perez2007},
    \textsc{MatplotLib} \citep{Hunter2007},
    \textsc{Numpy} \citep{vanderWalt2011, Harris2020},
    \textsc{PypeIt} \citep{Prochaska2019pypeit1, Prochaska2019pypeit2, Prochaska2020pypeit3},
    \textsc{Python} \citep{vanRossum1995, vanRossum2009},
    \textsc{SciPy} \citep{Virtanen2020}, and
    \textsc{Sculptor} \citep{Schindler2022}.
    }

\appendix

\section{The small black hole of J2229\texorpdfstring{$+$}{+}1457}\label{app:j2229}

%
The $z=6.15$ quasar J2229$+$1457 appears as an outlier in \autoref{fig:mbhlbol}.
This is the faintest quasar of the \textit{\xs{}/ALMA sample}, with a bolometric luminosity of $2.9\times10^{46}\,\ergs{}$ and a \civ{}--based black hole mass of only $3\times10^7\,\msun{}$.
For comparison, similarly faint high--redshift quasars typically have black hole masses at least one order of magnitude more massive \citep[see e.g., ][]{Willott2010BH, Onoue2019}.
This could imply that this quasar is accreting at a rate higher than Eddington, with $\Eratio\sim7.7$ (see \autoref{tab:mbh}).
The low mass estimate originates from the shape of the \civ{} line.
Which consists of a narrow (driving the measured relatively small $\mathrm{FWHM}_\mathrm{CIV}\sim900\,\kms{}$) and a broad distinct component (see \autoref{fig:fitciv}, row 4, panel 3).
Unfortunately, the quasar's redshift puts the \mgii{} line in a spectral region strongly affected by telluric absorption.
Therefore the \mgii{} measurements should be taken with caution. 
\citet{Eilers2020} measured a $\mathrm{FWHM}_\mathrm{MgII}=(5469\pm439)\,\kms{}$ from the \xs{} spectrum of this object, resulting in a black hole mass estimate of $\mbh{}\sim1.7\times10^{9}\,\msun{}$ (and an Eddington ratio of $\Eratio{}\sim0.1$).
While \citet{Willott2010BH} deduced  $\mathrm{FWHM}_\mathrm{MgII}=(1440\pm330)\,\kms{}$, $\mbh{}\sim1.4\times10^{8}\,\msun{}$, and $\Eratio{}\sim1.6$.
from their measurements on low resolution ($R\sim520$) \textit{Gemini}/NIRI spectroscopy.
Super--Eddington accretion episodes are often invoked to explain the rapid mass build up of the first SMBHs \citep[e.g.,][, see also \autoref{sec:bhseed} for details]{Volonteri2005} and this object could be an example of such accretion, 
especially considering that this is a potentially \textit{young} quasar \citep[][]{Eilers2017}.
However, the large uncertainties on \civ{}--based black hole masses (of order $\sim0.5$\,dex, see \autoref{sec:BlackHoleMasses}) do not allow us to derive solid conclusions based on this single object.
%

\section{Comparison with Shen et al.}\label{app:comparison}

\begin{figure}[tb]
    \centering
    \includegraphics[width=0.98\columnwidth]{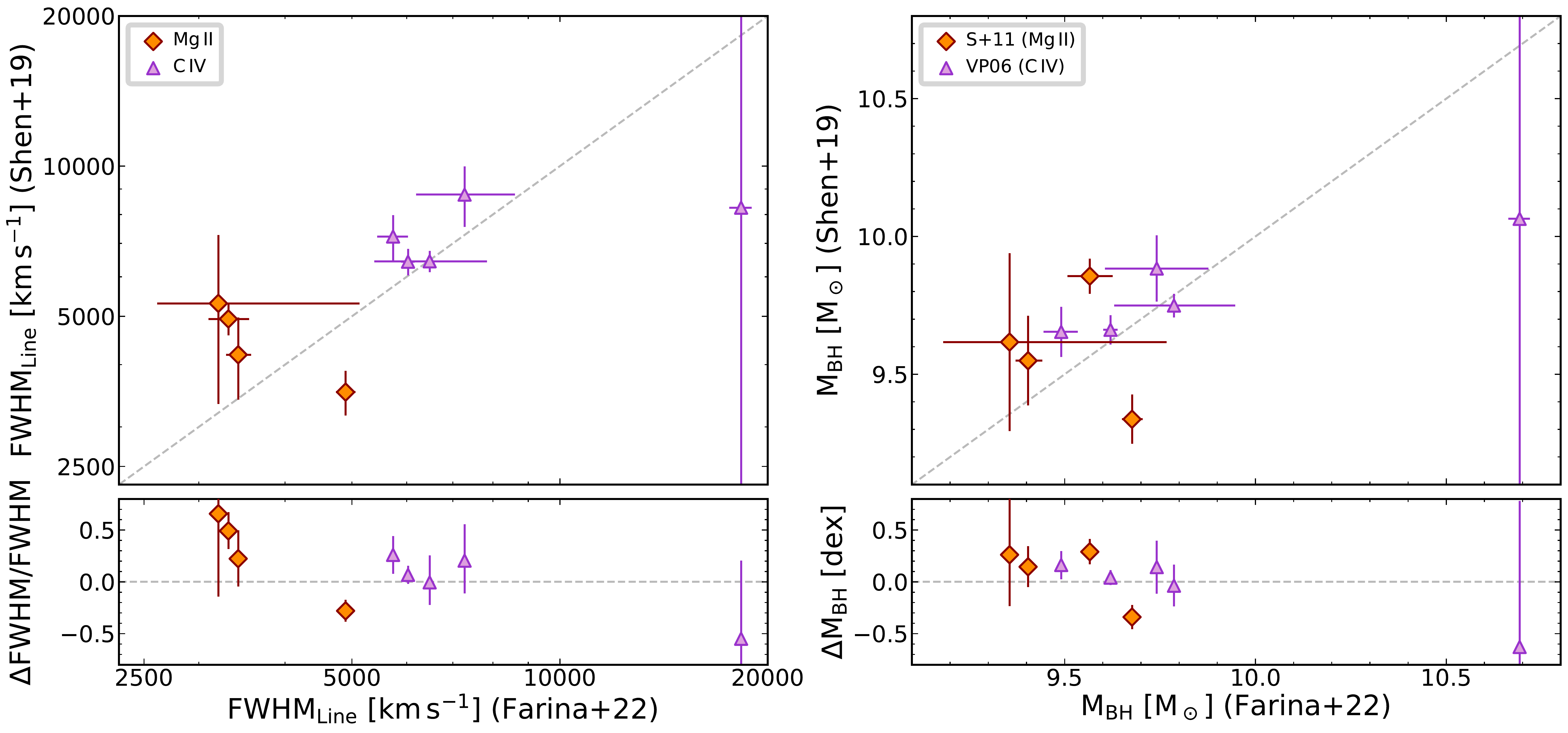}
    \caption{
    Comparison of FWHM measurements of the \mgii{} (orange diamonds) and \civ{} (purple triangles) line (left panel) and black hole mass estimates (right panel) for the five quasars in common between our \xs{} sample and the \citeauthor{Shen2019} one. 
    The relative differences between two measurements are showed in the bottom panels.
    }
    \label{fig:xs_vs_shen}
\end{figure}

\citet{Shen2019} presented black hole masses and Eddington ratios for a sample of 50 quasars at $z > 5.7$.
Data have been gathered with GNIRS on Gemini--North, that has a spectral resolution of $R\sim650$ (i.e., a factor $\sim10\times$ lower than \xs{}).
There are 5 quasars in common between ours and the \citeauthor{Shen2019} sample:
P007$+$04, J0842$+$1218, J1044$-$0125, J1148$+$0702, and J2310$+$1855.
While the overlap is small, it is worth to investigate if any significant discrepancy arises due to the different signal--to--noise, resolution, and fitting technique.
\autoref{fig:xs_vs_shen} summarizes the comparison.
Our measurements of the line FWHMs are on average $\sim130\,\kms{}$ (for \mgii{}) and $\sim610\,\kms{}$ (for \civ{}) smaller than the one measured by \citeauthor{Shen2019}.
The corresponding black hole masses are thus marginally larger but consistent within the uncertainties.

%

\section{Super--massive black holes seeds with different parameters}\label{app:test_seeds}

\begin{figure}[tbh]
    \centering
    \includegraphics[width=0.98\columnwidth]{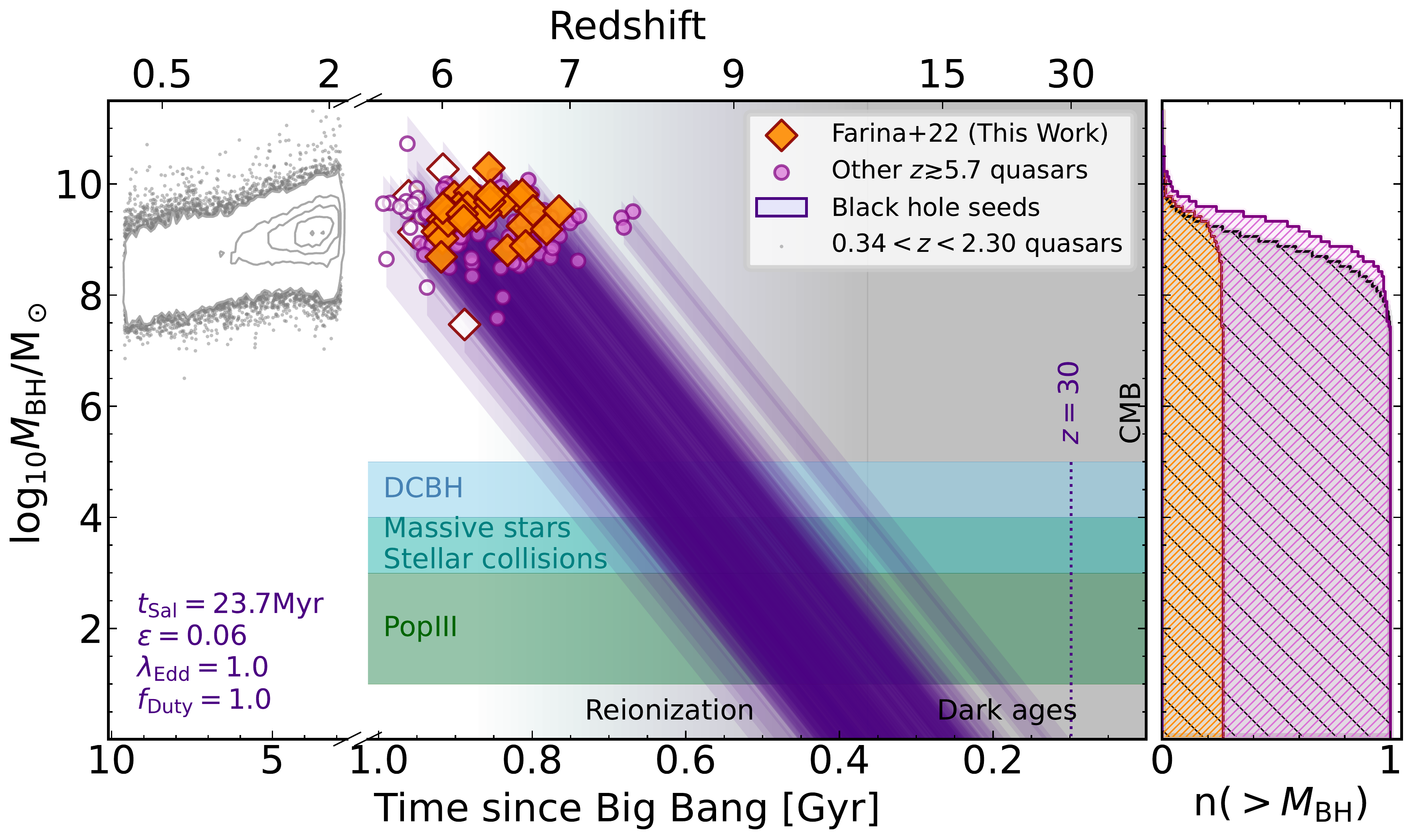}
    \includegraphics[width=0.98\columnwidth]{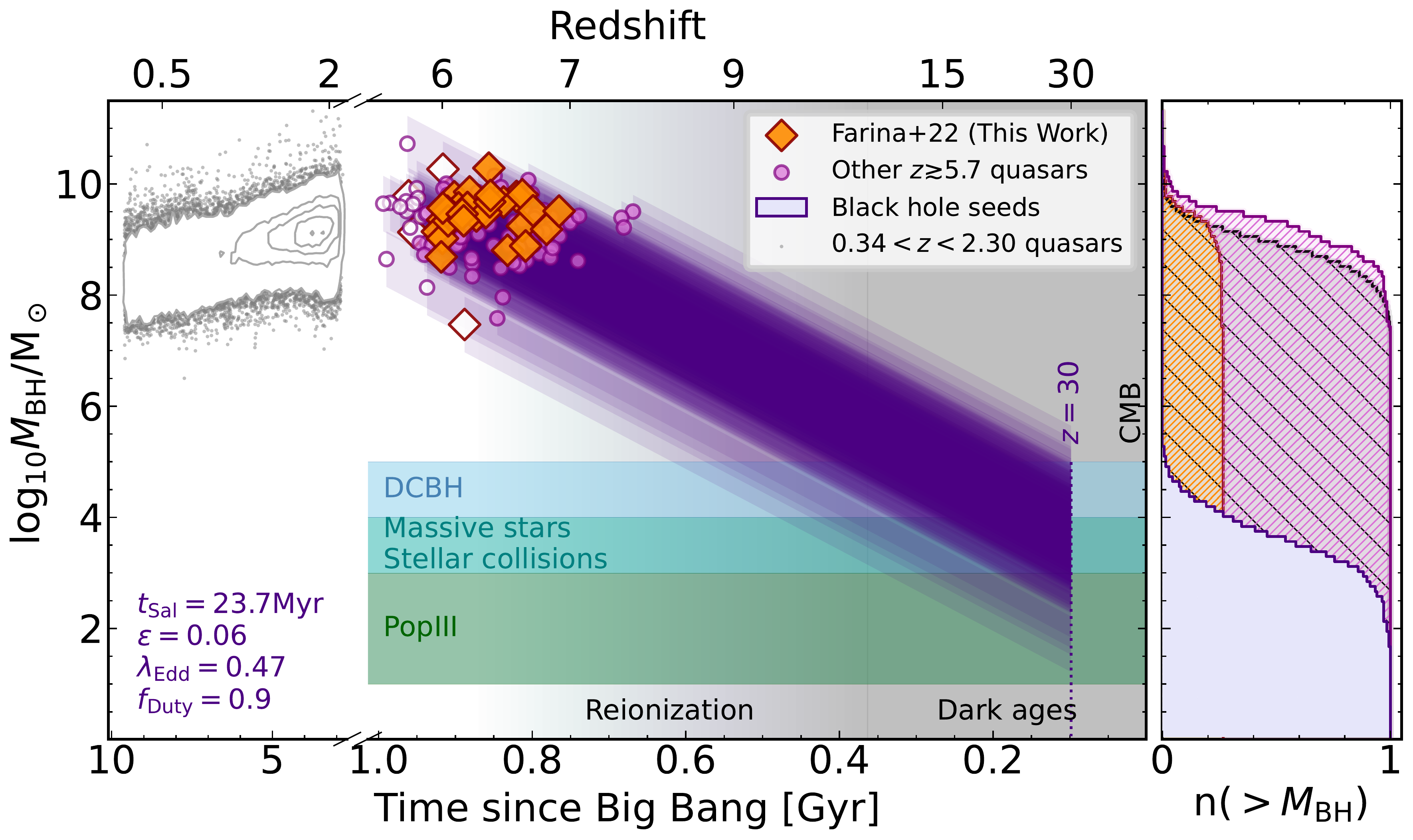}
    \caption{
    Same of \autoref{fig:bh_seeds} but for different black hole growth parameters.
    }
    \label{fig:bh_seeds_fast_slow}
\end{figure}

%
In \autoref{fig:bh_seeds_fast_slow} we show the different path for the black hole seed from redshift $z = 30$ to the observed masses at $z\sim6-7$.
As examples with consider to plug into \autoref{eq:massevolution} a Salpeter time of $t_\mathrm{Sal} \approx \epsilon\,395\,\mathrm{Myr}$ (estimated considering the presence of primordial Helium), a radiative efficiency of $\epsilon = 0.06$.
The two panels show two different cases: \textit{Top Panel} maximally accreting black holes ($\Eratio{} \sim 1.0$) for the entire history of the Universe ($f_\mathrm{duty} = 1.0$); and \textit{Bottom Panel} a more conservative scenario with $\Eratio{} \sim 0.48$ (the median value observed for $z\sim6$ quasars) and a duty cycle of $f_\mathrm{duty} = 0.9$.

\section{Quasars with \mgii--based black hole masses}\label{app:litsample}

%
In \autoref{tab:samplelit} we provide a summary of \mgii{} black hole mass estimates for $z \gtrsim 5.7$ quasars from the literature. 
All measurements are converted to the cosmology used in this paper.
Unless otherwise specified, the iron pseudo--continuum has been fitted using the \citet{Vestergaard2001} template.
The full table will be presented in the electronic version of the journal.
{\begin{deluxetable*}{lllLlCCCCCCl}
\rotate
\tabletypesize{\footnotesize} 
\tablecaption{List of quasars with \mgii{} based black hole masses \label{tab:samplelit}}
\tablehead{
\colhead{ID}                                                     &
\colhead{RA}                                                     &
\colhead{Dec}                                                    &
\colhead{$z_{\textrm{sys}}$}                                     &
\colhead{$z_{\textrm{sys}}$ Method}                              &
\colhead{$\MMCDL$}                                               &
\colhead{$\textrm{FWHM}_\textrm{MgII}$}                          &
\colhead{$3000\textrm{{\AA{}}}\,L_{\lambda,3000\textrm{\AA{}}}$} &
\colhead{$L_\textrm{bol}$}              &
\colhead{$\mbh^\textrm{S+11}$}          &
\colhead{$\Eratio^\textrm{S+11}$}       &
\colhead{Ref.}       \\
\colhead{}         &
\colhead{(J2000)}  &
\colhead{(J2000)}  &
\colhead{}         &
\colhead{}         &
\colhead{(mag)}    &
\colhead{($\kms{}$)}              &
\colhead{($10^{46}\,\ergs{}$)}    &
\colhead{($10^{46}\,\ergs{}$)}  &
\colhead{($10^9\,\msun{}$)}     &
\colhead{}                      &
\colhead{}            
}
\startdata
J0008$-$0626 & 00:08:25.77 & $-$06:26:04.6  & 5.929\phn    \pm   0.006\phn    & \ciii   & -26.32 & \phn2279_{-274\phn    }^{+274\phn       } & \phn2.07\phn_{-0.01\phn   }^{+0.01\phn   } &      \phn10.68 & \phn0.78 & 1.09 & \citet{Shen2019}  \\
J0020$-$3653 & 00:20:31.47 & $-$36:53:41.8  & 6.834\phn    \pm   0.0004       & \mgii   & -26.92 & \phn3800_{-360\phn    }^{+360\phn       } & \phn2.62\phn_{-0.05\phn   }^{+0.05\phn   } &      \phn13.49 & \phn2.51 & 0.43 & \citet{Reed2019}    \\
J0024$+$3913 & 00:24:29.77 & $+$39:13:19.0  & 6.621\phn    \pm   0.002\phn    & \ciimu  & -25.65 & \phn1783_{-38\phn\phn }^{+38\phn\phn    } & \phn1.69\phn_{-0.15\phn   }^{+0.15\phn   } &   \phn\phn8.70 & \phn0.42 & 1.65 & \citet{Yang2021}    \\
   P007$+$04 & 00:28:06.56 & $+$04:57:25.7  & 6.0015       \pm   0.0002       & \ciimu  & -26.51 & \phn3203_{-591\phn    }^{+1925          } & \phn3.91\phn_{-0.16\phn   }^{+0.13\phn   } &      \phn20.13 & \phn2.27 & 0.70 & Farina et al.\ (2022) \\
J0033$-$0125 & 00:33:11.40 & $-$01:25:24.9  & 6.019\phn    \pm   0.007\phn    & \mgii   & -25.14 & \phn6762_{-6808       }^{+6808          } & \phn0.63\phn_{-0.01\phn   }^{+0.01\phn   } &   \phn\phn3.24 & \phn3.28 & 0.08 & \citet{Shen2019} \\
J0038$-$1527 & 00:38:36.10 & $-$15:27:23.6  & 7.034\phn    \pm   0.0003       & \ciimu  & -27.13 & \phn3102_{-45\phn\phn }^{+45\phn\phn    } & \phn4.62\phn_{-0.19\phn   }^{+0.19\phn   } &      \phn23.79 & \phn2.37 & 0.80 & \citet{Yang2021} \\
   P011$+$09 & 00:45:33.57 & $+$09:01:57.0  & 6.4694       \pm   0.0025       & \ciimu  & -25.87 & \phn2238_{-373\phn    }^{+744\phn       } & \phn1.60\phn_{-0.03\phn   }^{+0.03\phn   } &   \phn\phn8.22 & \phn0.64 & 1.02 & Farina et al.\ (2022) \\
J0046$-$2837 & 00:46:23.65 & $-$28:37:47.3  & 5.9926       \pm   0.0028       & \mgii   & -25.09 & \phn1974_{-133\phn    }^{+134\phn       } & \phn1.55\phn_{-0.07\phn   }^{+0.06\phn   } &   \phn\phn7.98 & \phn0.49 & 1.30 & Farina et al.\ (2022) \\
J0050$+$3445 & 00:50:06.67 & $+$34:45:22.6  & 6.246\phn    \pm   0.006\phn    & \mgii   & -26.70 & \phn5860_{-491\phn    }^{+491\phn       } & \phn3.51\phn_{-0.03\phn   }^{+0.01\phn   } &      \phn18.10 & \phn7.15 & 0.20 & \citet{Shen2019} \\
J0100$+$2802 & 01:00:13.03 & $+$28:02:25.8  & 6.3269       \pm   0.0002       & \ciimu  & -29.02 & \phn4962_{-62\phn\phn }^{+65\phn\phn    } &    29.79\phn_{-0.03\phn   }^{+0.03\phn   } &         153.39 &    19.28 & 0.63 & Farina et al.\ (2022) \\
J0109$-$3047 & 01:09:53.13 & $-$30:47:26.3  & 6.7904       \pm   0.0003       & \ciimu  & -25.41 & \phn3530_{-620\phn    }^{+590\phn       } & \phn1.49\phn_{-0.04\phn   }^{+0.03\phn   } &   \phn\phn7.69 & \phn1.53 & 0.40 & Farina et al.\ (2022) \\
J0136$+$0226 & 01:36:03.17 & $+$02:26:05.7  & 6.206\phn    \pm   0.009\phn    & \mgii   & -24.66 & \phn2581_{-984\phn    }^{+984\phn       } & \phn0.43\phn_{-0.01\phn   }^{+0.01\phn   } &   \phn\phn2.15 & \phn0.37 & 0.46 & \citet{Shen2019} \\
J0142$-$3327 & 01:42:43.73 & $-$33:27:45.5  & 6.3373       \pm   0.0002       & \ciimu  & -27.76 & \phn4109_{\phm{-0000} }^{\phm{+0000}    } & \phn4.48\phn_{\phm{-0.000}}^{\phm{+0.000}} &      \phn23.07 & \phn4.09 & 0.45 & \citet{Chehade2018} \\
J0159$-$3633 & 01:59:57.97 & $-$36:33:56.6  & 6.027\phn\phn                   & \mgii   & -26.95 & \phn6510_{\phm{-0000} }^{\phm{+0000}    } & \phn4.40\phn_{\phm{-0.000}}^{\phm{+0.000}} &      \phn22.66 &    10.14 & 0.18 & \citet{Chehade2018} \\
J0210$-$0456 & 02:10:13.19 & $-$04:56:20.9  & 6.4323       \pm   0.0005       & \ciimu  & -24.28 & \phn1300_{-350\phn    }^{+350\phn       } & \phn0.40\phn_{-0.05\phn   }^{+0.05\phn   } &   \phn\phn2.06 & \phn0.09 & 1.79 & \citet{Willott2010BH} \\            
J0218$+$0007 & 02:18:47.04 & $+$00:07:15.2  & 6.7700       \pm   0.0013       & \ciimu  & -25.55 & \phn2745_{-42\phn\phn }^{+42\phn\phn    } & \phn1.24\phn_{-0.27\phn   }^{+0.27\phn   } &   \phn\phn6.39 & \phn0.82 & 0.62 & \citet{Yang2021} \\                 
J0221$-$0802 & 02:21:22.71 & $-$08:02:51.5  & 6.200\phn    \pm   0.036\phn    & \mgii   & -24.70 & \phn2297_{-8519       }^{+8519          } & \phn0.879   _{-0.026      }^{+0.027      } &   \phn\phn4.53 & \phn0.47 & 0.77 & \citet{Shen2019} \\                 
J0224$-$4711 & 02:24:26.54 & $-$47:11:29.4  & 6.527\phn    \pm   0.001\phn    & \mgii   & -26.94 & \phn2655_{-144\phn    }^{+144\phn       } & \phn6.27\phn_{-0.37\phn   }^{+0.37\phn   } &      \phn32.29 & \phn2.10 & 1.22 & \citet{WangFeige2021XRay} \\        
   P036$+$03 & 02:26:01.88 & $+$03:02:59.4  & 6.5405       \pm   0.0001       & \ciimu  & -27.15 & \phn4840_{-341\phn    }^{+364\phn       } & \phn4.847   _{-0.029      }^{+0.028      } &      \phn24.96 & \phn5.96 & 0.33 & Farina et al.\ (2022) \\            
J0227$-$0605 & 02:27:43.29 & $-$06:05:30.2  & 6.206\phn    \pm   0.008\phn    & \mgii   & -25.28 & \phn1969_{-709\phn    }^{+709\phn       } & \phn0.427   _{-0.009      }^{+0.009      } &   \phn\phn2.20 & \phn0.22 & 0.80 & \citet{Shen2019} \\                 
J0244$-$5008 & 02:44:01.02 & $-$50:08:53.7  & 6.7240       \pm   0.0008       & \mgii   & -26.72 & \phn3100_{-530\phn    }^{+530\phn       } & \phn2.79\phn_{-0.05\phn   }^{+0.05\phn   } &      \phn14.37 & \phn1.73 & 0.66 & \citet{Reed2019} \\                 
J0246$-$5219 & 02:46:55.90 & $-$52:19:49.9  & 6.8876       \pm   0.0003       & \ciimu  & -25.36 & \phn3319_{-693\phn    }^{+693\phn       } & \phn1.98\phn_{-0.19\phn   }^{+0.19\phn   } &      \phn10.20 & \phn1.61 & 0.50 & \citet{Yang2021} \\                 
J0252$-$0503 & 02:52:16.64 & $-$05:03:31.8  & 7.0006       \pm   0.0009       & \ciimu  & -26.63 & \phn3406_{-219\phn    }^{+219\phn       } & \phn2.56\phn_{-0.08\phn   }^{+0.08\phn   } &      \phn13.18 & \phn1.98 & 0.53 & \citet{Yang2021} \\                 
J0305$-$3150 & 03:05:16.92 & $-$31:50:55.9  & 6.6139       \pm   0.0001       & \ciimu  & -25.91 & \phn2347_{-280\phn    }^{+239\phn       } & \phn1.828   _{-0.024      }^{+0.02\phn   } &   \phn\phn9.42 & \phn0.77 & 0.98 & Farina et al.\ (2022) \\            
J0313$-$1806 & 03:13:43.84 & $-$18:06:36.4  & 7.6423       \pm   0.0013       & \ciimu  & -26.13 & \phn4219_{-465\phn    }^{+465\phn       } & \phn2.76\phn_{-0.14\phn   }^{+0.14\phn   } &      \phn14.21 & \phn3.19 & 0.35 & \citet{Yang2021} \\                 
J0319$-$1008 & 03:19:41.66 & $-$10:08:46.0  & 6.8275       \pm   0.0021       & \ciimu  & -25.36 & \phn2103_{-8\phm{000} }^{+8\phm{000}    } & \phn1.88\phn_{-0.27\phn   }^{+0.27\phn   } &   \phn\phn9.68 & \phn0.62 & 1.23 & \citet{Yang2021} \\                 
   P056$-$16 & 03:46:52.04 & $-$16:28:36.9  & 5.9670       \pm   0.0023       & \ciimu  & -26.26 & \phn3049_{-118\phn    }^{+121\phn       } & \phn2.056   _{-0.022      }^{+0.022      } &      \phn10.59 & \phn1.39 & 0.61 & Farina et al.\ (2022) \\            
J0353$+$0104 & 03:53:49.72 & $+$01:04:04.4  & 6.057\phn    \pm   0.005\phn    & \mgii   & -26.43 & \phn3229_{-316\phn    }^{+316\phn       } & \phn2.472   _{-0.013      }^{+0.013      } &      \phn12.73 & \phn1.75 & 0.58 & \citet{Shen2019} \\                 
   P060$+$24 & 04:02:12.69 & $+$24:51:24.4  & 6.174\phn    \pm   0.006\phn    & \mgii   & -26.95 & \phn3877_{-459\phn    }^{+459\phn       } & \phn2.736   _{-0.008      }^{+0.008      } &      \phn14.09 & \phn2.68 & 0.42 & \citet{Shen2019} \\                 
J0411$-$0907 & 04:11:28.63 & $-$09:07:49.8  & 6.8260       \pm   0.0007       & \ciimu  & -26.58 & \phn2837_{-75\phn\phn }^{+75\phn\phn    } & \phn3.11\phn_{-0.21\phn   }^{+0.21\phn   } &      \phn16.02 & \phn1.55 & 0.82 & \citet{Yang2021} \\                 
   P065$-$26 & 04:21:38.05 & $-$26:57:15.6  & 6.1871       \pm   0.0003       & \ciimu  & -26.94 & \phn5558_{-303\phn    }^{+352\phn       } & \phn4.212   _{-0.087      }^{+0.088      } &      \phn21.69 & \phn7.20 & 0.24 & Farina et al.\ (2022) \\            
   P065$-$19 & 04:22:01.00 & $-$19:27:28.7  & 6.1247       \pm   0.0006       & \ciimu  & -26.11 & \phn3861_{-103\phn    }^{+109\phn       } & \phn3.716   _{-0.077      }^{+0.073      } &      \phn19.14 & \phn3.21 & 0.47 & Farina et al.\ (2022) \\            
J0439$+$1634 & 04:39:47.08 & $+$16:34:15.7  & 6.5188       \pm   0.0004       & \ciimu  & -25.31 & \phn3041_{-14\phn\phn }^{+14\phn\phn    } & \phn0.99\phn_{-0.02\phn   }^{+0.02\phn   } &   \phn\phn5.10 & \phn0.88 & 0.46 & \citet{Yang2021} \\                 
J0525$-$2406 & 05:25:59.68 & $-$24:06:23.0  & 6.5397       \pm   0.0001       & \ciimu  & -25.47 & \phn2048_{-472\phn    }^{+472\phn       } & \phn1.5 \phn_{-0.76\phn   }^{+0.76\phn   } &   \phn\phn7.72 & \phn0.52 & 1.19 & \citet{Yang2021} \\                 
J0535$+$1150 & 05:35:20.89 & $+$11:50:53.6  & 6.3401       \pm   0.0004       & \ciimu  & -26.67 & \phn4140_{-430\phn    }^{+880\phn       } & \phn2.58\phn_{-0.06\phn   }^{+0.02\phn   } &      \phn13.29 & \phn2.95 & 0.36 & \citet{Andika2020} \\               
\enddata 
\end{deluxetable*}

\begin{deluxetable*}{lllLlCCCCCCl}
\rotate
\tablenum{5}
\tabletypesize{\footnotesize} 
\tablecaption{Continued}
\tablehead{
\colhead{ID}                                                     &
\colhead{RA}                                                     &
\colhead{Dec}                                                    &
\colhead{$z_{\textrm{sys}}$}                                     &
\colhead{$z_{\textrm{sys}}$ Method}                              &
\colhead{$\MMCDL$}                                               &
\colhead{$\textrm{FWHM}_\textrm{MgII}$}                          &
\colhead{$3000\textrm{{\AA{}}}\,L_{\lambda,3000\textrm{\AA{}}}$} &
\colhead{$L_\textrm{bol}$}              &
\colhead{$\mbh^\textrm{S+11}$}          &
\colhead{$\Eratio^\textrm{S+11}$}       &
\colhead{Ref.}       \\
\colhead{}         &
\colhead{(J2000)}  &
\colhead{(J2000)}  &
\colhead{}         &
\colhead{}         &
\colhead{(mag)}    &
\colhead{($\kms{}$)}              &
\colhead{($10^{46}\,\ergs{}$)}    &
\colhead{($10^{46}\,\ergs{}$)}  &
\colhead{($10^9\,\msun{}$)}     &
\colhead{}                      &
\colhead{}            
}
\startdata
J0706$+$2921 & 07:06:26.39 & $+$29:21:05.5  & 6.6037       \pm   0.0003       & \ciimu  & -27.44 & \phn3375_{-31\phn\phn }^{+31\phn\phn    } & \phn7.15\phn_{-0.31\phn   }^{+0.31\phn   } &      \phn36.82 & \phn3.68 & 0.80 & \citet{Yang2021} \\                 
J0803$+$3138 & 08:03:05.42 & $+$31:38:34.2  & 6.384\phn    \pm   0.004\phn    & \mgii   & -26.49 & \phn4432_{-119\phn    }^{+119\phn       } & \phn2.82\phn_{-0.25\phn   }^{+0.25\phn   } &      \phn14.52 & \phn3.57 & 0.32 & \citet{Yang2021} \\                 
J0829$+$4117 & 08:29:31.97 & $+$41:17:40.4  & 6.773\phn    \pm   0.007\phn    & \mgii   & -26.07 & \phn2869_{-82\phn\phn }^{+82\phn\phn    } & \phn2.52\phn_{-0.19\phn   }^{+0.19\phn   } &      \phn12.98 & \phn1.39 & 0.74 & \citet{Yang2021} \\                 
J0835$+$3217 & 08:35:25.76 & $+$32:17:52.6  & 5.902\phn    \pm   0.009\phn    & \mgii   & -25.81 & \phn4826_{-1049       }^{+1049          } & \phn0.587   _{-0.005      }^{+0.005      } &   \phn\phn3.02 & \phn1.60 & 0.15 & \citet{Shen2019} \\                 
J0837$+$4929 & 08:37:37.84 & $+$49:29:00.4  & 6.702\phn    \pm   0.001\phn    & \mgii   & -26.33 & \phn2565_{-48\phn\phn }^{+48\phn\phn    } & \phn3.38\phn_{-0.10\phn   }^{+0.1 \phn   } &      \phn17.41 & \phn1.34 & 1.04 & \citet{Yang2021} \\                 
J0839$+$3900 & 08:39:46.88 & $+$39:00:11.5  & 6.9046       \pm   0.0003       & \mgii   & -26.36 & \phn2332_{-47\phn\phn }^{+47\phn\phn    } & \phn3.63\phn_{-0.12\phn   }^{+0.12\phn   } &      \phn18.69 & \phn1.16 & 1.29 & \citet{Yang2021} \\                 
J0841$+$2905 & 08:41:19.52 & $+$29:05:04.4  & 5.954\phn    \pm   0.005\phn    & \mgii   & -26.50 & \phn3073_{-551\phn    }^{+551\phn       } & \phn2.111   _{-0.009      }^{+0.009      } &      \phn10.87 & \phn1.43 & 0.60 & \citet{Shen2019} \\                 
J0842$+$1218 & 08:42:29.43 & $+$12:18:50.5  & 6.0754       \pm   0.0005       & \ciimu  & -26.69 & \phn3422_{-135\phn    }^{+153\phn       } & \phn3.76\phn_{-0.039      }^{+0.038      } &      \phn19.37 & \phn2.54 & 0.61 & Farina et al.\ (2022) \\            
J0859$+$0022 & 08:59:07.19 & $+$00:22:55.9  & 6.3903       \pm   0.0005       & \ciimu  & -23.10 & \phn1280_{-410\phn    }^{+240\phn       } & \phn0.103   _{-0.100\phn  }^{+0.100      } &   \phn\phn0.53 & \phn0.04 & 1.10 & \citet{Onoue2019} \\                
J0910$+$1656 & 09:10:13.63 & $+$16:56:29.8  & 6.7289       \pm   0.0005       & \ciimu  & -25.34 & \phn2321_{-40\phn\phn }^{+40\phn\phn    } & \phn1.15\phn_{-0.12\phn   }^{+0.12\phn   } &   \phn\phn5.92 & \phn0.56 & 0.84 & \citet{Yang2021} \\                 
J0910$-$0414 & 09:10:54.53 & $-$04:14:06.8  & 6.6363       \pm   0.0003       & \ciimu  & -26.61 & \phn7825_{-844\phn    }^{+844\phn       } & \phn3.09\phn_{-0.17\phn   }^{+0.17\phn   } &      \phn15.91 &    11.77 & 0.11 & \citet{Yang2021} \\                 
J0921$+$0007 & 09:21:20.56 & $+$00:07:22.9  & 6.5646       \pm   0.0003       & \ciimu  & -25.19 & \phn1729_{-105\phn    }^{+105\phn       } & \phn1.28\phn_{-0.12\phn   }^{+0.12\phn   } &   \phn\phn6.59 & \phn0.33 & 1.58 & \citet{Yang2021} \\                 
J0923$+$0402 & 09:23:47.12 & $+$04:02:54.4  & 6.633\phn    \pm   0.0003       & \ciimu  & -26.68 & \phn3362_{-183\phn    }^{+183\phn       } & \phn4.99\phn_{-0.60\phn   }^{+0.60\phn   } &      \phn25.70 & \phn2.92 & 0.70 & \citet{Yang2021} \\                 
J0923$+$0753 & 09:23:59.00 & $+$07:53:49.1  & 6.6817       \pm   0.0005       & \ciimu  & -25.50 & \phn2800_{-475\phn    }^{+475\phn       } & \phn1.11\phn_{-0.35\phn   }^{+0.35\phn   } &   \phn\phn5.72 & \phn0.80 & 0.57 & \citet{Yang2021} \\                 
J1007$+$2115 & 10:07:58.26 & $+$21:15:29.2  & 7.5149       \pm   0.0004       & \ciimu  & -26.73 & \phn3321_{-170\phn    }^{+170\phn       } & \phn3.96\phn_{-0.27\phn   }^{+0.27\phn   } &      \phn20.39 & \phn2.47 & 0.66 & \citet{Yang2021} \\                 
J1030$+$0524 & 10:30:27.10 & $+$05:24:55.0  & 6.3048       \pm   0.0012       & \lya    & -26.76 & \phn3511_{-192\phn    }^{+209\phn       } & \phn4.655   _{-0.071      }^{+0.072      } &      \phn23.97 & \phn3.06 & 0.62 & Farina et al.\ (2022) \\            
   P158$-$14 & 10:34:46.51 & $-$14:25:15.9  & 6.0681       \pm   0.0024       & \ciimu  & -27.07 & \phn2794_{-194\phn    }^{+208\phn       } & \phn9.078   _{-0.201      }^{+0.209      } &      \phn46.75 & \phn2.93 & 1.27 & Farina et al.\ (2022) \\            
   P159$-$02 & 10:36:54.19 & $-$02:32:37.9  & 6.3809       \pm   0.0005       & \ciimu  & -26.47 & \phn4798_{-252\phn    }^{+286\phn       } & \phn3.535   _{-0.062      }^{+0.063      } &      \phn18.21 & \phn4.82 & 0.30 & Farina et al.\ (2022) \\            
J1048$-$0109 & 10:48:19.08 & $-$01:09:40.3  & 6.6759       \pm   0.0002       & \ciimu  & -26.20 & \phn4703_{-635\phn    }^{+620\phn       } & \phn2.048   _{-0.058      }^{+0.059      } &      \phn10.55 & \phn3.29 & 0.26 & Farina et al.\ (2022) \\            
J1058$+$2930 & 10:58:07.72 & $+$29:30:41.7  & 6.5846       \pm   0.0005       & \ciimu  & -25.68 & \phn2642_{-76\phn\phn }^{+76\phn\phn    } & \phn1.26\phn_{-0.29\phn   }^{+0.29\phn   } &   \phn\phn6.49 & \phn0.77 & 0.67 & \citet{Yang2021} \\                 
J1104$+$2134 & 11:04:21.59 & $+$21:34:28.8  & 6.7662       \pm   0.0009       & \ciimu  & -26.63 & \phn4198_{-55\phn\phn }^{+55\phn\phn    } & \phn3.07\phn_{-0.17\phn   }^{+0.17\phn   } &      \phn15.81 & \phn3.37 & 0.37 & \citet{Yang2021} \\                 
J1110$-$1329 & 11:10:33.96 & $-$13:29:45.6  & 6.5148       \pm   0.0005       & \ciimu  & -25.57 & \phn2071_{-354\phn    }^{+211\phn       } & \phn0.90\phn_{-0.04\phn   }^{+0.02\phn   } &   \phn\phn4.64 & \phn0.38 & 0.96 & \citet{Mazzucchelli2017} \\         
J1120$+$0641 & 11:20:01.48 & $+$06:41:24.3  & 7.0851       \pm   0.0005       & \ciimu  & -26.44 & \phn3928_{-344\phn    }^{+344\phn       } & \phn2.60\phn_{-0.21\phn   }^{+0.21\phn   } &      \phn13.39 & \phn2.66 & 0.40 & \citet{Yang2021} \\                 
J1129$+$1846 & 11:29:25.34 & $+$18:46:24.2  & 6.824\phn    \pm   0.001\phn    & \mgii   & -25.73 & \phn1862_{-45\phn\phn }^{+45\phn\phn    } & \phn1.77\phn_{-0.39\phn   }^{+0.39\phn   } &   \phn\phn9.12 & \phn0.47 & 1.54 & \citet{Yang2021} \\                 
J1135$+$5011 & 11:35:08.93 & $+$50:11:33.0  & 6.5851       \pm   0.0008       & \ciimu  & -26.16 & \phn3651_{-98\phn\phn }^{+98\phn\phn    } & \phn2.35\phn_{-0.16\phn   }^{+0.16\phn   } &      \phn12.10 & \phn2.16 & 0.45 & \citet{Yang2021} \\                 
J1137$+$3549 & 11:37:17.73 & $+$35:49:56.9  & 6.007\phn    \pm   0.014\phn    & \mgii   & -27.36 & \phn5947_{-390\phn    }^{+390\phn       } & \phn4.176   _{-0.019      }^{+0.019      } &      \phn21.50 & \phn8.19 & 0.21 & \citet{Shen2019} \\                 
J1148$+$0702 & 11:48:03.29 & $+$07:02:08.3  & 6.3337       \pm   0.0028       & \mgii   & -26.31 & \phn4895_{-154\phn    }^{+165\phn       } & \phn3.244   _{-0.037      }^{+0.042      } &      \phn16.71 & \phn4.75 & 0.28 & Farina et al.\ (2022) \\            
J1148$+$5251 & 11:48:16.64 & $+$52:51:50.3  & 6.416\phn    \pm   0.006\phn    & \ciimu  & -27.62 & \phn5302_{-115\phn    }^{+115\phn       } & \phn6.705   _{-0.029      }^{+0.029      } &      \phn34.53 & \phn8.73 & 0.31 & \citet{Shen2019} \\                 
J1152$+$0055 & 11:52:21.27 & $+$00:55:36.6  & 6.3637       \pm   0.0005       & \ciimu  & -25.08 & \phn3240_{-450\phn    }^{+280\phn       } & \phn0.677   _{-0.11\phn   }^{+0.11\phn   } &   \phn\phn3.49 & \phn0.79 & 0.35 & \citet{Onoue2019} \\                
J1205$-$0000 & 12:05:05.09 & $-$00:00:27.9  & 6.699\phn    \pm   0.004\phn    & \mgii   & -25.54 & \phn5620_{-990\phn    }^{+220\phn       } & \phn0.896   _{-0.66\phn   }^{+0.66\phn   } &   \phn\phn4.61 & \phn2.82 & 0.13 & \citet{Onoue2019} \\                
J1207$+$0630 & 12:07:37.43 & $+$06:30:10.1  & 6.028\phn    \pm   0.013\phn    & \ciimu  & -26.63 & \phn6801_{-467\phn    }^{+467\phn       } & \phn1.804   _{-0.014      }^{+0.014      } &   \phn\phn9.29 & \phn6.37 & 0.12 & \citet{Shen2019} \\                 
J1208$-$0200 & 12:08:59.23 & $-$02:00:34.8  & 6.1165       \pm   0.0002       & \ciimu  & -24.36 & \phn3850_{-1990       }^{+920\phn       } & \phn0.438   _{-0.04\phn   }^{+0.04\phn   } &   \phn\phn2.26 & \phn0.85 & 0.21 & \citet{Onoue2019} \\                
   P183$+$05 & 12:12:26.98 & $+$05:05:33.5  & 6.4386       \pm   0.0002       & \ciimu  & -26.87 & \phn4459_{-334\phn    }^{+311\phn       } & \phn4.356   _{-0.058      }^{+0.058      } &      \phn22.43 & \phn4.72 & 0.38 & Farina et al.\ (2022) \\            
J1216$+$4519 & 12:16:27.58 & $+$45:19:10.7  & 6.648\phn    \pm   0.003\phn    & \mgii   & -25.57 & \phn3102_{-652\phn    }^{+652\phn       } & \phn1.26\phn_{-0.23\phn   }^{+0.23\phn   } &   \phn\phn6.49 & \phn1.06 & 0.49 & \citet{Yang2021} \\                 
J1243$+$0100 & 12:43:53.93 & $+$01:00:38.5  & 7.0749       \pm   0.0001       & \ciimu  & -24.13 & \phn3100_{-900\phn    }^{+900\phn       } & \phn0.274   _{\phm{-0.000}}^{\phm{+0.000}} &   \phn\phn1.41 & \phn0.41 & 0.27 & \citet{Matsuoka2019} \\             
\enddata 
\end{deluxetable*}

\begin{deluxetable*}{lllLlCCCCCCl}
\rotate
\tablenum{5}
\tabletypesize{\footnotesize} 
\tablecaption{Continued}
\tablehead{
\colhead{ID}                                                     &
\colhead{RA}                                                     &
\colhead{Dec}                                                    &
\colhead{$z_{\textrm{sys}}$}                                     &
\colhead{$z_{\textrm{sys}}$ Method}                              &
\colhead{$\MMCDL$}                                               &
\colhead{$\textrm{FWHM}_\textrm{MgII}$}                          &
\colhead{$3000\textrm{{\AA{}}}\,L_{\lambda,3000\textrm{\AA{}}}$} &
\colhead{$L_\textrm{bol}$}              &
\colhead{$\mbh^\textrm{S+11}$}          &
\colhead{$\Eratio^\textrm{S+11}$}       &
\colhead{Ref.}       \\
\colhead{}         &
\colhead{(J2000)}  &
\colhead{(J2000)}  &
\colhead{}         &
\colhead{}         &
\colhead{(mag)}    &
\colhead{($\kms{}$)}              &
\colhead{($10^{46}\,\ergs{}$)}    &
\colhead{($10^{46}\,\ergs{}$)}  &
\colhead{($10^9\,\msun{}$)}     &
\colhead{}                      &
\colhead{}            
}
\startdata
J1250$+$3130 & 12:50:51.93 & $+$31:30:22.9  & 6.138\phn    \pm   0.005\phn    & \mgii   & -26.53 & \phn2580_{-277\phn    }^{+277\phn       } & \phn2.535   _{-0.013      }^{+0.013      } &      \phn13.06 & \phn1.13 & 0.92 & \citet{Shen2019} \\                 
J1257$+$6349 & 12:57:57.47 & $+$63:49:37.2  & 5.977\phn    \pm   0.015\phn    & \mgii   & -26.27 & \phn5563_{-488\phn    }^{+488\phn       } & \phn1.273   _{-0.011      }^{+0.012      } &   \phn\phn6.56 & \phn3.43 & 0.15 & \citet{Shen2019} \\                 
J1306$+$0356 & 13:06:08.26 & $+$03:56:26.3  & 6.033\phn    \pm   0.0002       & \ciimu  & -26.70 & \phn3895_{-58\phn\phn }^{+69\phn\phn    } & \phn3.639   _{-0.022      }^{+0.023      } &      \phn18.74 & \phn3.23 & 0.46 & Farina et al.\ (2022) \\            
J1316$+$1028 & 13:16:08.14 & $+$10:28:32.8  & 6.329\phn    \pm   0.005\phn    & \mgii   & -25.67 & \phn3168_{-1314       }^{+1314          } & \phn2.95\phn_{-0.56\phn   }^{+0.56\phn   } &      \phn15.19 & \phn1.87 & 0.64 & \citet{Yang2021} \\                 
J1319$+$0950 & 13:19:11.30 & $+$09:50:51.5  & 6.1347       \pm   0.0005       & \ciimu  & -26.80 & \phn3747_{-113\phn    }^{+112\phn       } & \phn3.431   _{-0.012      }^{+0.013      } &      \phn17.67 & \phn2.88 & 0.49 & Farina et al.\ (2022) \\            
J1335$+$3533 & 13:35:50.81 & $+$35:33:15.8  & 5.87\phn\phn \pm   0.02\phn\phn & \co     & -26.67 & \phn3917_{-2728       }^{+2728          } & \phn3.112   _{-0.015      }^{+0.015      } &      \phn16.03 & \phn2.96 & 0.43 & \citet{Shen2019} \\                 
J1342$+$0928 & 13:42:08.10 & $+$09:28:38.6  & 7.5413       \pm   0.0007       & \ciimu  & -26.67 & \phn3094_{-195\phn    }^{+195\phn       } & \phn2.58\phn_{-0.21\phn   }^{+0.21\phn   } &      \phn13.29 & \phn1.65 & 0.64 & \citet{Yang2021} \\                 
   P210$+$27 & 14:01:47.34 & $+$27:49:35.0  & 6.167\phn    \pm   0.010\phn    & \mgii   & -26.54 & \phn4705_{-1025       }^{+1025          } & \phn1.79\phn_{-0.018      }^{+0.018      } &   \phn\phn9.22 & \phn3.03 & 0.24 & \citet{Shen2019} \\                 
J1411$+$1217 & 14:11:11.30 & $+$12:17:37.0  & 5.854\phn    \pm   0.003\phn    & \mgii   & -26.62 & \phn2208_{-317\phn    }^{+317\phn       } & \phn2.70\phn_{-0.07\phn   }^{+0.07\phn   } &      \phn13.91 & \phn0.86 & 1.28 & \citet{DeRosa2011} \\               
J1427$+$3312 & 14:27:38.59 & $+$33:12:41.0  & 6.121\phn    \pm   0.005\phn    & \mgii   & -26.10 & \phn2670_{-225\phn    }^{+225\phn       } & \phn2.16\phn_{-0.017      }^{+0.017      } &      \phn11.12 & \phn1.10 & 0.81 & \citet{Shen2019} \\                 
J1429$+$5447 & 14:29:52.17 & $+$54:47:17.7  & 6.119\phn    \pm   0.008\phn    & \co     & -26.10 & \phn3739_{-1599       }^{+1599          } & \phn1.678   _{-0.01\phn   }^{+0.01\phn   } &   \phn\phn8.64 & \phn1.84 & 0.37 & \citet{Shen2019} \\                 
J1509$-$1749 & 15:09:41.78 & $-$17:49:26.8  & 6.1225       \pm   0.0007       & \ciimu  & -26.56 & \phn4067_{-153\phn    }^{+156\phn       } & \phn4.871   _{-0.047      }^{+0.043      } &      \phn25.09 & \phn4.22 & 0.47 & Farina et al.\ (2022) \\            
   P228$+$21 & 15:14:44.91 & $+$21:14:19.8  & 5.892\phn    \pm   0.014\phn    & \mgii   & -26.11 & \phn4057_{-905\phn    }^{+905\phn       } & \phn1.039   _{-0.017      }^{+0.018      } &   \phn\phn5.35 & \phn1.61 & 0.26 & \citet{Shen2019} \\                 
   P231$-$20 & 15:26:37.84 & $-$20:50:00.7  & 6.5869       \pm   0.0004       & \ciimu  & -27.07 & \phn5196_{-493\phn    }^{+588\phn       } & \phn4.337   _{-0.102      }^{+0.104      } &      \phn22.34 & \phn6.40 & 0.28 & Farina et al.\ (2022) \\            
J1535$+$1943 & 15:35:32.87 & $+$19:43:20.1  & 6.370\phn    \pm   0.001\phn    & \mgii   & -27.09 & \phn6577_{-464\phn    }^{+464\phn       } & \phn7.17\phn_{-0.35\phn   }^{+0.35\phn   } &      \phn36.93 &    14.01 & 0.21 & \citet{Yang2021} \\                 
   P239$-$07 & 15:58:50.99 & $-$07:24:09.6  & 6.1097       \pm   0.0024       & \ciimu  & -27.07 & \phn3611_{-150\phn    }^{+135\phn       } & \phn5.896   _{-0.084      }^{+0.075      } &      \phn30.37 & \phn3.73 & 0.65 & Farina et al.\ (2022) \\            
J1602$+$4228 & 16:02:53.98 & $+$42:28:24.9  & 6.079\phn    \pm   0.005\phn    & \mgii   & -26.94 & \phn3359_{-160\phn    }^{+160\phn       } & \phn3.597   _{-0.017      }^{+0.017      } &      \phn18.53 & \phn2.38 & 0.62 & \citet{Shen2019} \\                 
J1609$+$3041 & 16:09:37.27 & $+$30:41:47.6  & 6.142\phn    \pm   0.006\phn    & \mgii   & -26.39 & \phn6365_{-257\phn    }^{+257\phn       } & \phn1.189   _{-0.007      }^{+0.007      } &   \phn\phn6.12 & \phn4.31 & 0.11 & \citet{Shen2019} \\                 
J1623$+$3112 & 16:23:31.81 & $+$31:12:00.5  & 6.254\phn    \pm   0.006\phn    & \ciimu  & -26.55 & \phn3841_{-207\phn    }^{+207\phn       } & \phn2.54\phn_{-0.022      }^{+0.022      } &      \phn13.08 & \phn2.51 & 0.41 & \citet{Shen2019} \\                 
J1629$+$2407 & 16:29:11.30 & $+$24:07:39.7  & 6.476\phn    \pm   0.004\phn    & \mgii   & -26.53 & \phn1975_{-288\phn    }^{+312\phn       } & \phn3.40\phn_{-0.15\phn   }^{+0.01\phn   } &      \phn17.51 & \phn0.80 & 1.75 & \citet{Mazzucchelli2017} \\         
J1630$+$4012 & 16:30:33.90 & $+$40:12:09.6  & 6.063\phn    \pm   0.006\phn    & \mgii   & -26.19 & \phn4472_{-401\phn    }^{+401\phn       } & \phn1.37\phn_{-0.009      }^{+0.009      } &   \phn\phn7.06 & \phn2.32 & 0.24 & \citet{Shen2019} \\                 
J1641$+$3755 & 16:41:21.64 & $+$37:55:20.5  & 6.047\phn    \pm   0.003\phn    & \mgii   & -25.19 & \phn1740_{-190\phn    }^{+190\phn       } & \phn1.14\phn_{-0.12\phn   }^{+0.12\phn   } &   \phn\phn5.87 & \phn0.31 & 1.49 & \citet{Willott2010BH} \\            
J1724$+$1901 & 17:24:08.74 & $+$19:01:43.0  & 6.480\phn    \pm   0.001\phn    & \mgii   & -25.55 & \phn2927_{-50\phn\phn }^{+50\phn\phn    } & \phn1.67\phn_{-0.25\phn   }^{+0.25\phn   } &   \phn\phn8.60 & \phn1.12 & 0.61 & \citet{Yang2021} \\                 
J2002$-$3013 & 20:02:41.59 & $-$30:13:21.7  & 6.6876       \pm   0.0004       & \ciimu  & -26.90 & \phn3501_{-250\phn    }^{+250\phn       } & \phn3.20\phn_{-0.37\phn   }^{+0.37\phn   } &      \phn16.48 & \phn2.41 & 0.54 & \citet{Yang2021} \\                 
   P308$-$21 & 20:32:09.99 & $-$21:14:02.3  & 6.2355       \pm   0.0003       & \ciimu  & -26.27 & \phn3355_{-361\phn    }^{+203\phn       } & \phn4.25\phn_{-0.038      }^{+0.037      } &      \phn21.89 & \phn2.65 & 0.66 & Farina et al.\ (2022) \\            
J2054$-$0005 & 20:54:06.49 & $-$00:05:14.8  & 6.0389       \pm   0.0001       & \ciimu  & -26.15 & \phn3633_{-230\phn    }^{+215\phn       } & \phn2.402   _{-0.061      }^{+0.072      } &      \phn12.37 & \phn2.17 & 0.45 & Farina et al.\ (2022) \\            
J2100$-$1715 & 21:00:54.62 & $-$17:15:22.5  & 6.0807       \pm   0.0004       & \ciimu  & -24.63 & \phn8012_{-1466       }^{+814\phn       } & \phn0.931   _{-0.021      }^{+0.02\phn   } &   \phn\phn4.80 & \phn5.88 & 0.06 & Farina et al.\ (2022) \\            
J2102$-$1458 & 21:02:19.22 & $-$14:58:54.0  & 6.6645       \pm   0.0002       & \ciimu  & -25.53 & \phn2882_{-203\phn    }^{+203\phn       } & \phn1.22\phn_{-0.12\phn   }^{+0.12\phn   } &   \phn\phn6.28 & \phn0.90 & 0.56 & \citet{Yang2021} \\                 
   P323$+$12 & 21:32:33.19 & $+$12:17:55.3  & 6.5872       \pm   0.0004       & \ciimu  & -26.89 & \phn2803_{-184\phn    }^{+174\phn       } & \phn3.891   _{-0.04\phn   }^{+0.039      } &      \phn20.04 & \phn1.75 & 0.92 & Farina et al.\ (2022) \\            
J2211$-$6320 & 22:11:00.60 & $-$63:20:55.8  & 6.8449       \pm   0.0003       & \ciimu  & -25.38 & \phn2613_{-819\phn    }^{+819\phn       } & \phn1.13\phn_{-0.04\phn   }^{+0.04\phn   } &   \phn\phn5.82 & \phn0.70 & 0.66 & \citet{Yang2021} \\                 
J2211$-$3206 & 22:11:12.39 & $-$32:06:12.9  & 6.3394       \pm   0.001\phn    & \ciimu  & -27.09 & \phn4666_{-156\phn    }^{+196\phn       } & \phn5.819   _{-0.118      }^{+0.112      } &      \phn29.97 & \phn6.21 & 0.38 & Farina et al.\ (2022) \\            
   P333$+$26 & 22:15:56.63 & $+$26:06:29.4  & 6.027\phn    \pm   0.006\phn    & \mgii   & -26.44 & \phn3069_{-605\phn    }^{+605\phn       } & \phn1.534   _{-0.014      }^{+0.014      } &   \phn\phn7.90 & \phn1.17 & 0.54 & \citet{Shen2019} \\                 
J2216$-$0016 & 22:16:44.47 & $-$00:16:50.1  & 6.0962       \pm   0.0003       & \ciimu  & -23.65 & \phn4320_{-1060       }^{+830\phn       } & \phn0.266   _{-0.05\phn   }^{+0.05\phn   } &   \phn\phn1.37 & \phn0.78 & 0.14 & \citet{Onoue2019} \\                
J2232$+$2930 & 22:32:55.15 & $+$29:30:32.0  & 6.666\phn    \pm   0.004\phn    & \ciimu  & -26.26 & \phn6705_{-810\phn    }^{+810\phn       } & \phn2.02\phn_{-0.31\phn   }^{+0.31\phn   } &      \phn10.40 & \phn6.64 & 0.12 & \citet{Yang2021} \\                 
J2239$+$0207 & 22:39:47.47 & $+$02:07:47.5  & 6.2499       \pm   0.0004       & \ciimu  & -24.60 & \phn4670_{-700\phn    }^{+910\phn       } & \phn0.444   _{-0.08\phn   }^{+0.08\phn   } &   \phn\phn2.29 & \phn1.26 & 0.14 & \citet{Onoue2019} \\                
\enddata 
\end{deluxetable*}

\begin{deluxetable*}{lllLlCCCCCCl}
\rotate
\tablenum{5}
\tabletypesize{\footnotesize} 
\tablecaption{Continued}
\tablehead{
\colhead{ID}                                                     &
\colhead{RA}                                                     &
\colhead{Dec}                                                    &
\colhead{$z_{\textrm{sys}}$}                                     &
\colhead{$z_{\textrm{sys}}$ Method}                              &
\colhead{$\MMCDL$}                                               &
\colhead{$\textrm{FWHM}_\textrm{MgII}$}                          &
\colhead{$3000\textrm{{\AA{}}}\,L_{\lambda,3000\textrm{\AA{}}}$} &
\colhead{$L_\textrm{bol}$}              &
\colhead{$\mbh^\textrm{S+11}$}          &
\colhead{$\Eratio^\textrm{S+11}$}       &
\colhead{Ref.}       \\
\colhead{}         &
\colhead{(J2000)}  &
\colhead{(J2000)}  &
\colhead{}         &
\colhead{}         &
\colhead{(mag)}    &
\colhead{($\kms{}$)}              &
\colhead{($10^{46}\,\ergs{}$)}    &
\colhead{($10^{46}\,\ergs{}$)}  &
\colhead{($10^9\,\msun{}$)}     &
\colhead{}                      &
\colhead{}            
}
\startdata
   P340$-$18 & 22:40:49.00 & $-$18:39:43.8  & 6.0007       \pm   0.002\phn    & \lya    & -26.23 & \phn2468_{-144\phn    }^{+190\phn       } & \phn2.579   _{-0.03\phn   }^{+0.03\phn   } &      \phn13.28 & \phn1.05 & 1.01 & Farina et al.\ (2022) \\            
J2310$+$1855 & 23:10:38.88 & $+$18:55:19.7  & 6.0031       \pm   0.0002       & \ciimu  & -27.22 & \phn3312_{-213\phn    }^{+238\phn       } & \phn7.599   _{-0.086      }^{+0.091      } &      \phn39.13 & \phn3.68 & 0.85 & Farina et al.\ (2022) \\            
J2318$-$3029 & 23:18:33.10 & $-$30:29:33.4  & 6.1456       \pm   0.0002       & \ciimu  & -26.11 & \phn3460_{-160\phn    }^{+177\phn       } & \phn2.82\phn_{-0.042      }^{+0.039      } &      \phn14.53 & \phn2.17 & 0.53 & Farina et al.\ (2022) \\            
J2329$-$0301 & 23:29:08.28 & $-$03:01:58.8  & 6.4164       \pm   0.0008       & \ciimu  & -25.00 & \phn2020_{-110\phn    }^{+110\phn       } & \phn0.68\phn_{-0.08\phn   }^{+0.08\phn   } &   \phn\phn3.50 & \phn0.31 & 0.91 & \citet{Willott2010BH} \\            
J2329$-$0403 & 23:29:14.46 & $-$04:03:24.1  & 5.883\phn    \pm   0.007\phn    & \ciii   & -24.65 & \phn3019_{-1942       }^{+1942          } & \phn0.451   _{-0.007      }^{+0.007      } &   \phn\phn2.32 & \phn0.53 & 0.35 & \citet{Shen2019} \\                 
J2338$+$2143 & 23:38:07.03 & $+$21:43:58.2  & 6.565\phn    \pm   0.009\phn    & \mgii   & -26.00 & \phn2434_{-113\phn    }^{+113\phn       } & \phn1.48\phn_{-0.25\phn   }^{+0.25\phn   } &   \phn\phn7.62 & \phn0.72 & 0.84 & \citet{Yang2021} \\                 
J2348$-$3054 & 23:48:33.34 & $-$30:54:10.2  & 6.9007       \pm   0.0005       & \ciimu  & -25.79 & \phn5182_{-811\phn    }^{+843\phn       } & \phn1.459   _{-0.038      }^{+0.039      } &   \phn\phn7.51 & \phn3.25 & 0.18 & Farina et al.\ (2022) \\            
   P359$-$06 & 23:56:32.45 & $-$06:22:59.3  & 6.1719       \pm   0.0002       & \ciimu  & -26.62 & \phn3825_{-370\phn    }^{+427\phn       } & \phn4.98\phn_{-0.061      }^{+0.057      } &      \phn25.65 & \phn3.78 & 0.54 & Farina et al.\ (2022) \\            
J2356$+$0023 & 23:56:51.58 & $+$00:23:33.3  & 6.044\phn    \pm   0.046\phn    & \mgii   & -24.56 &    10126_{-7758       }^{+7758          } & \phn0.268   _{-0.007      }^{+0.007      } &   \phn\phn1.38 & \phn4.33 & 0.03 & \citet{Shen2019} \                  
\enddata 
\tablecomments{\scriptsize
No iron pseudo continuum has been subtracted in the estimates from \citet{Chehade2018}. 
Measures from \citet{WangFeige2021XRay}, \citet{Andika2020}, and \citet{Matsuoka2019} has been derived from a combination of the \citet{Vestergaard2001} and the \citet{Tsuzuki2006} iron templates, while those from \citet{Willott2010BH} from the \citet{McLure2004} one.
}
\end{deluxetable*}}

\bibliography{bib_xshooter_highz}{}
\bibliographystyle{aasjournal}



\end{document}